\documentclass[]{jfm}

\usepackage{graphicx}
\usepackage{newtxtext}
\usepackage{newtxmath}
\usepackage{natbib}
\usepackage{subfig} 
\usepackage{amsmath} 
\usepackage{bigints}
\usepackage{comment}
\usepackage{hyperref}
\hypersetup{
    colorlinks = true,
    urlcolor   = blue,
    citecolor  = blue,
}

\newcommand{\RomanNumeralCaps}[1]
\linenumbers

\usepackage{siunitx}
\usepackage{booktabs}
\usepackage{hyperref}
\usepackage[toc,page]{appendix}

\title{Blade secondary vortex interaction noise in hovering rotors}

\author{Jordon Won\aff{1}
 \and Seongkyu Lee\aff{1}
 \corresp{\email{skulee@ucdavis.edu}}}

\affiliation{\aff{1}Department of Mechanical and Aerospace Engineering, University of California, Davis, CA 95616, USA}

\begin{document}
\maketitle

\begin{abstract}
Recent indoor experiments on small-scale hovering rotors have reported mid-frequency quasi-tonal peaks of uncertain origin in the residual (aperiodic) far-field sound spectrum, even without wake recirculation. This study investigates their physical origin using hybrid Reynolds-averaged Navier–Stokes/large-eddy simulations coupled with a Ffowcs Williams–Hawkings acoustic analogy, applied to a four-bladed ideally twisted rotor in hover. The high-resolution simulation reproduces the measured higher-harmonic blade-passage-frequency tones, confirming that they arise from the rotor flow itself and not from the enclosed-chamber test environment. Flow-field analysis attributes the tones to \emph{blade secondary vortex interaction} (BSVI): the outboard blade sections, between approximately $r/R = 0.88$ and $0.96$, are repeatedly impinged upon by coherent, S-shaped secondary vortex worms formed by entrainment of the wake shear layer into the primary tip vortices of preceding blades. These aperiodic secondary vortices produce blade-to-blade correlated loading, yielding a quasi-tonal signature distinct from both purely periodic blade-vortex interaction and mostly stochastic blade-wake interaction noise. Spectral proper orthogonal decomposition of the zero-mean upwash confirms that the mid-frequency ($3$--$10$~kHz) content is dominated by a single, low-rank, spatially coherent mode associated with the braids, whereas a lower-frequency mode near $1$~kHz is linked to the primary tip vortex core. BSVI is thus identified as a distinct and previously uncharacterized source of aperiodic tonal noise in hovering rotors, with direct implications for the acoustic design of small- to medium-scale rotorcraft and electric vertical take-off and landing propulsors.
\end{abstract}

\section{Introduction} \label{sec: Introduction}

Aerodynamic noise remains a significant challenge in modern rotor and propeller design, an issue that has become especially prominent with the emergence of electric vertical take-off and landing (eVTOL) aircraft such as small-scale unmanned aerial systems (UAS) and medium-scale urban air mobility (UAM) vehicles. These aircraft carry broad implications across several sectors, including civilian transportation in the form of air taxis \citep{Johnson2018-AirTaxi}, humanitarian aid via delivery drones to remote areas \citep{EmadAlfaris2024}, and defense \citep{Karpowicz2021}. Unlike larger helicopters, eVTOL platforms are dominated primarily by higher-frequency ($f \ge 500$ Hz) turbulence-induced noise, to which humans are more sensitive \citep{Christian&Cabell2017}. Thus, concerns ranging from acoustic pollution in urban centers to reducing the detection radius of military drones have placed rotorcraft aeroacoustics at the forefront of current research, motivating further investigation into these intermittent noise sources.
 
This study identifies and examines one such source in detail: \emph{blade secondary vortex interaction (BSVI)} noise, a previously uncharacterized noise-generation mechanism responsible for non-deterministic acoustic tones in hovering rotors. The following subsection reviews rotor secondary vortices, which form the physical basis of the BSVI noise mechanism. This is followed by a review of previously identified rotor aeroacoustic sources and recent experimental measurements that together highlight the significance and distinct characteristics of BSVI noise.
 
\subsection{Rotor Secondary Vortices}
 
Secondary vortices were first discovered by \citet{Chaderjian2011} through hybrid Reynolds-averaged Navier--Stokes/large-eddy simulations (HRLES) of a $1/4$-scale V-22 Osprey tiltrotor operating in hover. Later, experiments by \citet{Wolf2019} confirmed their existence via particle image velocimetry (PIV) measurements using the ``shake-the-box'' method \citep{Schanz2016}. They investigated a rotor with two untwisted, untapered Black Belt 685 blades and the Align T-Rex 800 model helicopter body, operating under hover conditions at tip Reynolds and Mach numbers of approximately $0.42\times10^6$ and $0.29$, respectively. Coherent vortex braids were identified, revealing distinct vortices with approximately equal distributions of positive and negative rotation. They also noted that the occurrence of these structures was intermittent, reporting no obvious spatial organization despite observing coherent secondary vortices in each temporal snapshot across every revolution. 
 
\begin{figure} 
    \centering
    \subfloat[\label{subfig: Secondary Vortex Diagram}]{\includegraphics[width=0.53\textwidth]{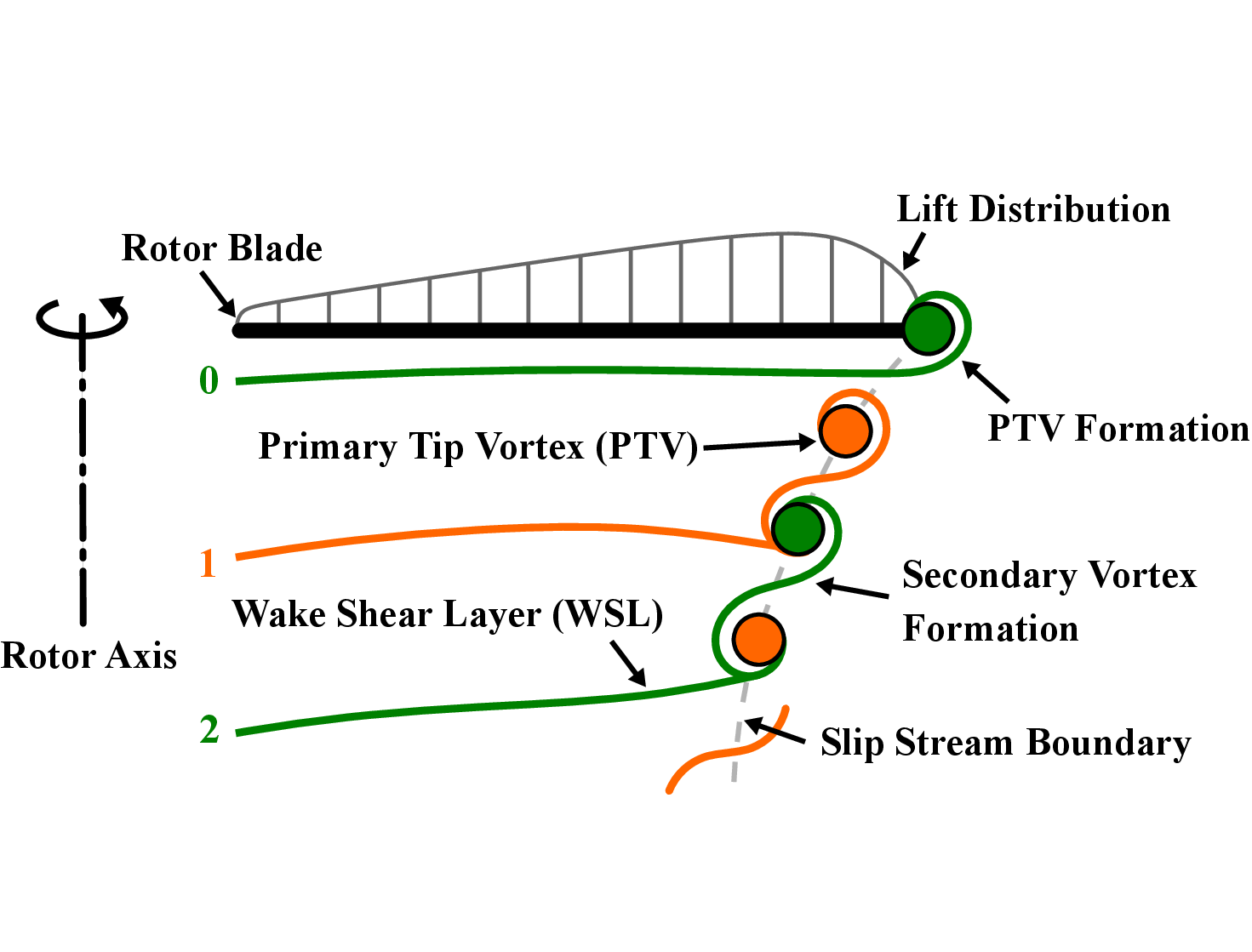}}
    \hspace{0.25 cm}
    \subfloat[\label{subfig: Secondary Vortex Isosurface}]{\includegraphics[width = 0.40\textwidth]{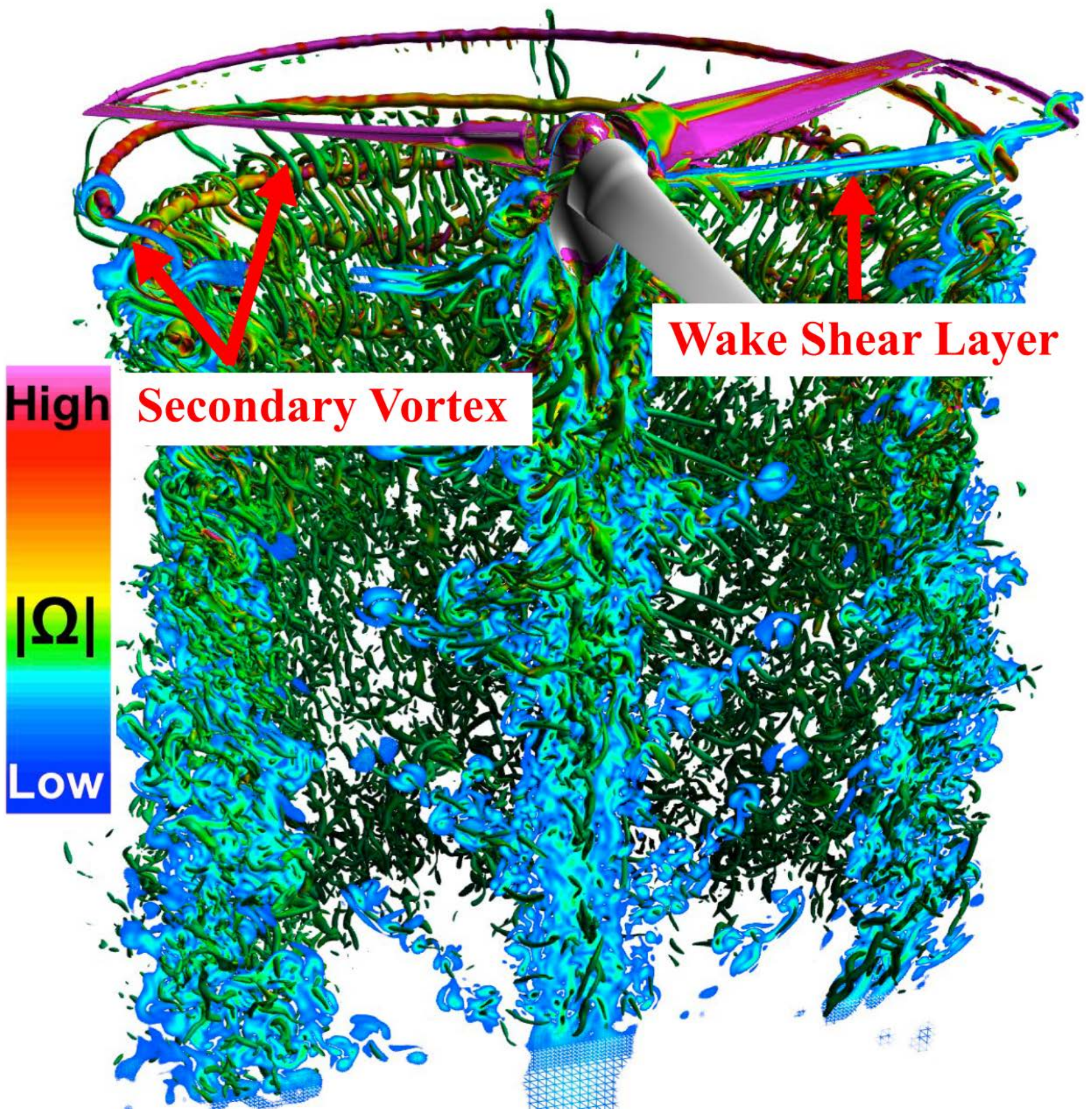}}
    \caption{(a) Conceptual diagram of secondary vortex formation for a two-bladed rotor, modified from \citet{Wolf2019}. (b) An iso-surface of q-criterion colored by vorticity magnitude for the $1/4\text{-scale}$ V-22 tiltrotor, modified from \citet{Chaderjian2011}.}
    \label{fig: Secondary Vortex Examples}
\end{figure}

Figure \ref{subfig: Secondary Vortex Diagram}, modified from \citet{Wolf2019}, sketches the formation of secondary vortices for a hovering two-bladed rotor. Each blade sheds a primary tip vortex (PTV) and a wake shear layer (WSL), which arise from the blade's finite span and the viscous boundary layers, respectively. Each wake structure is labeled corresponding to its blade passage: the $0^{\mathrm{th}}$ passage marks a newly shed WSL and the formation of a PTV, the $1^{\mathrm{st}}$ blade passage indicates the wake structures corresponding to a wake age of $180^\circ$, and so on. For clarity, the colors denote whether the structures originate from blade 1 (green) or blade 2 (orange). Due to the hovering rotor's induced velocity, these wake structures convect downward. However, because the inboard regions of the WSL convect downstream much faster than the blade's PTV, the WSL surpasses and interacts with the previous blade's PTV, entraining into both PTV cores. This entrainment process yields distinct S-shaped secondary vortices that migrate upward toward the rotor disk. A three-dimensional (3-D) rendering from the CFD analysis of \citet{Chaderjian2011} is presented in Fig. \ref{subfig: Secondary Vortex Isosurface} to further illustrate these structures, with a Q-criterion isovalue highlighting small, S-shaped, worm-like tubes.
 
These distinct vortices result from the axial stretching of the WSL, which can be conceptualized as a vortex tube that elongates in the spanwise and vertical directions during its entrainment \citep{Chaderjian2012}. Following the vortex dynamics discussed by \citet{Tennekes&Lumley-1972}, when an incompressible vortex is stretched axially, its cross-sectional area is reduced, thereby intensifying its rotational velocity through conservation of angular momentum. This increased rotation rate enhances the strength and prominence of these structures. In turbulent flows, fluctuating vorticity and strain rates play a central role in this stretching process, explaining why secondary vortices are resolved only by modern scale-resolving simulations. Moreover, \citet{Chaderjian2011} and \citet{Wolf2019} identified similarities between the formation of rotor secondary vortices and the Kelvin--Helmholtz instability mechanisms observed in mixing layers \citep{Bernal&Roshko-1986}, highlighting the complex dynamics of these structures and their broader relevance to aerodynamics research.
 
Since the foundational studies of \citet{Chaderjian2011} and \citet{Wolf2019}, several numerical \citep{Chaderjian2012, Jian2018, Abras2019, Lee&Baeder2021, Bodling&Potsdam-2022, Chaderjian2023, Bodling2024} and experimental \citep{Schwarz2022, Heintz2025} investigations have prioritized the investigation of secondary vortex formation. \citet{Bodling&Potsdam-2022} performed CFD simulations to analyze the effect of various numerical settings on the development of secondary vortices, and quantitatively assessed how temporal convergence (i.e., the subiteration convergence associated with each physical time step) influenced the number of detected secondary vortices. Investigating the same rotor and helicopter model as \citet{Wolf2019}, they found that the number of secondary vortices converged with increasing subiterations. Thus, although early studies raised concerns about the sensitivity of the secondary vortex instability---prompting speculation that the structures were numerical artifacts---the conclusive findings of \citet{Bodling&Potsdam-2022} demonstrated that a consistent solution emerges.
 
 
Following these findings and adopting a similar numerical approach, \citet{Schwarz2022} conducted both HRLES and experimental PIV analysis. The key finding was that the simulations and measurements agreed on the average number of detected secondary vortices immediately below the rotor, between $z/R = 0.1$--$0.6$, where $R$ is the rotor radius. This quantitative agreement firmly established that secondary vortices are physical phenomena rather than numerical artifacts, and that HRLES retains sufficient fidelity to accurately predict the secondary vortex instability. However, an underprediction of the number of smaller secondary vortices, associated with reduced swirl velocity and circulation, indicated that the CFD grid resolution was too coarse to resolve the smallest structures. \citet{Bodling2024} subsequently increased the grid resolution, achieving markedly improved agreement in the predicted number of small secondary vortices. Although it should be noted that discrepancies concerning rotor wake breakdown---hypothesized to be influenced by vortex pairing and interactions with secondary vortex structures \citep{Jian2018, Abras2019, Bodling2024}---remained and were attributed to experimental uncertainties such as blade-to-blade pitch offsets.
 
Recently, \citet{Heintz2025} addressed this speculation through an extensive parametric experimental study on the same rotor investigated in the aforementioned studies \citep{Wolf2019, Schwarz2022}. Additionally, they also examined a similar-sized rotor with twist and a cambered airfoil, contrasting the untwisted, symmetric airfoil rotor blades investigated previously. \citet{Heintz2025} varied the number of blades, rotation rate, blade pitch offset, and collective pitch. Importantly, they found that the blade passage frequency (BPF) is linearly related to the number of secondary vortices and influences vortex decay. However, blade-to-blade pitch offset and vortex pairing did not appear to significantly influence wake breakdown. These new findings, which contradict earlier hypotheses, underscore that the behavior of rotor secondary vortices is not yet fully understood and warrants further investigation.

\subsection{Rotor Aeroacoustics}
 
It is now the prevailing consensus that secondary vortices are a genuine and notable feature of rotorcraft wakes; however, their aeroacoustic role has not yet been established or characterized in detail. Accordingly, noise generated by blade interactions with these intermittent vortex structures has not been thoroughly investigated, and the literature remains limited. This subsection therefore reviews previously identified flow-induced rotor noise mechanisms that are relevant to yet notably different from BSVI. Indeed, these differences ultimately motivate the present investigation of BSVI as a distinct source of stochastic hovering-rotor noise.
 
 
\citet{paterson&amiet1979} provide a summary of the early investigations of rotor turbulent ingestion (TI) noise. The investigation by \citet{SHARLAND1964} is generally regarded as the first study of rotor noise generated by interactions with ingested turbulence, with subsequent experiments revealing that its acoustic signature is characterized by BPF-aligned humps, referred to as ``haystacks'' \citep{kazin&Volk1971, Sevik1974}. However, it was not until the seminal work of \citet{Hanson1974} that TI noise and the haystacking phenomenon were better understood. \citet{Hanson1974} conducted outdoor experiments with an aircraft engine fan to characterize ingested atmospheric turbulence, while simultaneously formulating a theoretical model to predict the noise it generates. Anisotropy was found to be a critical feature of atmospheric TI noise: ingested eddies are axially stretched as they convect into the rotor, yielding eddies with large streamwise-to-transverse integral length-scale ratios. This elongation produces blade-to-blade correlation, as the same eddy is chopped by multiple blades, explaining why TI produces haystacks aligned with BPF harmonics rather than a completely smooth broadband spectrum.
 
Later, \citet{paterson&amiet1979} conducted both indoor and outdoor experiments on an open rotor, in which the outdoor experiments under hover conditions were subject to large-scale anisotropic turbulent structures. The acoustic measurements for this case revealed quasi-tonal peaks (narrowband peaks caused by intermittent rather than periodic or steady loading) aligned with the first 25 BPF harmonics. These higher-frequency tones dominated the aeroacoustic signature of their model rotor, leading to the conclusion that a hovering rotor's interactions with turbulence can constitute a major source of flow-induced noise. \citet{paterson&amiet1979} also developed a theoretical model to predict this noise. The model correctly predicted BPF-aligned peaks, demonstrating good qualitative agreement with the measurements, but also exhibited a mismatch with the measured sound pressure level (SPL). This discrepancy was attributed to the model's neglect of the anisotropy in the turbulence field encountered by each blade. More recent work on TI noise includes \citet{Murray2018, Piccolo2025, Raposo&Azarpeyvand-2024}. The study by \citet{Murray2018} is of particular relevance to the current work, as they experimentally showed that blade interactions with coherent vortices embedded in planar boundary-layer turbulence produced sharper quasi-tones, in contrast to the smooth broadband haystacks exhibited by configurations without such vortices.
 
In addition to TI, interactions between a rotor blade and its own wake-generated turbulence---commonly referred to as blade-wake interaction (BWI) noise---constitute another stochastic aeroacoustic source. Early work by \citet{Brooksetal1987} identified BWI noise as a distinct mechanism that produces a smooth broadband acoustic signature. They showed that BWI noise is most pronounced in the vicinity of perpendicular blade-vortex interaction (BVI) events, while remaining acoustically distinct from the impulsive loading and tonal noise associated with parallel BVI. Motivated by this broadband noise mechanism, and by the analytical work of \citet{AMIET1975} on the noise produced by airfoils subjected to a turbulent upwash spectrum, \citet{Glegg1991} and \citet{Glegg_etal_1999} proposed a semi-analytical framework to model the noise produced by a rotor blade undergoing perpendicular BVI. More recent investigations of BWI noise include the numerical studies of \citet{Thurman&Baeder2023} and \citet{Ghimire&Li-2026}, as well as modifications to Glegg's model proposed by \citet{Naseer-etal-2026}.
 
\begin{figure}
    \centering
    \subfloat[\label{subfig: ITR Sound Spectrum}]{\includegraphics[width = 0.435\textwidth]{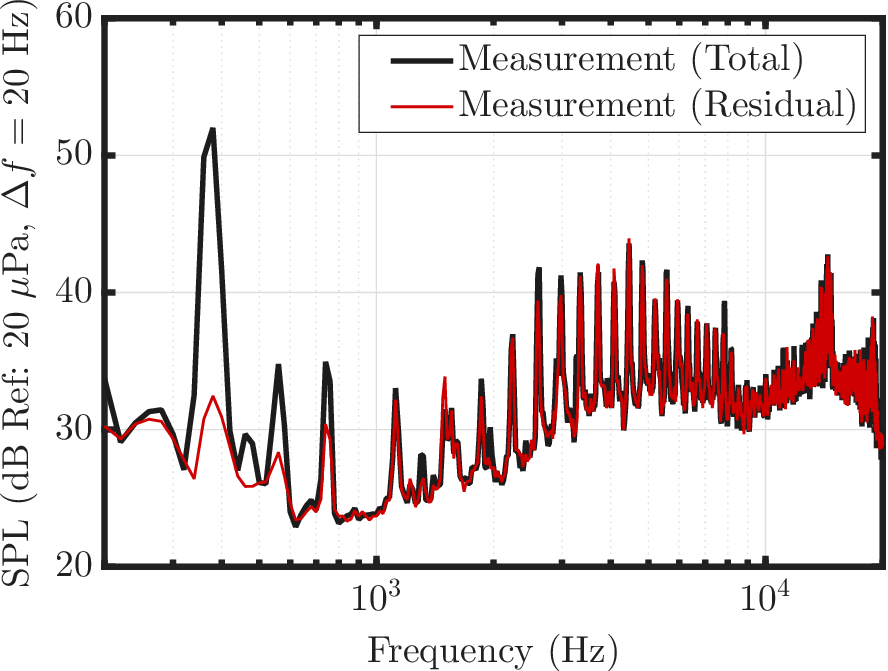}}
    \hspace{0.5 cm}
    \subfloat[\label{subfig: ITR Spectrogram}]{\includegraphics[width = 0.515\textwidth]{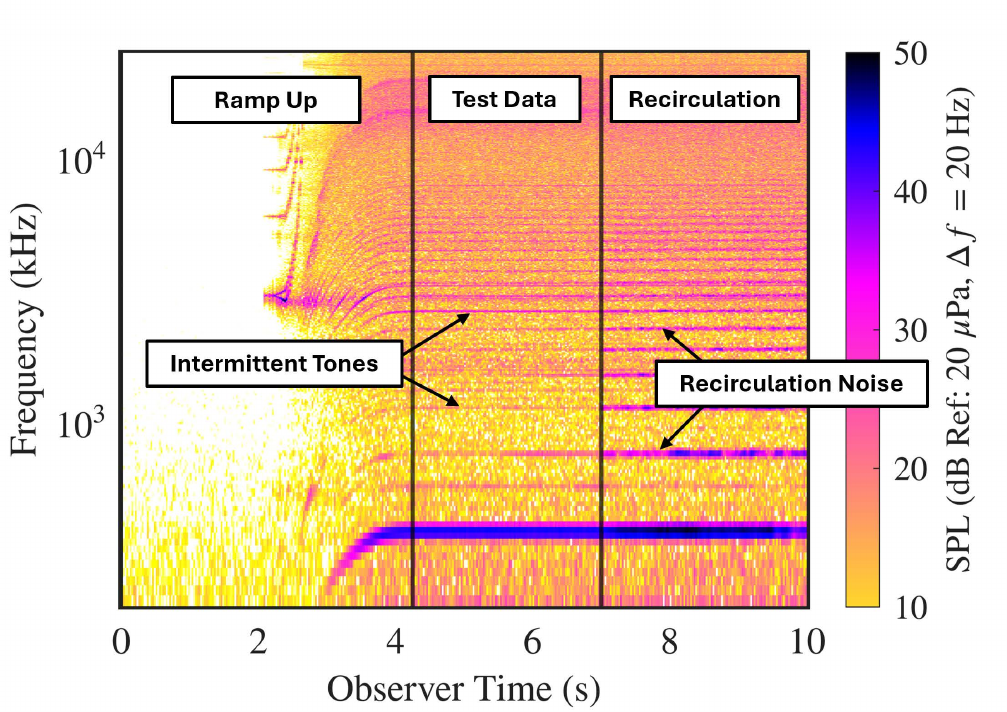}}
    \caption{Measurement data of a small-scale ITR from \citet{pettingill2021}: (a) the far-field sound spectrum of the total and residual signals and (b) the far-field acoustic spectrogram.}
    \label{fig: ITR Experiments}
\end{figure}

Beyond these previously identified stochastic aeroacoustic mechanisms, recent studies on indoor experiments of hovering rotors have observed unexpected BPF-aligned quasi-tonal peaks in far-field acoustic measurements \citep{pettingill2019-SUI, pettingill2021, Zawodny2023, Tinney&Valdez2025, Santamaria2026Serration, Brooks2026RapidAcoustic}. For instance, the sound spectrum provided by \citet{pettingill2021} in Fig. \ref{subfig: ITR Sound Spectrum} reveals that mid-frequency quasi-tones dominate the residual (total minus phase-averaged) acoustic signal for an ideally twisted rotor (ITR), despite the test being conducted in an enclosed anechoic chamber with static air. While such quasi-tones might be expected due to wake recirculation in such environments, \citet{pettingill2021} excluded this possibility using the approach of \citet{Weitsman2020}. Moreover, the spectrogram in Fig. \ref{subfig: ITR Spectrogram} shows that significant intermittent tones are observed even before the rotor's wake has time to recirculate.
 
HRLES computations by \citet{Thurman2024} and \citet{Won&Lee-2026-BSVI} were conducted on the same small-scale ITR, omitting the enclosed anechoic chamber walls from their simulations. Both groups successfully predicted these peaks, further demonstrating that the higher-harmonic noise in Fig. \ref{fig: ITR Experiments} is not an experimental artifact. \citet{Won&Lee-2026-BSVI} hypothesized that the repeated impingement of coherent secondary vortices may be responsible for the quasi-tonal noise observed in the aforementioned small-scale hovering-rotor experiments. This characterization of BSVI as a dominant source of mid-frequency quasi-tones forms the basis of the present investigation, which seeks to elucidate the BSVI noise mechanism and assess its distinction from other thoroughly studied sources. 

Previous investigations have attributed secondary-vortex-related structures to mid-frequency noise within the broader framework of BWI noise, including works by \citet{Thurman&Baeder2023, Casalino2023, Thurman2024, Thurman2025}. However, we believe that this interpretation is incomplete, as characteristic differences---in both the organization of the physical wake structures and their resulting acoustic signature---indicate that BSVI represents a separate mechanism rather than simply a subset of BWI. Attributing these mid-frequency tones broadly to BWI can obscure the underlying source physics and may have potential implications for future modeling frameworks. Importantly, \citet{Won&Lee-2026-BSVI} also showed that the mid-frequency quasi-tonal signature can persist without a strong preceding perpendicular BVI, further distinguishing BSVI from conventional BWI interpretations. 
 
\begin{figure}
    \centering
    {\includegraphics[width = 0.55\textwidth]{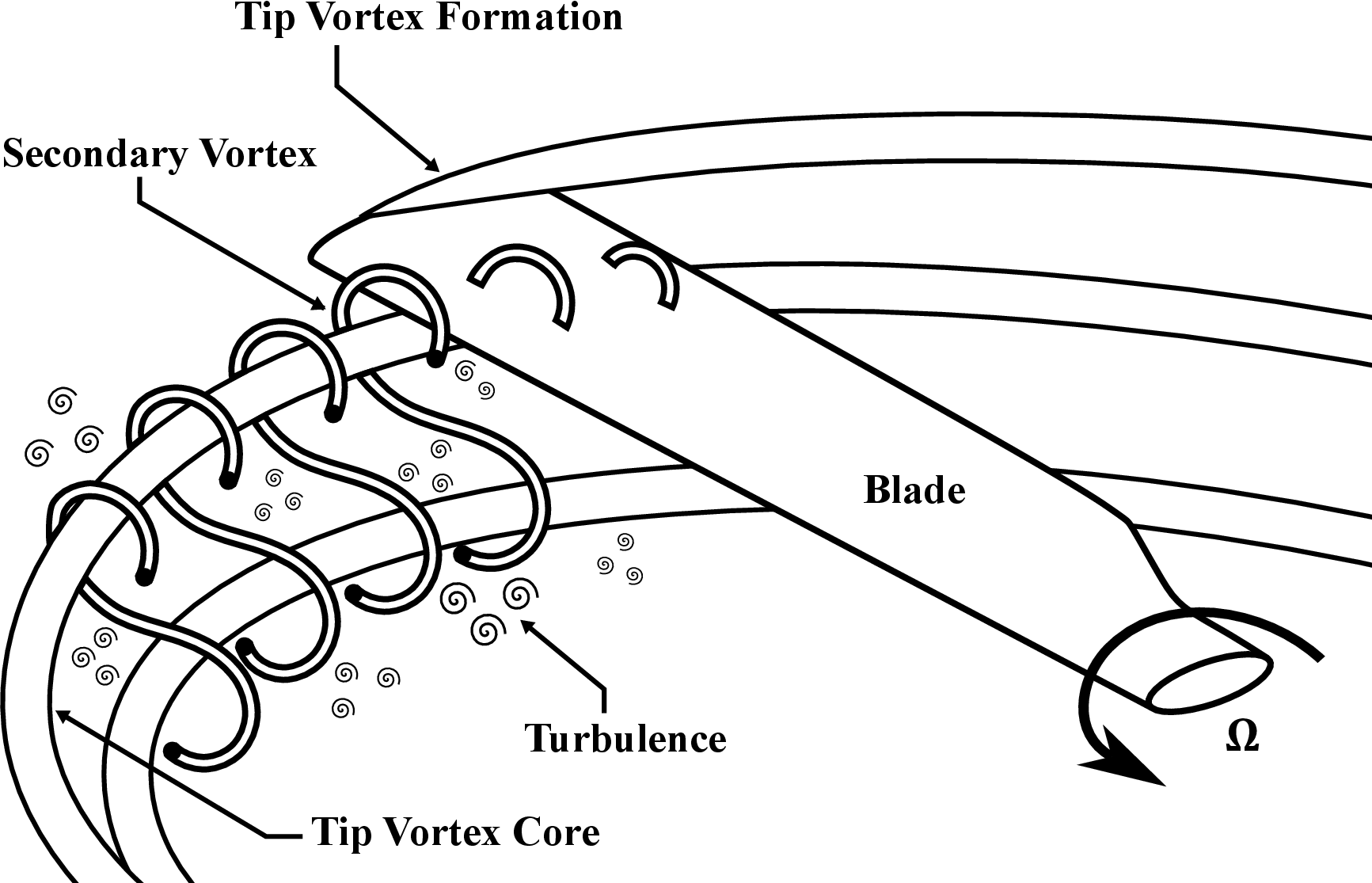}}
    \caption{Conceptual diagram of blade secondary vortex interaction (BSVI) noise.}
    \label{fig: BSVI Conceptual Diagram}
\end{figure}

Figure \ref{fig: BSVI Conceptual Diagram} provides a conceptual diagram of BSVI. In an idealized sense, if secondary vortices were identical and uniformly spaced, blade rotation would produce periodic loading through the successive chopping of discrete vortex structures. In practice, however, secondary vortex production is stochastic, such that these structures vary in both spatial distribution and strength, giving rise to the intermittent tonal behavior observed in the spectrogram shown in Fig.~\ref{subfig: ITR Spectrogram}. BSVI can also be interpreted as a localized recirculation process, wherein the WSL is reingested by the rotor disk through the upward migration of coherent secondary vortices, affecting only the outboard sections of the rotor blades. This perspective is consistent with why these quasi-tones resemble noise generated by wake recirculation \citep{Weitsman2020}; in particular, with previous numerical findings by \citet{Nardari2019}. 

The remainder of this paper is organized as follows. Section~\ref{sec: Methodology} describes the numerical setup for the HRLES computations, far-field acoustic predictions, and Spectral Proper Orthogonal Decomposition (SPOD) \citep{Towne2018} analysis used in this investigation. Results are presented in Section~\ref{sec: Results}, which is subdivided into three parts: (1) aerodynamics and flow visualization, (2) far-field acoustic prediction and validation against experiments, and (3) flow statistics and SPOD analysis. Finally, Section~\ref{sec: Conclusions} provides concluding remarks and summaries, discussing the broader implications for rotorcraft noise prediction and mitigation.


\section{Methodology} \label{sec: Methodology}

\subsection{Geometry and Microphone Locations}

The small-scale ITR studied by \citet{pettingill2021} is investigated numerically with HRLES. The ITR contains four blades with a radius $R = 0.1588$ m. The chord length $C_\mathrm{tip}$ is a constant 0.0318 m (20\% radius), using the NACA 0012 airfoil along its entire span. Note that the trailing-edge bluntness is $0.50$ mm. The blade's radial twist distribution is given by

\begin{equation}
    \theta(r) = \theta_\text{tip} \frac{R}{r},
\end{equation}

\noindent where the pitch at the blade tip is $\theta_\mathrm{tip} = 6.9^\circ$. \citet{pettingill2021} evaluated multiple operating conditions, including different collective pitch settings and rotation rates. In the current numerical analysis, only the baseline case operating at $\Omega =576~\mathrm{rad/s}$ with no increase in collective pitch is studied. This operating condition corresponds to tip Reynolds and Mach numbers of $Re_\mathrm{tip} = 0.2\times10^6$ and $M_\mathrm{tip} = 0.267$, respectively, where the speed of sound, fluid density, and dynamic viscosity are $340~\mathrm{m/s}$, $1.225~\mathrm{kg/m^3}$, and $1.789\times10^{-5}~\mathrm{Pa\cdot s}$, respectively.

\begin{figure}
    \centering
    \includegraphics[width=0.95\linewidth]{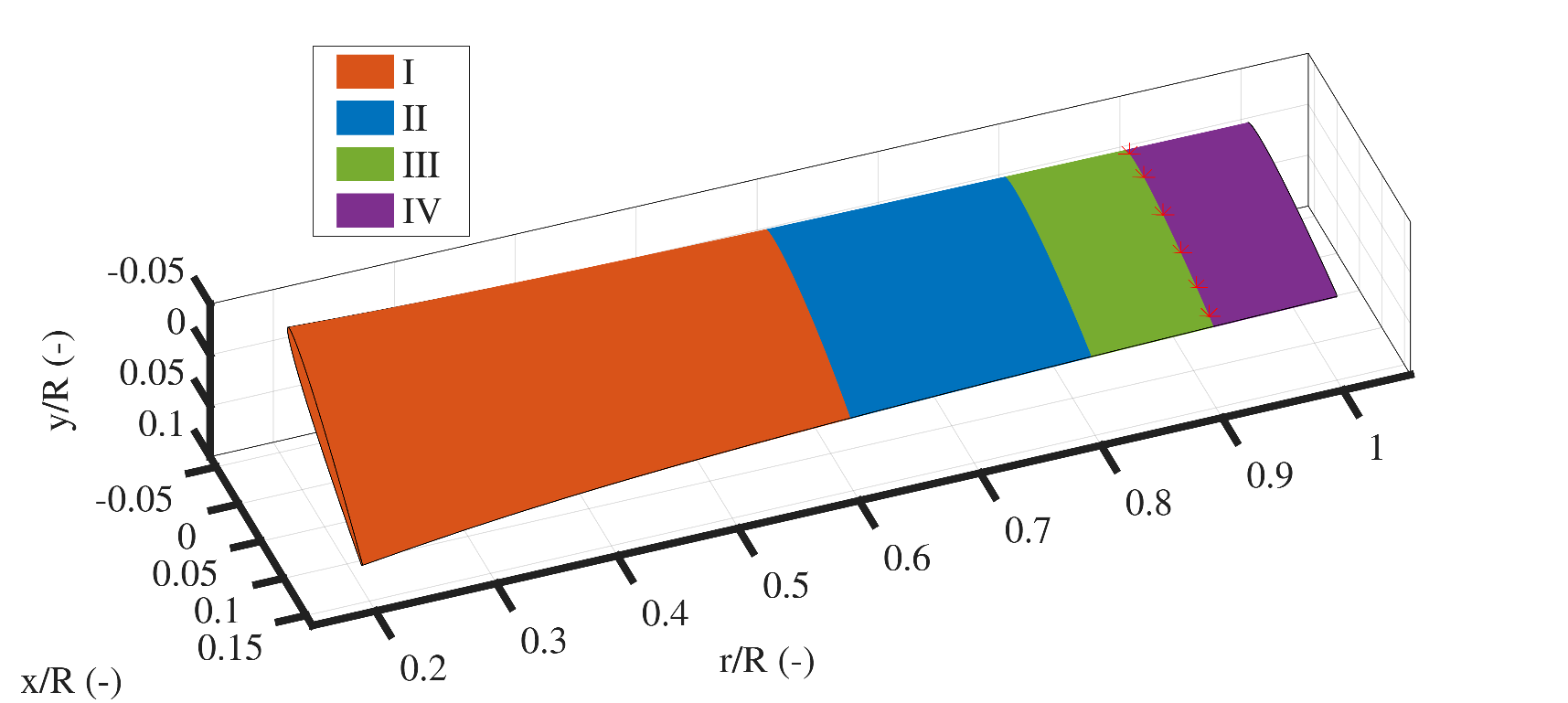}
    \caption{ITR blade geometry partitioned into four spanwise sections for acoustic analysis with probes highlighted along the upper surface at the $r/R=0.90$ location.}
    \label{fig: ITR Geometry Blade}
\end{figure}

The geometric model for a single blade is shown in Fig. \ref{fig: ITR Geometry Blade}. The coordinate system in this figure is defined such that the $x$-direction is streamwise, the $y$-direction points downward, and the radial direction is relative to the rotor hub. The origin is aligned with the blade's $1/4$-chord, and the modeled geometry omits the rotor hub and extended root section; as a result, the blade geometry begins at $r/R = 0.2034$. The blade is further partitioned into four sections, labeled I, II, III, and IV, corresponding to $r/R = 0.2034$--$0.60$, $0.60$--$0.80$, $0.80$--$0.90$, and $0.90$--$1.00$, respectively. An acoustic comparison between these sections is presented later to evaluate their relative contributions to the overall blade noise signature. In addition, multiple probes are placed along the spanwise location $r/R = 0.90$, at normalized chord locations of $x/c = [0.05,\ 0.20,\ 0.40,\ 0.60,\ 0.80,\ 0.95]$ on the upper surface, to compute the wall-pressure spectrum (WPS).

\begin{figure}
    \centering
    \subfloat[\label{subfig: ITR Geometry Rotor}]{\includegraphics[width = 0.52\textwidth]{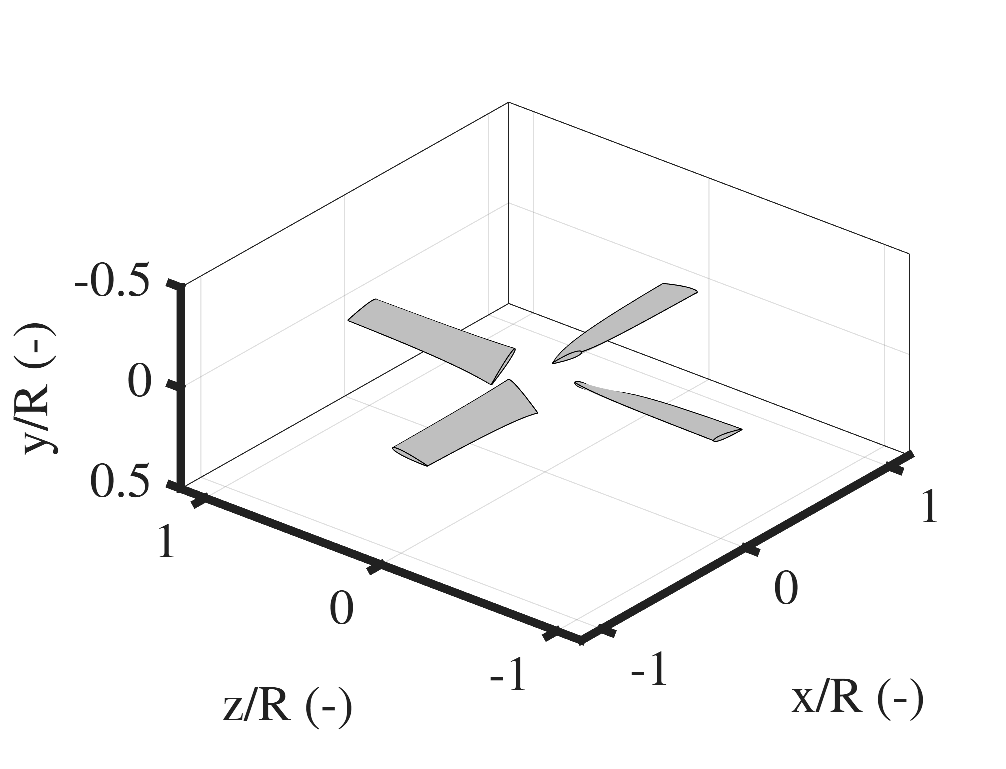}}
    \hspace{0.25 cm}
    \subfloat[\label{subfig: ITR Experimental Setup}]{\includegraphics[width = 0.45\textwidth]{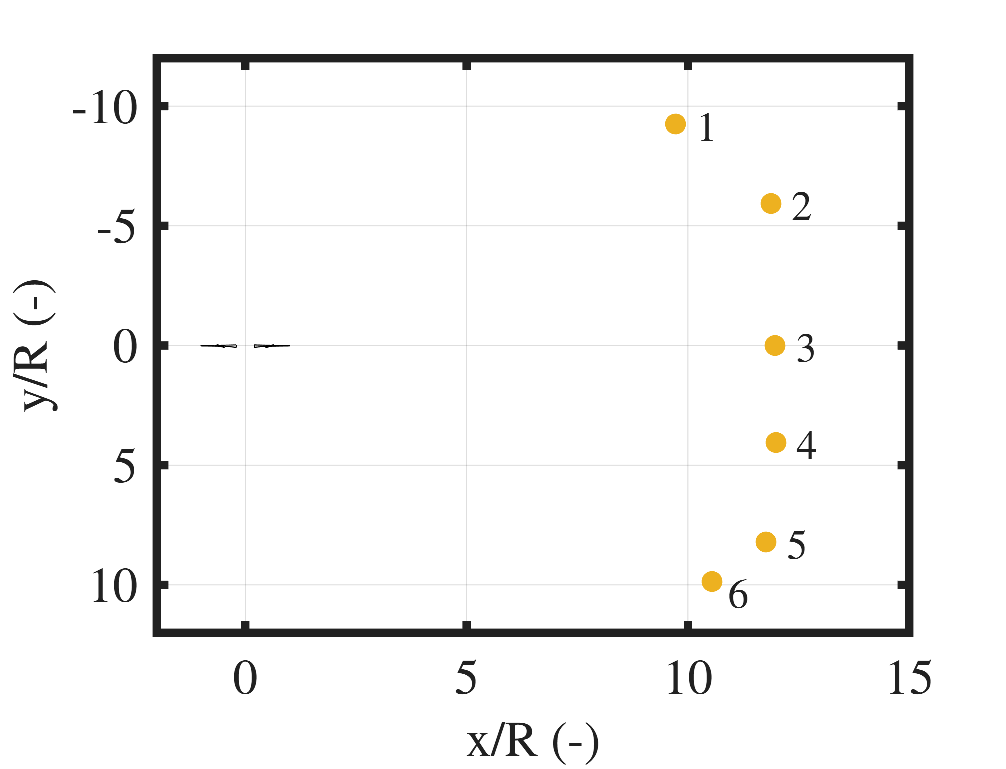}}
    \caption{(a) The modeled rotor geometry and (b) the six far-field microphone locations in global coordinates.}
    \label{fig: ITR Setup}
\end{figure}

Figure \ref{subfig: ITR Geometry Rotor} displays the geometry of the entire rotor used in this study. The motor, rotor hub, and the extended root section connecting the hub are not included in the current numerical analysis. As a result, the mechanical vibrations and aerodynamic noise generated by these components during the experiments are not captured in the simulations. These noise sources are assumed to be minor in comparison to the flow-induced noise at outboard sections of the rotor experiencing much greater Mach numbers; the validity of this assumption is evaluated later in Section \ref{sec: Results}. Figure \ref{subfig: ITR Experimental Setup} shows the six observer locations at which the measurements were obtained, with the exact coordinates listed in Table \ref{tab: microphone_locations}. All microphone locations are at least $10R$ from the rotor hub and are considered sufficiently far away that they are not influenced by hydrodynamic pressure fluctuations. The experiments of \citet{pettingill2021} used B\&K Type 4939 microphones with a baseline $\pm 2$~dB uncertainty, to which a further $\pm 1$~dB is added from post-processing of the experimental data with Welch's paradigm \citep{welch1967, Schmidt2019}. Therefore, all measurement data shown in the current study carry a $\pm 3$~dB uncertainty bound.

\begin{table} 
    \centering
    \small
    \begin{tabular}{ 
        c
        S[table-format=2.5]
        S[table-format=2.5]
        S[table-format=-2.5]
        S[table-format=2.5]
    }
        \toprule
        {Mic.} &
        {$d$ (\si{\meter})} &
        {$x$ (\si{\meter})} &
        {$y$ (\si{\meter})} &
        {$z$ (\si{\meter})} \\
        \midrule
        1 & 2.1240 & 1.5393 & -1.4638 & 0.0000 \\
        2 & 2.1020 & 1.8800 & -0.9398 & 0.0000 \\
        3 & 1.8960 & 1.8956 & -0.0000 & 0.0000 \\
        4 & 2.0040 & 1.8981 &  0.6414 & 0.0000 \\
        5 & 2.2730 & 1.8638 &  1.3016 & 0.0000 \\
        6 & 2.2860 & 1.6692 &  1.5620 & 0.0000 \\
        \bottomrule
    \end{tabular} 
    \caption{Far-field microphone locations for the ITR experiments \citep{pettingill2021}.}
\end{table} \label{tab: microphone_locations}

\subsection{Computational Grid}

\begin{figure} 
    \centering
    \subfloat[\label{subfig: ITR NB Geometry}]{\includegraphics[width = 0.45\textwidth]{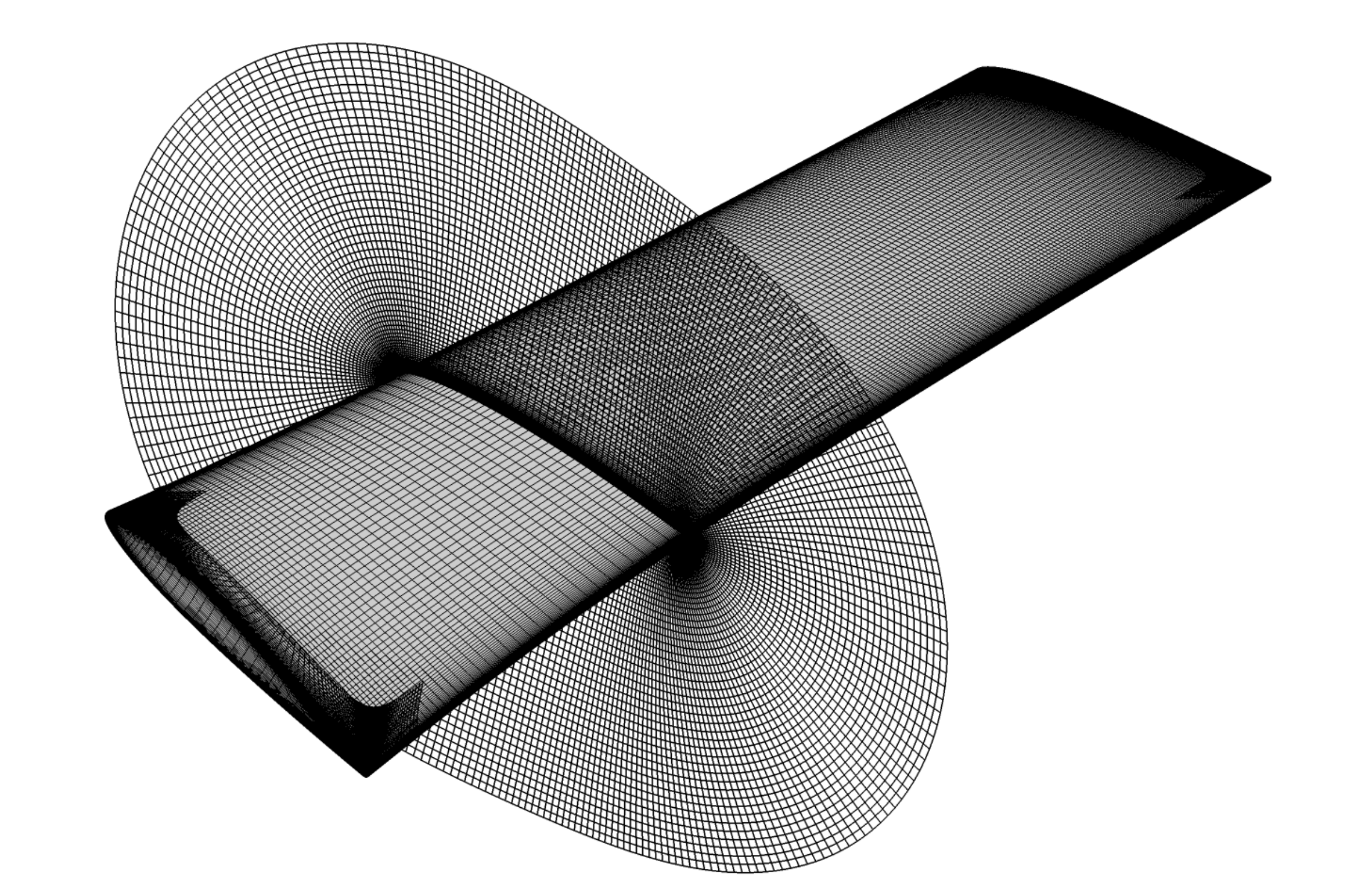}}
    \hspace{0.25 cm}
    \subfloat[\label{subfig: ITR O-mesh}]{\includegraphics[width = 0.45\textwidth]{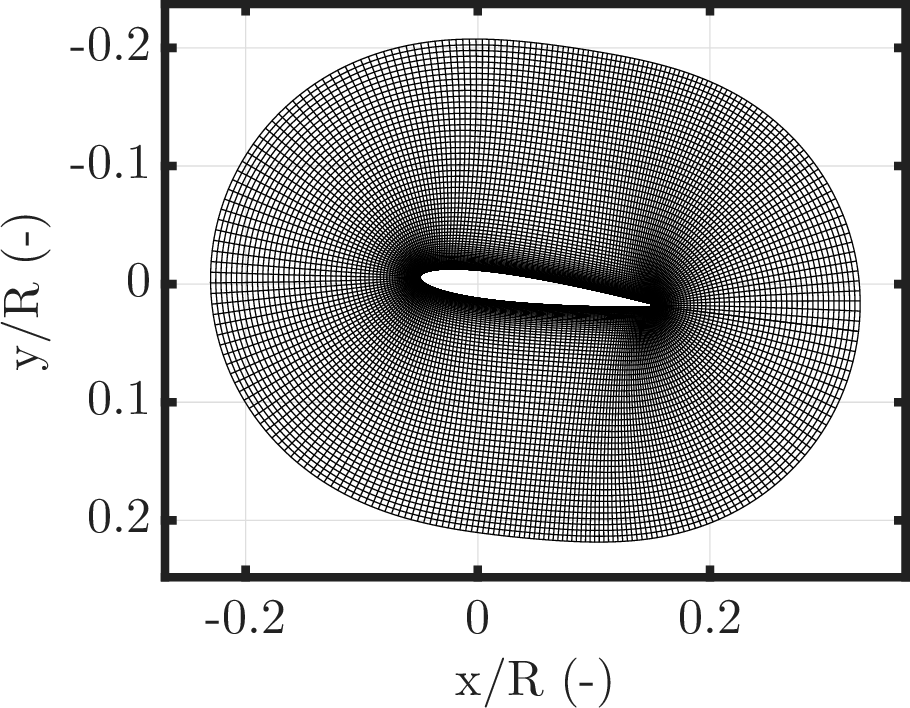}}
    \caption{NB grids: (a) the primary, root cap, and tip cap grids and (b) the O-mesh cross-section of the primary grid.}
    \label{fig: ITR NB Grids}
\end{figure}

The CFD software adopted here is NASA's OVERFLOW 2.4c \citep{OVERFLOW}, with structured curvilinear grids generated using OVERGRID within the Chimera Grid Tools v2.1 software package \citep{CHAN2009}. Each blade consists of three separate grids: a primary blade grid, a root cap grid, and a tip cap grid, shown in Fig. \ref{subfig: ITR NB Geometry}. The primary blade grid is an O-mesh containing $257\times239\times99$ grid points in the chordwise, spanwise, and wall-normal directions, respectively. Owing to their aerodynamic and acoustic significance, the spanwise sections between $r/R = 0.65$--$1.0$ use a maximum grid spacing of $2.5\%\, C_{\mathrm{tip}}$ ($7.94\times10^{-4}~\mathrm{m}$) in all directions; this resolution is essential to accurately resolve secondary vortices \citep{Chaderjian2023, Bodling2024} and stochastic flow-induced noise \citep{Thurman2024}. The root and tip cap grids are included to produce a fully enclosed mesh.

After each component's surface grid is generated, the volume grid is constructed through hyperbolic marching in the wall-normal direction \citep{CHAN2009}. A cross-section of the primary blade grid is shown in Fig. \ref{subfig: ITR O-mesh}, in which the wall-normal volume comprises three layers: a uniform wall layer, a stretching layer, and a uniform outer layer. The wall layer comprises five grid points with a uniform spacing of $1.1 \times 10^{-6}$~m, corresponding to $y^{+} = 0.33$ based on the tip velocity ($\Omega R$). The stretching layer extends a distance of $50\%\,C_{\mathrm{tip}}$, with the grid spacing increasing at a maximum stretching ratio of $1.149$ to a final spacing of $2.5\%\,C_{\mathrm{tip}}$. The outer layer then extends an additional $50\%\,C_{\mathrm{tip}}$ with a uniform grid spacing of $2.5\%\,C_{\mathrm{tip}}$. Because OVERFLOW is formulated for overlapping, overset grids, hole cutting and grid-to-grid interpolation are performed within this outer layer. Together, the primary blade, root cap, and tip cap grids form a complete near-body (NB) mesh for one blade comprising approximately 12 million grid points, so the entire four-bladed rotor mesh contains a total of 36 million points.

\begin{figure} 
    \centering
    {\includegraphics[width = 0.85\textwidth]{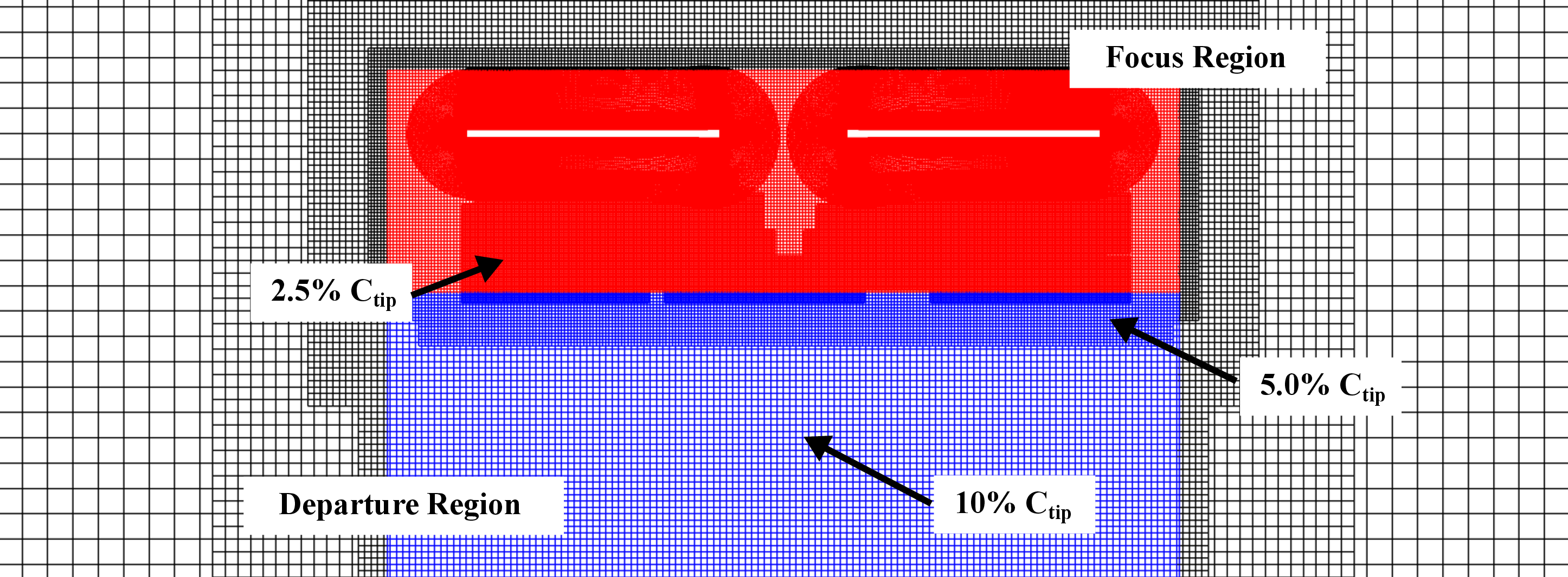}}
    \caption{OB Cartesian grids with the focus region highlighted in red (AMR included) and the departure region highlighted in blue.}
    \label{fig: ITR OB Grids}
\end{figure}

Figure \ref{fig: ITR OB Grids} displays a cross-section of the off-body (OB) Cartesian grids, which are generated automatically by OVERFLOW. The region highlighted in red is referred to as the focus region, in which adaptive mesh refinement (AMR) \citep{Chaderjian2012DetachedES} is used to achieve a $2.5\%\,C_{\mathrm{tip}}$ grid spacing from a baseline spacing of $5.0\%\,C_{\mathrm{tip}}$. This focus region extends $0.50R$ below the rotor, $0.20R$ above it, and $1.25R$ radially outward to accurately resolve near-wake flow structures that can significantly influence the aeroacoustic signature of this relatively low-Mach-number rotor. While \citet{Chaderjian2012} demonstrated that a $5.0\%\,C_{\mathrm{tip}}$ spacing is sufficient to resolve the formation of secondary vortices, subsequent work showed that $2.5\%\,C_{\mathrm{tip}}$ provides far better predictions of PTV diameter \citep{Chaderjian2023} and the average number of secondary vortices that form, with \citet{Bodling2024} reporting excellent agreement between experimental measurements of the number of small vortex structures and HRLES results at a $3.0\%\,C_{\mathrm{tip}}$ refinement level.

Below the focus region, Fig. \ref{fig: ITR OB Grids} highlights the departure region in blue, extending $3.0R$ below the rotor. This region uses a uniform grid spacing of $10.0\%~C_{\mathrm{tip}}$, as far-wake structures are not expected to significantly affect the rotor's aeroacoustic signature; at the present low Mach number, volume sources are negligible \citep{Curle1955}, and the far wake is unlikely to appreciably affect blade loading. \citet{Chaderjian2023} demonstrated this latter point directly, showing that refinements in the rotor's far wake yielded negligible differences in the predicted thrust and torque fluctuations. Outside the focus and departure regions, the OB Cartesian grids are coarsened by a factor of $2$, with each refinement level containing at least 8 grid points. Note that this coarsening dissipates any acoustic waves that might otherwise reflect off the domain boundaries. The full computational domain extends $20R$ in all directions to prevent wake recirculation, which can be an issue for hovering rotors within enclosed testing environments \citep{Weitsman2020, Nardari2019, Casalino2023}. The total grid resolution of the completed simulation is approximately 72 million grid points.

\subsection{Numerical Setup}

NASA's OVERFLOW 2.4c is a compressible Navier--Stokes solver formulated for structured overset grids. Overset grids accommodate complex motions with six degrees of freedom, making them well suited to rotorcraft aerodynamic and aeroacoustic predictions. The HRLES approach employed in this study is the delayed detached-eddy simulation (DDES) formulation of \citet{Gritskevich2012}, built on the two-equation $k$--$\omega$ SST model with a rotation/curvature correction. The SST model provides closure within the RANS boundary layer while also acting as a subgrid-scale model in the LES regions of the domain. OVERFLOW's DDES implementation has been well validated for rotorcraft aerodynamics \citep{Chaderjian2023}, notably resolving the secondary vortex structures \citep{Schwarz2022, Bodling2024} of particular interest here. Further validation, with a focus on stochastic aeroacoustic sources, is presented later in Section \ref{sec: Results}.

OVERFLOW supports a variety of high-order central and upwind discretization schemes. In this work, viscous fluxes are discretized using a second-order central-differencing scheme, while inviscid fluxes are treated with a fifth-order mapped WENO (WENOM) interpolation \citep{Nichols2007} paired with the HLLE++ approximate Riemann solver \citep{Tramel2009}. \citet{Chaderjian2023} demonstrated that this upwind approach for inviscid fluxes yields reduced numerical dissipation relative to earlier results obtained with high-order central-differencing schemes and artificial viscosity. Specifically, compared with previous predictions at the same $2.5\%\,C_{\mathrm{tip}}$ grid resolution using a sixth-order central-differencing scheme \citep{Chaderjian2012}, the fifth-order WENOM/HLLE++ pairing predicted a smaller and more accurate PTV core diameter across multiple wake ages. Furthermore, \citet{Thurman2024_Aeroacoustics, Thurman2025} demonstrated the importance of these reduced-dissipation methods for predicting stochastic rotorcraft noise sources, in comparisons against alternative CFD codes and lower-order schemes.

Time integration is performed using a dual-time-stepping approach, in which each physical time step is advanced with an optimized second-order backward-difference scheme. Between physical time steps, pseudo-time steps (sub-iterations) are used to reduce the $L_2$-norm residuals by at least two orders of magnitude (OOM), after which the solution advances to the next physical time step. This approach is consistent with the findings of \citet{Bodling&Potsdam-2022}, who demonstrated that the average number of secondary vortex structures converges with this OOM drop. The improved symmetric successive over-relaxation solver of \citet{Derlaga2020} is adopted for its superior convergence statistics, requiring only $12$--$15$ sub-iterations to achieve this $2.0$ OOM reduction criterion.

The hovering-rotor simulation is conducted in two distinct phases: a ``quick-start'' procedure and a ``time-accurate'' phase. The quick-start procedure is run for 30 revolutions using a relatively coarse time step of $7.58\times 10^{-5}~\mathrm{s}$, corresponding to a blade azimuthal change of $\Delta\Psi=2.50^\circ$ per time step. This lengthy procedure bypasses the flow-development phase associated with an impulsive start (i.e., from a static-airflow initial condition to the rotor immediately rotating at its operating speed). Following this procedure, the temporal resolution is refined to $\Delta\Psi=0.25^\circ$ per time step for the time-accurate phase. A total of 25 time-accurate revolutions are simulated, with the first 5 omitted from the analysis to avoid numerical transients caused by the coarse-to-fine time-step transition. Accordingly, all results presented in this paper are sampled from the final 20 revolutions. With the target grid spacing ($\Delta s=7.9\times 10^{-4}~\mathrm{m}$) that dominates the focus region and the time-accurate time step ($\Delta t=7.58\times10^{-6}~\mathrm{s}$), the Courant number with respect to tip speed ($\Omega R = 91.5~\mathrm{m/s}$) is $\mathrm{CFL} \approx 0.878$. 

The computational cost is approximately $8{,}240$ CPU hours per time-accurate revolution (CPU~h/rev). The simulation was conducted on the UC Davis College of Engineering HPC2 cluster using eight nodes, each equipped with AMD EPYC 7532 64-core processors and 256 GB of memory. AMR is performed, and surface pressures from all NB grids are saved every $\Delta\Psi=0.50^\circ$, corresponding to a sampling frequency of $66.0$ kHz and a Nyquist frequency of $33.0$ kHz. These surface pressures are used to predict the blade's far-field acoustic signature, aerodynamic performance, and wall-pressure spectrum. Likewise, the full volumetric solution on a single blade's primary grid is saved at the same frequency, consisting the conserved variables---i.e., density, momentum, and total energy---at every grid point within the $257\times239\times99$ O-mesh. This enables detailed analysis of the NB flow's evolution from the perspective of the blade as it rotates within the larger computational domain.

\subsection{Acoustic Processing} \label{Acoustic Processing}

Far-field acoustics are predicted using PSU-WOPWOP \citep{Bres:2004:Maneuver,Lee-2009,psuwopwop_user_manual}. Impermeable surfaces are constructed from the surface solutions of each NB grid, thereby omitting volume-source contributions from the analysis. This simplification is justified for the low-Mach-number operating condition investigated here, where volume sources associated with vortex breakdown and shocks are expected to be negligible according to the scaling arguments of \citet{Curle1955}. PSU-WOPWOP computes Farassat's Formulation 1A (FF1A) \citep{farassat_succi, farassat2007derivation} solution of the Ffowcs Williams--Hawkings (FW--H) equation \citep{Ffowcs-1969}, which generalizes Lighthill's acoustic analogy \citep{Lighthill1952} to account for moving surfaces. In FF1A, the acoustic pressure at an observer location $\boldsymbol{x} = [x, y, z]$ and time $t$ is expressed as

\begin{equation} \label{eq: FF1A_total}
    p(\boldsymbol{x},t) = p_T(\boldsymbol{x},t) + p_L(\boldsymbol{x},t).
\end{equation}

\noindent Here, $p$ denotes the acoustic pressure defined by $p = P-p_\infty$, where $p_\infty$ is the ambient pressure. The thickness and loading contributions are  $p_T$ and $p_L$, respectively, given by

\begin{equation}\label{eq: FF1A_thickness}
    \begin{aligned}
        4\pi p_T(\boldsymbol{x},t)
        &=
        \int_{f=0}\left[\frac{\rho_\infty(\dot{v}_n + v_{\dot{n}})}{d \left| 1 - M_d \right|^2}\right]_{\mathrm{ret}} dS
        \\
        &\qquad +
        \int_{f=0}\left[\frac{\rho_\infty v_n(r\dot{M}_d + a(M_d - M^2))}{d^2 \left| 1 - M_d \right|^3}\right]_{\mathrm{ret}} dS
    \end{aligned}
\end{equation}

\noindent
and

\begin{equation} \label{eq: FF1A_loading}
    \begin{aligned}
        4\pi p_L(\boldsymbol{x},t)
        &=
        \frac{1}{a} \int_{f=0}\left[\frac{\dot{L}_d}{d \left| 1 - M_d \right|^2}\right]_{\mathrm{ret}} dS
        +
        \int_{f=0}\left[\frac{L_d - L_M}{d^2 \left| 1 - M_d \right|^2}\right]_{\mathrm{ret}} dS
        \\
        &\qquad +
        \frac{1}{a}\int_{f=0}\left[\frac{L_d(d\dot{M}_d + a(M_d - M^2))}{d^2 \left| 1 - M_d \right|^3}\right]_{\mathrm{ret}} dS .
    \end{aligned}
\end{equation}

In these equations, $\rho_\infty$ is the freestream density, $M$ is the source Mach number, $d$ is the distance between the source and observer, and $a$ is the freestream speed of sound. The overdot notation $\dot{(\ )}$ denotes differentiation with respect to source time, and subscripts indicate a dot product with corresponding unit vectors: $n$ denotes the surface unit normal vector, $d$ the unit radiation vector directed from the source to the observer, and $M$ the surface Mach vector. In Eq.~\eqref{eq: FF1A_thickness}, $v$ denotes the surface velocity associated with the thickness contribution, whereas in Eq.~\eqref{eq: FF1A_loading}, $l$ represents the loading vector derived from the surface pressure distribution. Finally, the integral is evaluated over $f = 0$, which represents the impermeable blade surfaces. 

Because this study focuses on BSVI noise and its quasi-tonal signature, phase-averaging of the far-field acoustic pressure is applied to separate purely periodic signals from stochastic ones. The total acoustic pressure is decomposed into multiple blocks, each spanning one rotor revolution, such that each block comprises a periodic (tonal) and a stochastic (residual) component.
\begin{equation} \label{eq:PressureDecomposition}
    p_i(t) = p_{\mathrm{tonal}}(t) + p_{\mathrm{res},i}(t)
\end{equation}
\noindent Here, the subscript $i$ denotes an individual block. The periodic component, which is expected to be identical across all blocks, is approximated as the ensemble average
\begin{equation} \label{eq:EnsembleAverage}
    p_{\mathrm{tonal}}(t) \approx \frac{1}{N_{\mathrm{BLK}}} \sum_{i=1}^{N_{\mathrm{BLK}}} p_i(t),
\end{equation}
\noindent where $N_{\mathrm{BLK}}$ is the total number of blocks, and the approximation improves as $N_{\mathrm{BLK}}$ increases. In this study,  $N_{\mathrm{BLK}}=20$ is used. The residual component is then obtained by subtracting the periodic component from the original signal,
\begin{equation} \label{eq:ResidualPressure}
    p_{\mathrm{res},i}(t) = p_i(t) - p_{\mathrm{tonal}}(t).
\end{equation}
\noindent This residual component captures the stochastic contributions to the predicted acoustic pressure, encompassing both purely broadband noise sources and the quasi-tonal noise of primary interest in this study.

Finally, the RANS-based boundary-layer treatment cannot resolve the turbulent pressure fluctuations or pressure scattering responsible for TBL-TE noise. Nevertheless, TBL-TE noise remains a significant contributor to stochastic rotor noise, as it establishes a minimum noise floor \citep{Lee-2021}. Accounting for this source is therefore important both for validating the predictions against real-world measurements and for characterizing the relative contribution of each noise mechanism for small-scale hovering rotors. Accordingly, the PSU-WOPWOP predictions are supplemented with UCD-QuietFly estimates of far-field TBL-TE noise, following the methodology of \citet{Won2025-JASA}.

UCD-QuietFly combines the analytical TBL-TE framework of \citet{Amiet1976} with the empirical wall-pressure spectrum (WPS) model of \citet{Lee:2018:AIAAJ}. UCD-QuietFly has been extensively validated for a variety of drone-scale rotors operating in hover and forward flight \citep{Li-2020-JAHS, Li-2021-JAHS, Li-2022-JAHS}. For a given blade strip of width $b$ at azimuth angle $\Psi$, UCD-QuietFly evaluates Amiet's theory as

\begin{equation} \label{eq:Amiet}
    S_{pp,\Psi}(f) = \left(\frac{2\pi f}{a}\right)^2 C^2 b \left(\frac{1}{32\pi^2}\right) D_\Psi \left|L_\Psi(f)\right|^2 l_{r,\Psi}(f)\, \Phi_{pp,\Psi}(f).
\end{equation}

\noindent Here, the WPS $\Phi_{pp}(f)$ near the trailing edge is the most critical term. Additionally, $a$ is the speed of sound, $D_\Psi$ represents the blade's distance and directionality with respect to a fixed observer, $L$ is the airfoil's loading response function \citep{MOREAU2009397}, $C$ is the local chord length, and $l_r$ is the spanwise correlation. The far-field two-sided PSD $S_{pp}(f)$ is obtained by azimuthally averaging $S_{pp,\Psi}(f)$.

The WPS for an arbitrary airfoil and flow is not easily obtained, typically requiring experimental measurements, wall-resolved LES, or direct numerical simulation (DNS). UCD-QuietFly therefore adopts the empirical WPS model of \citet{Lee:2018:AIAAJ}, which predicts the WPS from time-averaged flow quantities near the trailing edge, including the boundary-layer thickness, displacement thickness, momentum thickness, friction coefficient, pressure gradient, and edge velocity. \citet{Won2025-JASA} described several approaches for combining HRLES rotor simulations with UCD-QuietFly, consistent with the methods of \citet{Jung2023} and \citet{Marques2024}. Azimuthally averaged sectional loads extracted from the CFD solution are used to inform XFOIL \citep{Drela:XFOIL} inputs, including the effective angle of attack and local freestream velocity; the resulting XFOIL boundary-layer solution provides the flow quantities required by Lee's WPS model, from which the WPS and far-field TBL-TE noise are subsequently predicted. Once the rotor's TBL-TE noise is predicted using Eq.~\eqref{eq:Amiet}, its acoustic energy is added to the PSU-WOPWOP predictions discussed earlier in this section.




\subsection{Spectral Proper Orthogonal Decomposition} \label{sec: SPOD Methodology}

The Spectral Proper Orthogonal Decomposition (SPOD) algorithm of \citet{Towne2018} is employed to investigate the spatiotemporal behavior of the secondary vortices. Derived from the space-time POD framework of \citet{Lumley1970} and analogous to principal component analysis, SPOD identifies an optimal set of deterministic basis functions (modes) representing a zero-mean stationary process. Each mode corresponds to a coherent spatial structure associated with the fluctuating flow field $q^\prime(\boldsymbol{x},t)$, where

\begin{equation} \label{FlowVariableDefinition}
    q^\prime(\boldsymbol{x},t) = q_\mathrm{res}(\boldsymbol{x},t) - \bar{q}_\mathrm{res}(\boldsymbol{x},t).
\end{equation}

\noindent Note that $q_\mathrm{res}(\boldsymbol{x},t)$ denotes the residual components of the flow field, as desribed in Section \ref{Acoustic Processing}, and $\bar{q}_\mathrm{res}(\boldsymbol{x},t)$ is its time-averaged value. Thus, this definition of $q^\prime(\boldsymbol{x},t)$ omits the flow periodicity induced by blade rotation, representing the zero-mean residual flow solution rather than the zero-mean variables in traditional wakes.

Mathematically, the SPOD modes are the eigenvectors of the cross-spectral density (CSD) matrix at each frequency, with the associated eigenvalues representing the modal energy content. Consequently, a reduced-order representation of the stochastic flow field is obtained through an accurate approximation of the CSD tensor. Because each mode isolates a coherent structure at a specific frequency, SPOD facilitates the identification of coherent flow structures and their interactions with the rotor blades, providing insight into the mechanisms responsible for loading-noise generation.

Following the work by \citet{Towne2018} and the guide provided by \citet{Schmidt2019}, the formulation of SPOD begins with the definition of the covariance tensor

\begin{equation} \label{eq: POD Covarience}
    C(\boldsymbol{x}_1, \boldsymbol{x}_2, t_1, t_2) = \mathbb{E}\left[q^\prime(\boldsymbol{x}_1,t_1)q^{\prime*}(\boldsymbol{x}_2,t_2)\right].
\end{equation}

\noindent Here, $\mathbb{E}$ is the expectation operator, and $^*$ denotes the complex conjugate. It is important to note that the variance of a flow variable is often associated with some form of energy, depending on the choice of $q^\prime(x,t)$. For example, the variance of velocity fluctuations is twice the turbulent kinetic energy. Therefore, the eigenvalues $\lambda_k$ of the covariance tensor quantify the portion of energy contained in their corresponding eigenmodes $\phi_k$, and importantly, they can be ranked in descending order of importance ($\lambda_1 \ge \lambda_2 \ge \dots \ge 0$).

Under the assumption of a stationary process, the covariance tensor is reduced to $C(\boldsymbol{x}_1, \boldsymbol{x}_2, \tau)$, where $\tau = t_1 - t_2$ is the time shift. Furthermore, we can take the Fourier transform with respect to $\tau$ to acquire the CSD tensor

\begin{equation} \label{eq: SPOD Covarience}
    \hat{C}(\boldsymbol{x}_1, \boldsymbol{x}_2, f) = \int_{-\infty}^\infty C(\boldsymbol{x}_1, \boldsymbol{x}_2, \tau)e^{-2\pi i f\tau} d\tau.
\end{equation}

\noindent Welch's periodogram \citep{welch1967} is employed to approximate a CSD matrix for each frequency. Given $N$ spatial points and a total of $N_t$ recording temporal snapshots, the data can be partitioned into $N_{\text{BLK}}$ blocks. Each block contains all $N$ spatial points and spans $N_{\text{FFT}}$ snapshots, with blocks potentially overlapping by an amount specified as $N_{\text{OVLP}}$.

The resulting frequency resolution $\Delta f$ of the CSD matrices is as follows:

\begin{equation} \label{eq: SPOD Bin Size}
    \Delta f = \frac{f_s}{N_{\text{FFT}}}.
\end{equation}

\noindent Thus, larger $N_{\text{FFT}}$ values result in finer spectral resolution but require more time snapshots for an accurate estimate. The sampling frequency considered here is $66$ kHz. Each block extends one physical rotor revolution ($N_{\text{FFT}} = 720$), yielding a frequency resolution of approximately $92$ Hz. Data spanning 20 time-accurate revolutions is sampled ($N_t = 20\times 720$) with a $50\%$ overlap between blocks. This results in $39$ blocks for approximating each CSD matrix. 

A step-by-step algorithm to perform SPOD is described by \citet{Schmidt2019}. For each frequency, one must solve the following eigenvalue problem:

\begin{equation} \label{eq: SPOD Eigenvalue Problem}
    \hat{C}W\hat{\Phi} = \hat{\Phi}\Lambda.
\end{equation}

\noindent The columns of $\hat{\Phi}$ contain the SPOD mode shapes with associated modal energies on the diagonal of $\Lambda$. The weight matrix $W$ defines both the numerical quadrature and the energy norm. Here, cell areas and volumes are used, calculated using the trapezoidal rule. Finally, the first $40$ body-normal grid points are omitted in the SPOD analysis, where the domain beyond the $40^\mathrm{{th}}$ grid point consists of LES-resolved regions.

\section{Results and discussion} \label{sec: Results}

\subsection{Aerodynamics and Flow Fields}\label{sec: aerodynamics}

\begin{figure}
    \centering
    \includegraphics[width=0.99\linewidth]{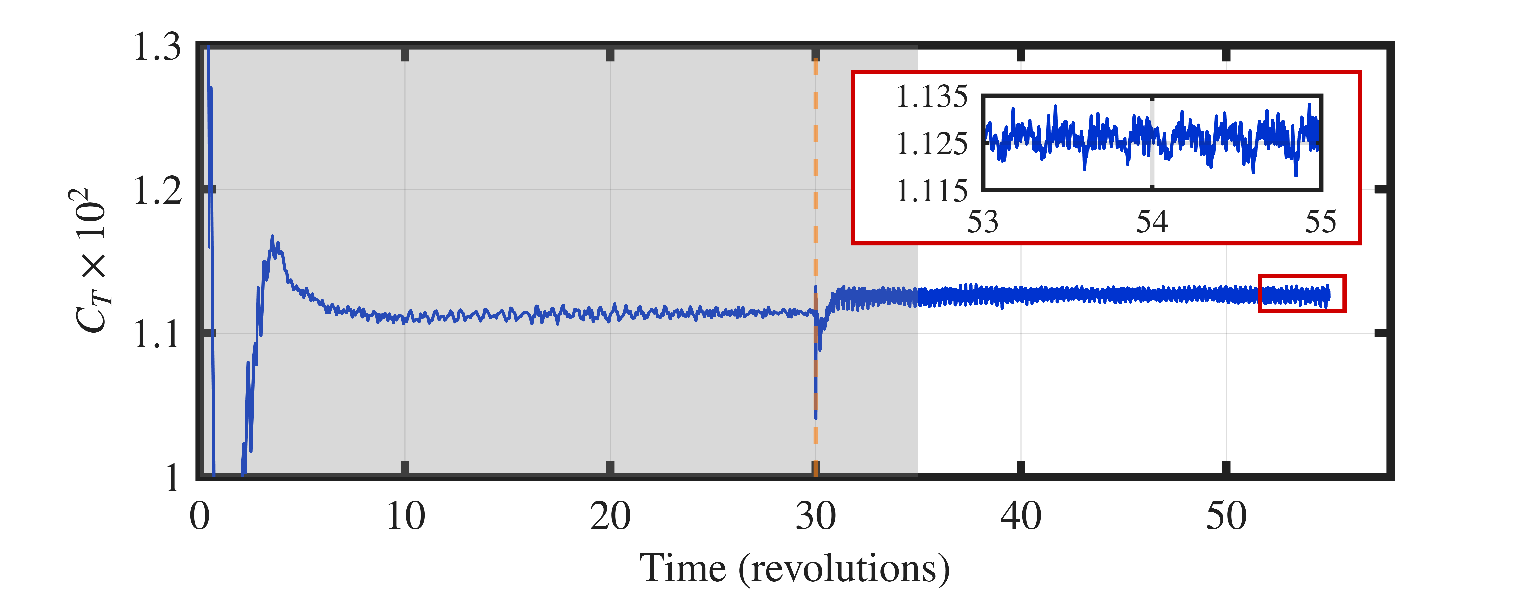}
    \caption{Numerical thrust convergence as a function of nondimensional time.}
    \label{fig: thrust convergence}
\end{figure}

The simulation's thrust convergence is shown in Fig. \ref{fig: thrust convergence}. Time is nondimensionalized by the rotor period ($T_0=2\pi/\Omega~\mathrm{s}$), and only the last $20$ revolutions are included in this study's analysis. The thrust converges to a mean value of $1.126 \times 10^{-2}$ after the time step is refined beyond the quick-start phase. The sudden increase in temporal resolution produces numerical transients that resolve within one or two revolutions, although this study conservatively omits the first five time-accurate revolutions. The inset of Fig. \ref{fig: thrust convergence} reveals significant thrust fluctuations about the mean, indicative of unsteady loading that significantly contributes to the rotor's acoustic signature.

\begin{figure}
    \centering
    \subfloat[\label{subfig: dFdPsi_Rev54}]{\includegraphics[width = 0.48\textwidth]{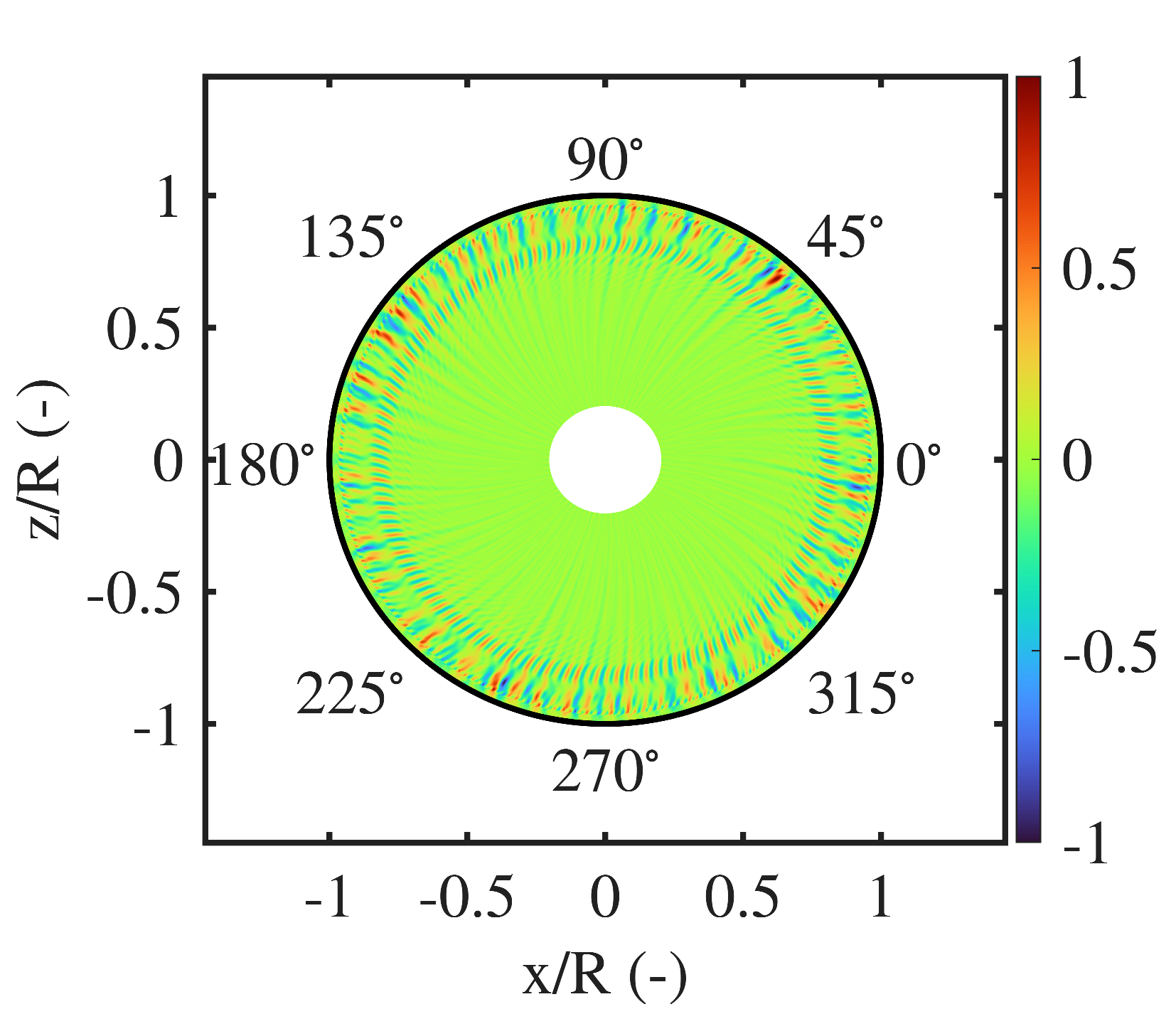}}
    \hspace{0.05 cm}
    \subfloat[\label{subfig: dFdPsi_Rev55}]{\includegraphics[width = 0.48\textwidth]{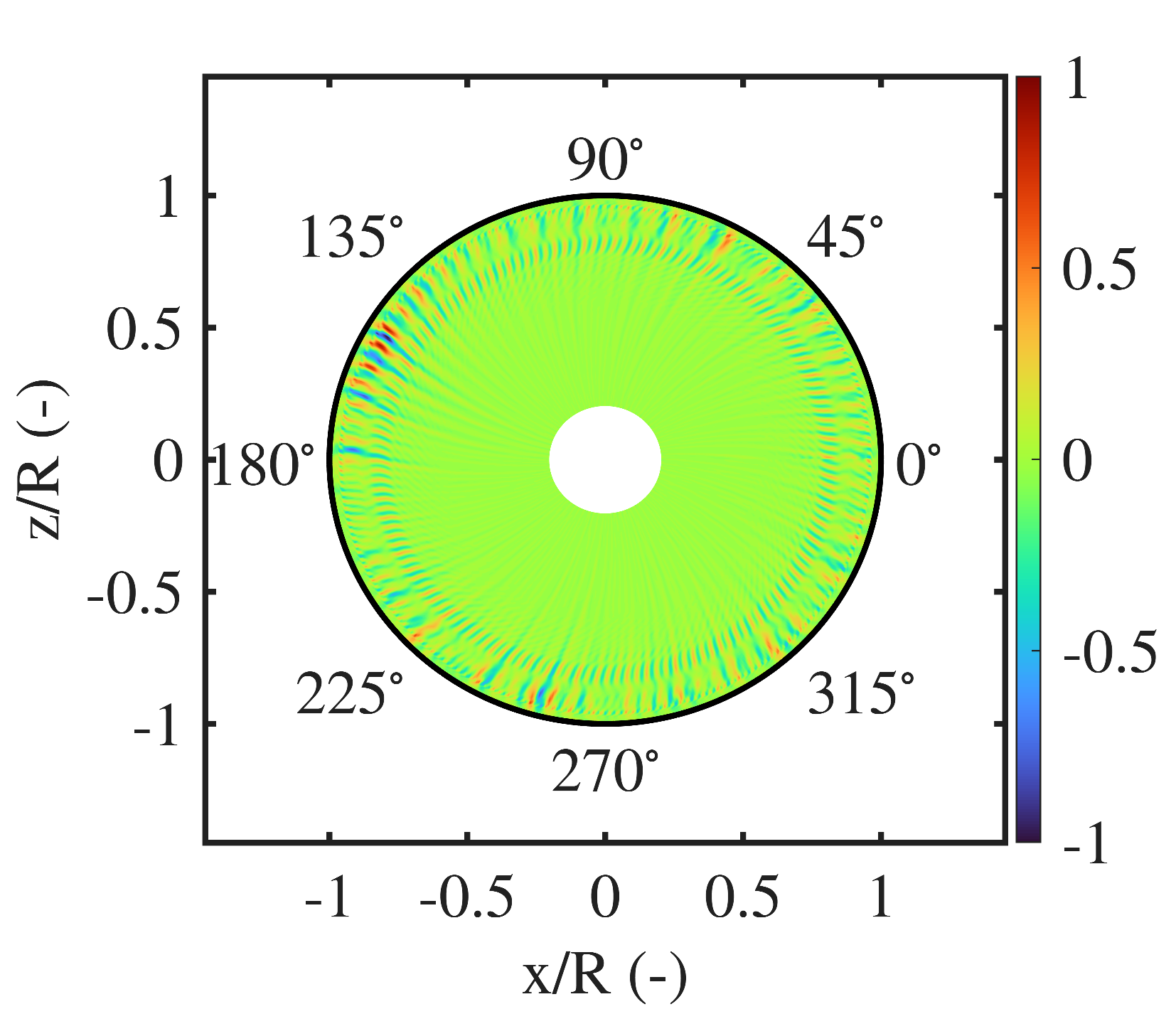}}
    \caption{Blade loading contours colored by instantaneous $dC_n/d\Psi$ ($\mathrm{deg}^{-1}$) normalized by $\mathrm{max}(|dC_n/d\Psi|_i)$ for (a) revolution 54 and (b) revolution 55. Subscript $i$ denotes revolution index.}
    \label{fig: dCndpsi_instant}
\end{figure}

\begin{figure}
    \centering
    {\includegraphics[width = 0.99\textwidth]{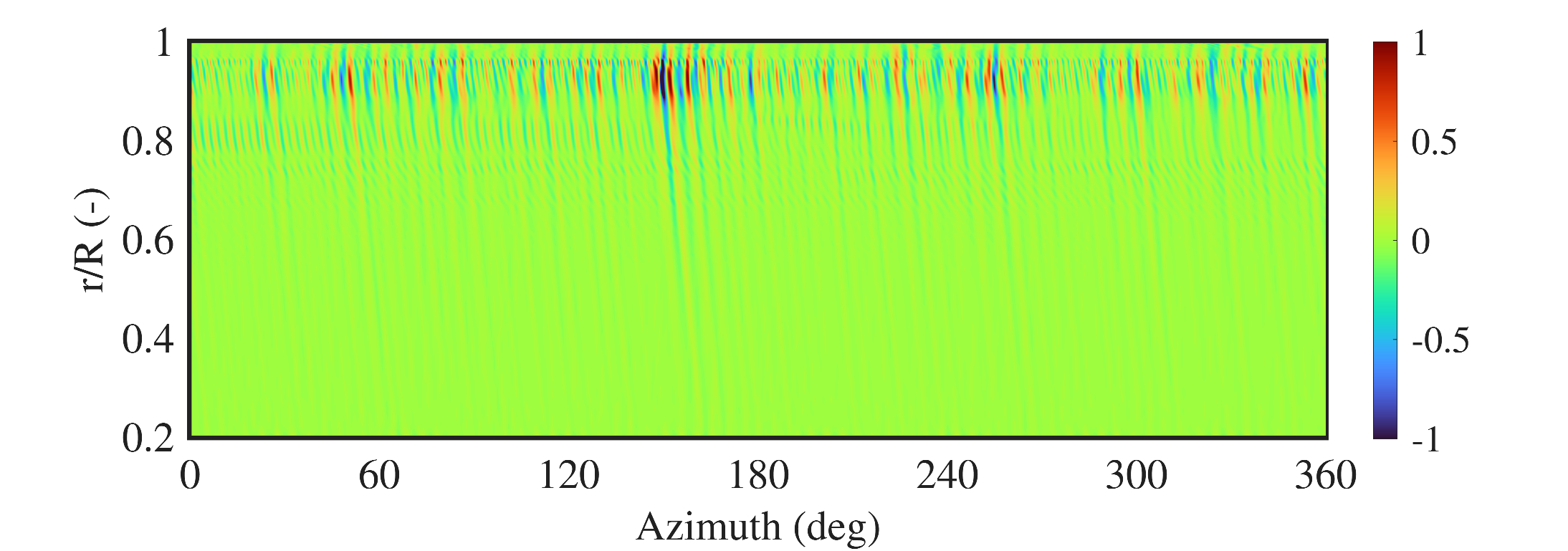}}

    \caption{Revolution-to-revolution difference in instantaneous blade loading, $(dC_n/d\Psi)_{55} - (dC_n/d\Psi)_{54}$, normalized by $\mathrm{max}(|dC_n/d\Psi|_{55})$, as a function of blade radius and azimuth angle.}
    \label{fig: dCndpsi_difference}
\end{figure}

Figure \ref{fig: dCndpsi_instant} presents two contours colored by the instantaneous $dC_n/d\Psi$ for a single blade as a function of blade radius and azimuth angle, where $C_n$ denotes the sectional force coefficient normal to the airfoil's chord. The time derivative of this force is directly related to the loading-noise terms discussed in Section \ref{sec: Methodology}, and $dC_n/d\Psi$ is computed by finite differences with $\Delta\Psi = 0.50^\circ$. Figure \ref{subfig: dFdPsi_Rev54} shows the loading contours for revolution $54$, and Fig. \ref{subfig: dFdPsi_Rev55} shows revolution $55$. The prevalence of unsteady loading at the outboard radial locations is clearly the main contributor to the unsteady thrust fluctuations shown in Fig. \ref{fig: thrust convergence}. Moreover, the $135^\circ$--$180^\circ$ azimuth locations reveal drastic revolution-to-revolution differences in loading, with Fig. \ref{subfig: dFdPsi_Rev54} (revolution 54) exhibiting a markedly different pattern from Fig. \ref{subfig: dFdPsi_Rev55} (revolution 55) in this region. Figure \ref{fig: dCndpsi_difference} highlights this variation further, showing a contour of $(dC_n/d\Psi)_{55} - (dC_n/d\Psi)_{54}$, where the subscript denotes the revolution from which the data is sampled. These notable revolution-to-revolution differences reveal highly aperiodic behavior that produces the stochastic loading noise of present focus.

\begin{figure}
    \centering
    \includegraphics[width=0.85\linewidth]{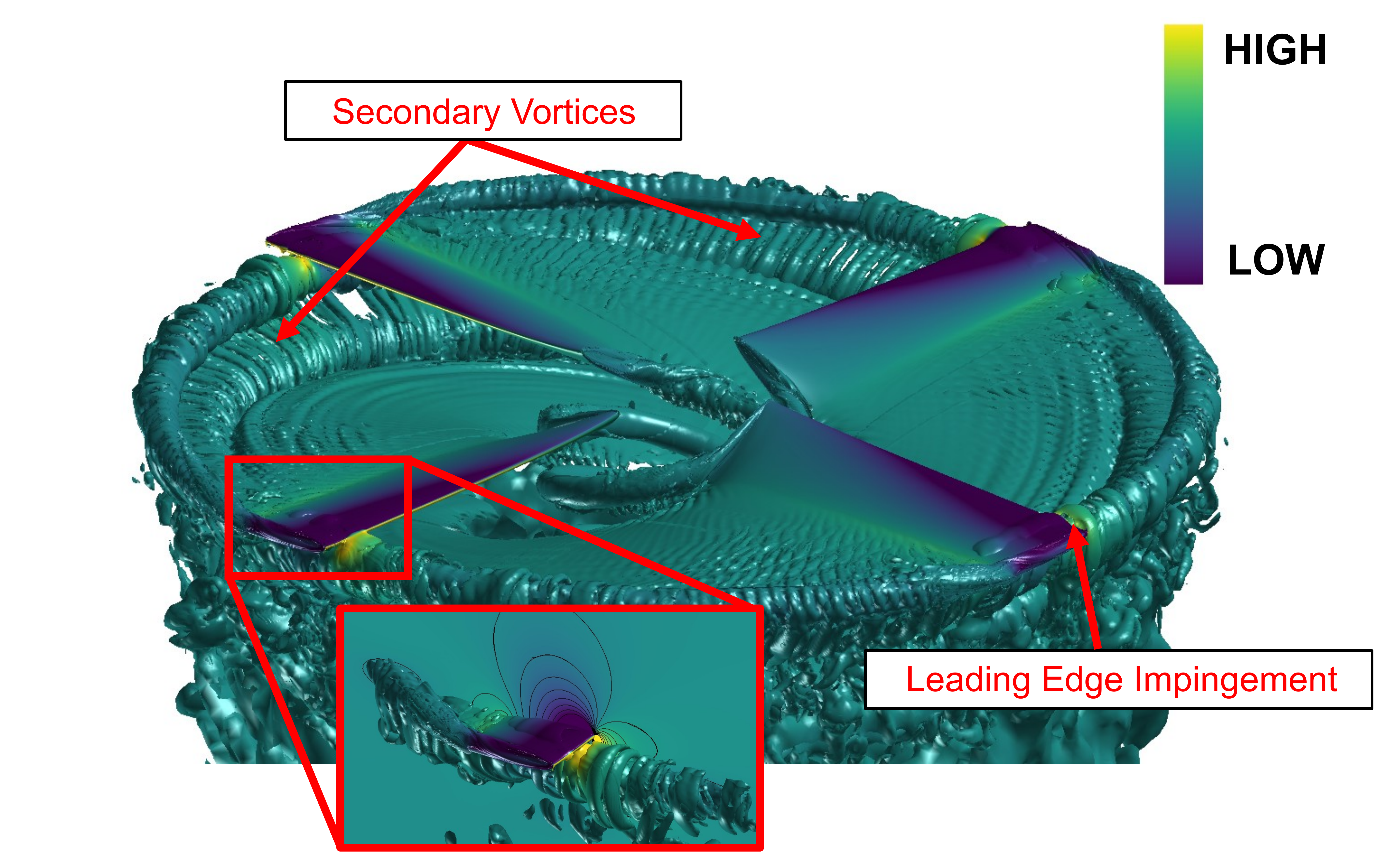}
    \caption{Iso-surface of vorticity magnitude ($|\omega| = 1000$ s$^{-1}$) colored by pressure coefficient, with color bar limits of $-0.25$ and $0.25$. Animation available at \url{https://ucdavis.box.com/s/gakguazno0p76ehxs4vlbhyumwaf09fp}.}
    \label{fig: ITR Vorticity Iso Surface}
\end{figure}

Figure~\ref{fig: ITR Vorticity Iso Surface} displays an isosurface of instantaneous vorticity magnitude ($|\omega| = 1000~\mathrm{s}^{-1}$) colored by pressure coefficient. The isosurface reveals tightly braided structures forming around the PTVs, closely resembling the conceptual diagram of the BSVI noise mechanism illustrated in Fig.~\ref{fig: BSVI Conceptual Diagram}. These secondary vortices impinge on the outboard regions of the following blade, inducing the unsteady loading fluctuations seen in Fig. \ref{fig: thrust convergence}. An animation corresponding to Fig.~\ref{fig: ITR Vorticity Iso Surface} (available online) shows that the secondary vortices originate in the rotor wake and migrate upward into the rotor disk despite the rotor's downwash. Qualitatively, this behavior resembles rotor wake recirculation \citep{Weitsman2020, Nardari2019}, in which large, elongated vortex structures are reingested when hovering rotors operate within a confined space or in proximity to the ground. As an approaching blade chops through coherent vortex structures, loading noise is generated at the leading edge of each blade, producing higher-harmonic BPF tones. However, unlike wake recirculation, BSVI involves much smaller vortices that are inherent to hovering-rotor wakes, independent of confinement \cite{Chaderjian2011, Wolf2019, Schwarz2022}..

The interaction of a blade with well-defined, coherent vortices is what makes BSVI distinct from conventional BWI, i.e., blade interactions with random turbulence. Even though the secondary vortex braids in Fig. \ref{fig: ITR Vorticity Iso Surface} are understood to be stochastic and highly intermittent \citep{Wolf2019}, the structures form clear spatial patterns in this small-scale rotor. This spatial coherence produces the alternating loading observed in Figs. \ref{fig: dCndpsi_instant} and \ref{fig: dCndpsi_difference}, where random turbulence would instead appear more irregular and scattered---explaining why BSVI produces intermittent tones while BWI produces a broadband acoustic spectrum. This spatial coherence is analyzed in more depth using SPOD in Section \ref{sec: statistics}.

Previous studies have reported similar BPF tones arising from blade interactions with intermittent coherent vortices. \citet{Murray2018}, investigating a propeller ingesting turbulence from a wind tunnel's boundary layer, found that haystacked humps associated with blade interactions with turbulence narrowed into distinct BPF tones when coherent vortices were chopped by the propeller blades. Similarly, for TI noise, as turbulent eddies elongate and form long coherent vortex structures, they also produce BPF-aligned peaks \citep{paterson&amiet1979}. A key distinction between BSVI and TI noise, however, is that a hovering rotor produces its own secondary vortex structures through WSL entrained in multiple PTVs, without any influence from external turbulence.


\begin{figure}
    \centering
    \includegraphics[width=0.85\linewidth]{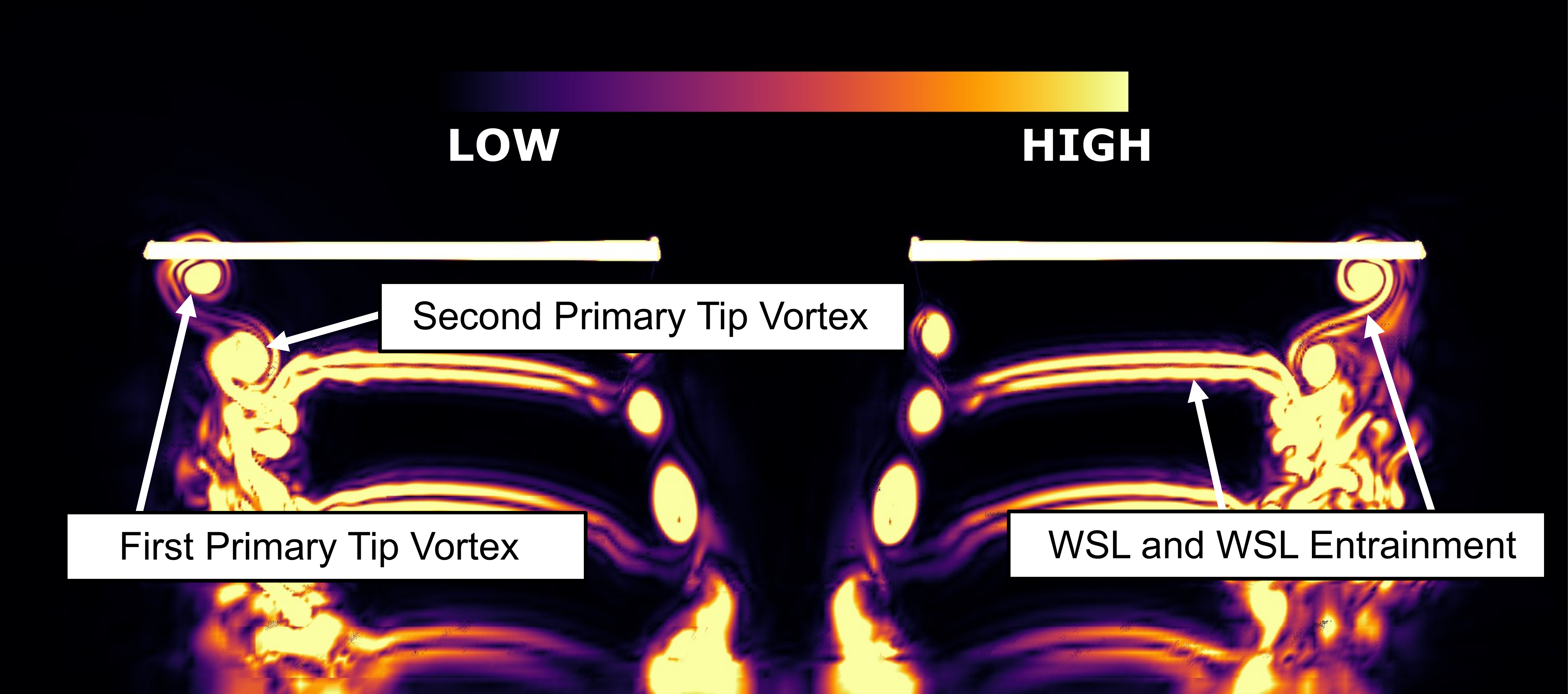}
    \caption{Cutting plane of the ITR in hover, colored by instantaneous nondimensional vorticity magnitude, with color bar limits of $0.0$ and $5.0$. Animation can be found at the link: \url{https://ucdavis.box.com/s/3310s2p6bsej2nw062lw95qqvw8x5cnk}}
    \label{fig: ITR Vorticity CrossSection}
\end{figure}

Figure \ref{fig: ITR Vorticity CrossSection} presents a cutting plane colored by instantaneous nondimensional vorticity magnitude ($\omega^*$), highlighting the WSL and its entrainment into the PTVs. Here, $\omega^* = \omega/(aL)$, where $L = 1$ m and $a$ is the speed of sound. The entrainment process produces the secondary vortices that wrap around the first and second PTVs, forming distinct S-shaped structures. Importantly, it is these secondary vortices---rather than the PTV core passing beneath each blade---that predominantly impinge on the leading edge of each blade between approximately $r/R = 0.88$ and $0.96$. These are the same outboard radial sections that were aperiodically loaded in Fig. \ref{fig: dCndpsi_instant}, further confirming that this loading arises from blade interactions with the coherent secondary vortices.

\begin{figure}
    \centering
    \subfloat[\label{subfig: Secondary vortex system A}]{\includegraphics[width = 0.45\textwidth]{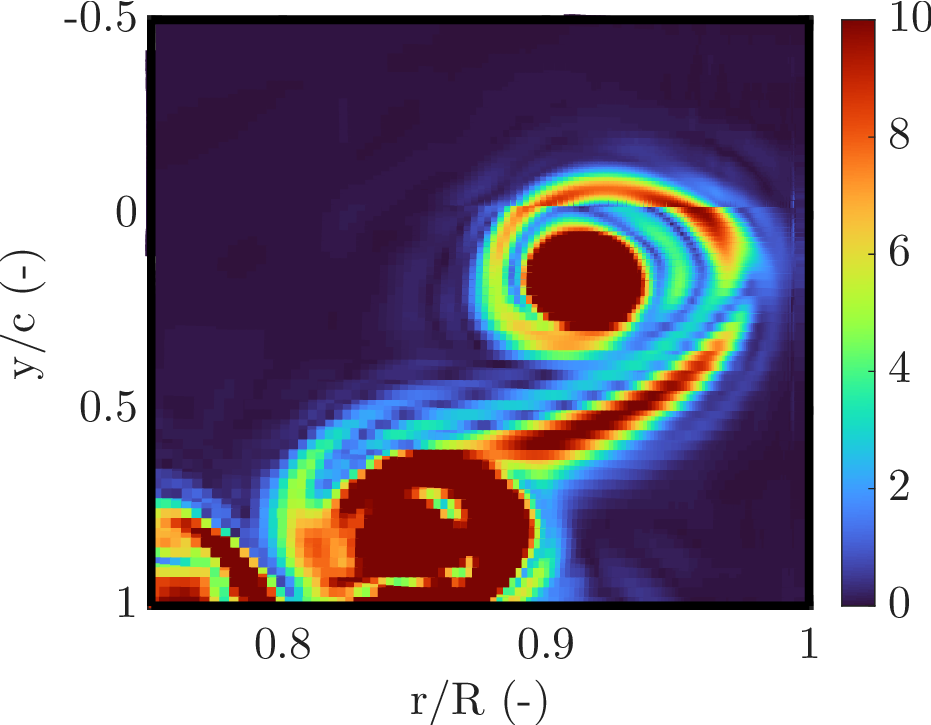}}
    \hspace{0.25 cm}
    \subfloat[\label{subfig: Secondary vortex system B}]{\includegraphics[width = 0.45\textwidth]{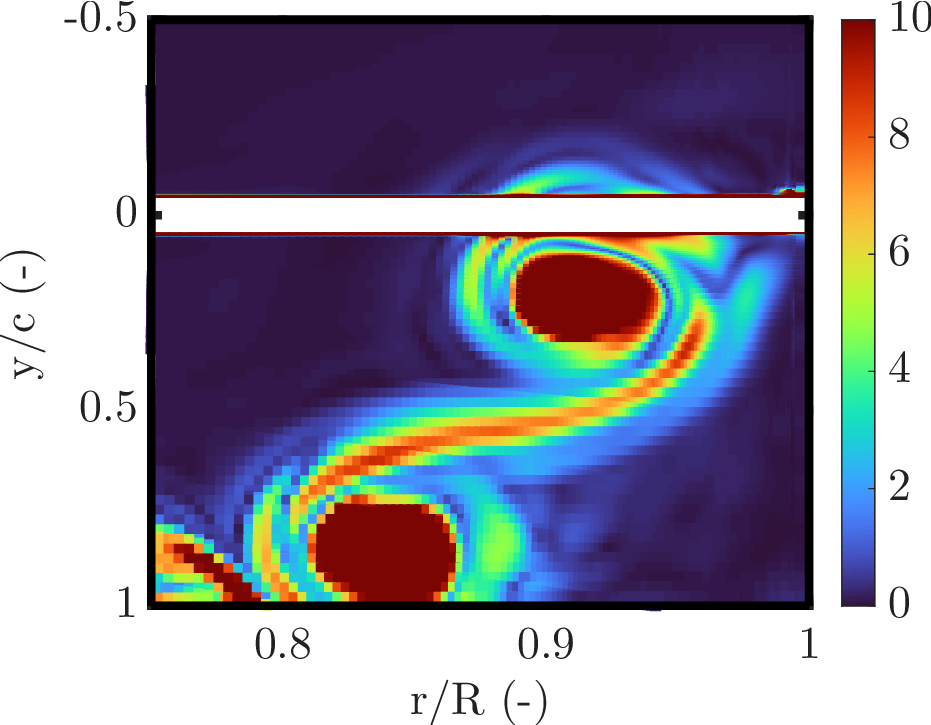}}
    
    \subfloat[\label{subfig: Secondary vortex system C}]{\includegraphics[width = 0.45\textwidth]{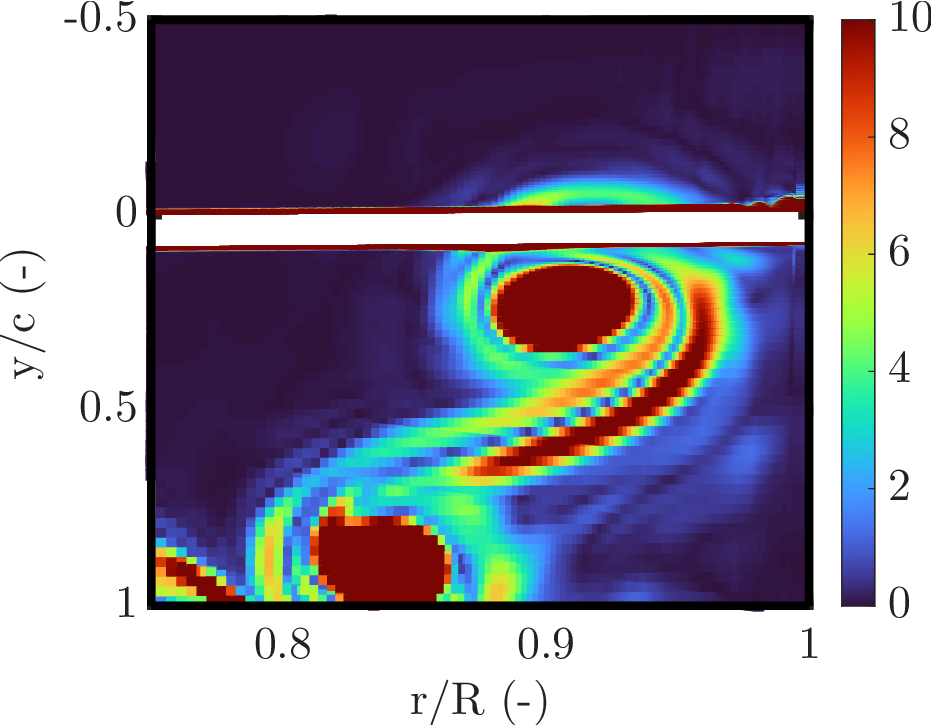}}
    \hspace{0.25 cm}
    \subfloat[\label{subfig: Secondary vortex system D}]{\includegraphics[width = 0.45\textwidth]{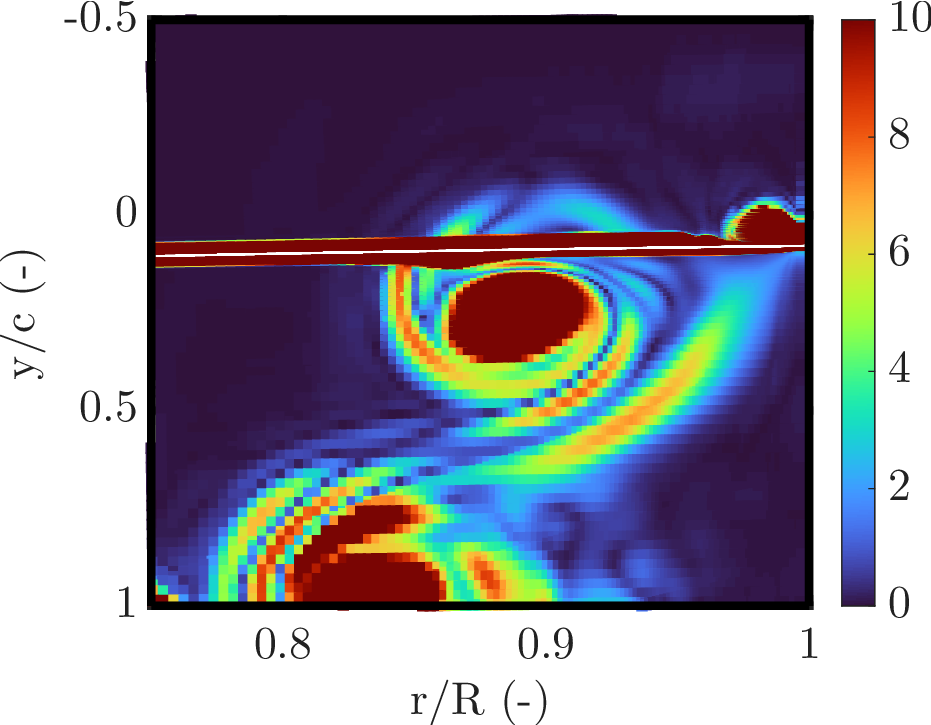}}
    \caption{Instantaneous cutting-plane contours of nondimensional vorticity magnitude at (a) $0.05c$ forward of the leading edge, (b) $0.25c$, (c) $0.50c$ and (d) $0.90c$.}
    \label{fig: Secondary Vortex System}
\end{figure}

Figure \ref{fig: Secondary Vortex System} displays instantaneous cutting-plane contours colored by the nondimensional vorticity magnitude, highlighting the interaction between a given blade and the preceding blade's PTV and surrounding secondary vortex system. Figure \ref{subfig: Secondary vortex system A} shows a cutting plane located $0.05c$ forward of the blade's leading edge, with the reference blade aligned at the zero-degree azimuthal location. Similarly, Figs. \ref{subfig: Secondary vortex system B}, \ref{subfig: Secondary vortex system C}, and \ref{subfig: Secondary vortex system D} are also $x$-normal cutting planes, but at locations progressively aft of the leading edge, cutting through the blade. In all panels, the first and second PTVs are located approximately between $0.0 \le y/c \le 0.3$ and $0.7 \le y/c \le 1.0$, respectively, although the more aft cutting planes show a slight downward displacement due to the blade's induced velocity. The S-shaped secondary vortex braids are also apparent, exhibiting lower but still significant vorticity magnitudes. The two PTVs together with their surrounding secondary vortex braids are hereafter referred to as a secondary vortex system.

Qualitatively, the blade interacts primarily with secondary vortices surrounding the PTVs, indicating that BSVI dominates the leading-edge noise under the present operating condition. We therefore argue that BSVI can be more impactful than perpendicular BVI for hovering rotors at these scales. The first PTV appears only to graze the blade and would therefore not produce significant unsteady loading noise. We also note that \citet{Won&Lee-2026-BSVI} demonstrated that the vortex miss distance separating the PTV from the blade is greatly increased for medium-scale UAM tiltrotors owing to their reduced BPF, further highlighting the significance of secondary vortex impingement for rotorcraft aeroacoustics. By the same argument, however, the balance between perpendicular BVI and BSVI is likely to depend on operating conditions such as rotation rate and collective pitch, and warrants further investigation. A more detailed quantitative analysis, computing the cross-correlation between these points and the pressure fluctuation on the blade's leading edge, is presented in Section \ref{sec: statistics} to evaluate the relation between the structures shown in Fig. \ref{fig: Secondary Vortex System} and loading-noise generation.

\subsection{Aeroacoustics} \label{sec: Acoustics}

\begin{figure}
    \centering
    \includegraphics[width=0.95\linewidth]{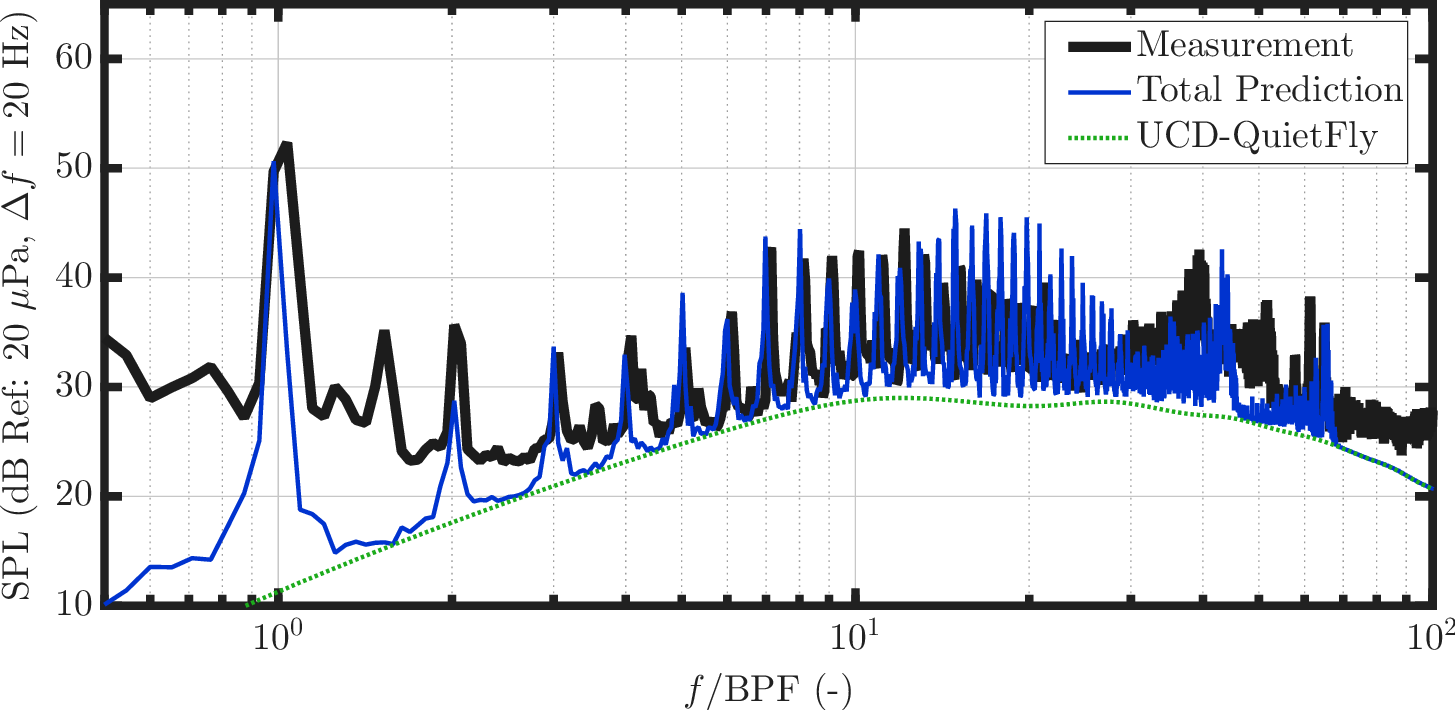}
    \caption{Narrowband ($\Delta f = 20$~Hz) far-field acoustic spectrum at mic. $5$: comparison of the combined HRLES/FW-H and UCD-QuietFly total prediction against the experimental measurement of \citet{pettingill2021}.}
    \label{fig: ITR_Narrowband_Prediction}
\end{figure}

A validation of the current HRLES/FW-H acoustic predictions is presented against the $\Omega = 576~\mathrm{rad/s}$ measurements of \citet{pettingill2021}. Narrowband ($\Delta f = 20$~Hz) acoustic predictions are shown in Fig.~\ref{fig: ITR_Narrowband_Prediction}, where frequency is normalized by $\mathrm{BPF} = 366~\mathrm{Hz}$. TBL-TE noise predictions from UCD-QuietFly are also included to supplement the CFD-based predictions, since RANS boundary-layer modeling does not resolve this noise source. This TBL-TE noise yields a broadband noise floor that closely aligns with the minimum SPL measurements. Note that Appendix \ref{Appendix: Validation Cont.} includes our narrowband predictions without the TBL-TE noise modeling contributions. 

Mid-frequency BPF tones generated from BSVIs are predicted up to the $30^\mathrm{th}$ harmonic, closely resembling the experimental measurements; however, above the $13^{\mathrm{th}}$ BPF the magnitude is overpredicted. We speculate that this overprediction may be associated, at least in part, with limitations of the DDES approach employed in the present study. In particular, the RANS-modeled boundary layer, which is subsequently shed and stretched to form the elongated secondary vortex structures, may remain smoother and more coherent than in the experiments. Such differences in the upstream boundary layer could lead to overly coherent secondary vortical structures and, consequently, elevated noise levels at higher frequencies. This may explain the remaining overprediction despite the use of a sufficiently fine grid, which has been shown to resolve both the number and strength of the secondary vortices \citep{Bodling2024}. Nevertheless, Fig. \ref{fig: ITR_Narrowband_Prediction} shows overall good agreement with the measurements and reproduces the dominant noise characteristics of the rotor.

\begin{figure}
    \centering
    \subfloat[\label{subfig: }]{\includegraphics[width = 0.32\textwidth]{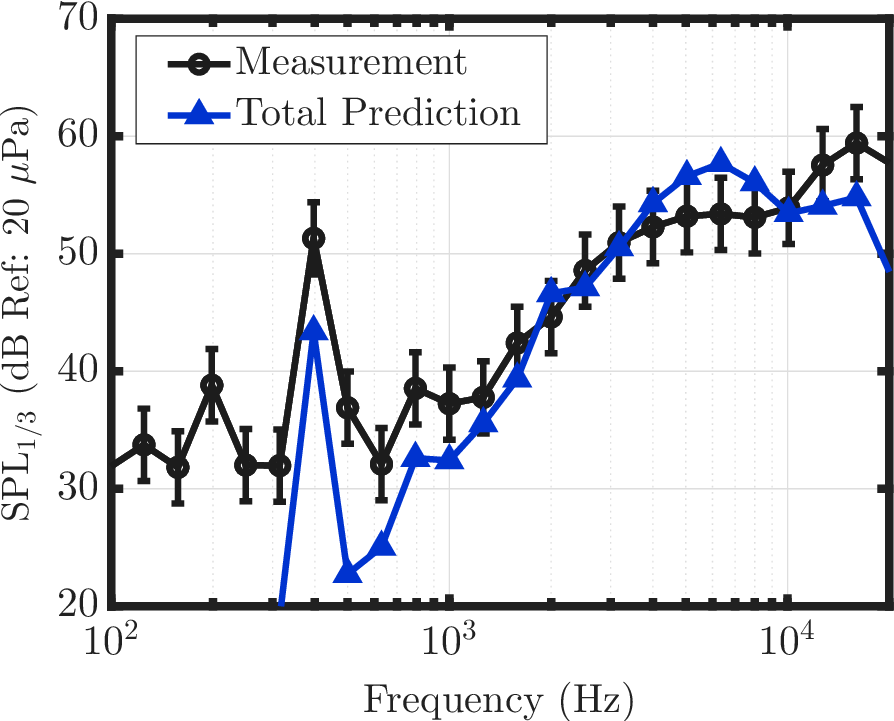}}
    \hspace{0.15 cm}
    \subfloat[\label{subfig: }]{\includegraphics[width = 0.32\textwidth]{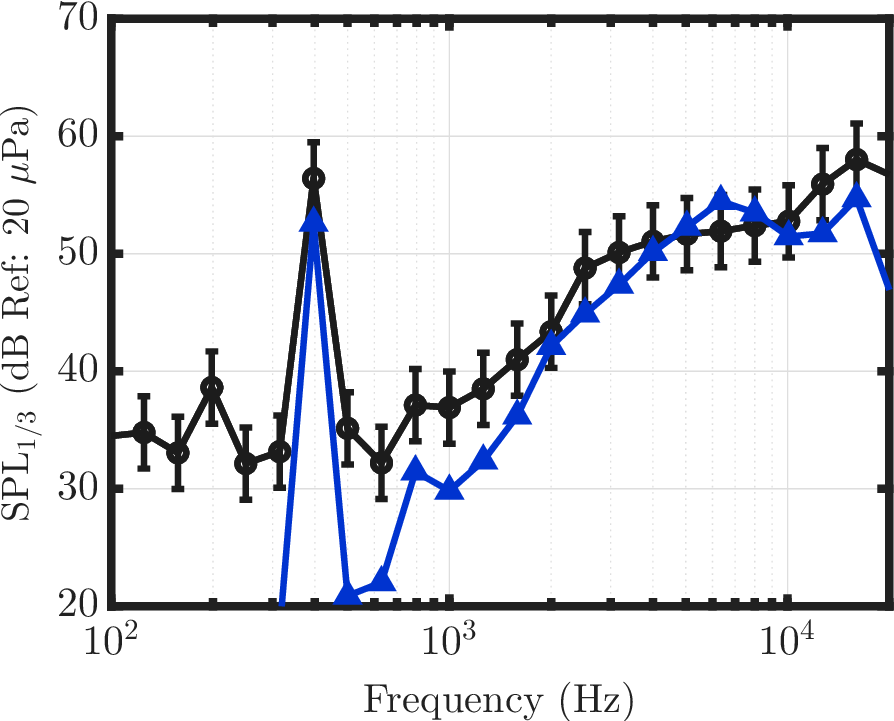}}
    \hspace{0.15 cm}
    \subfloat[\label{subfig: }]{\includegraphics[width = 0.32\textwidth]{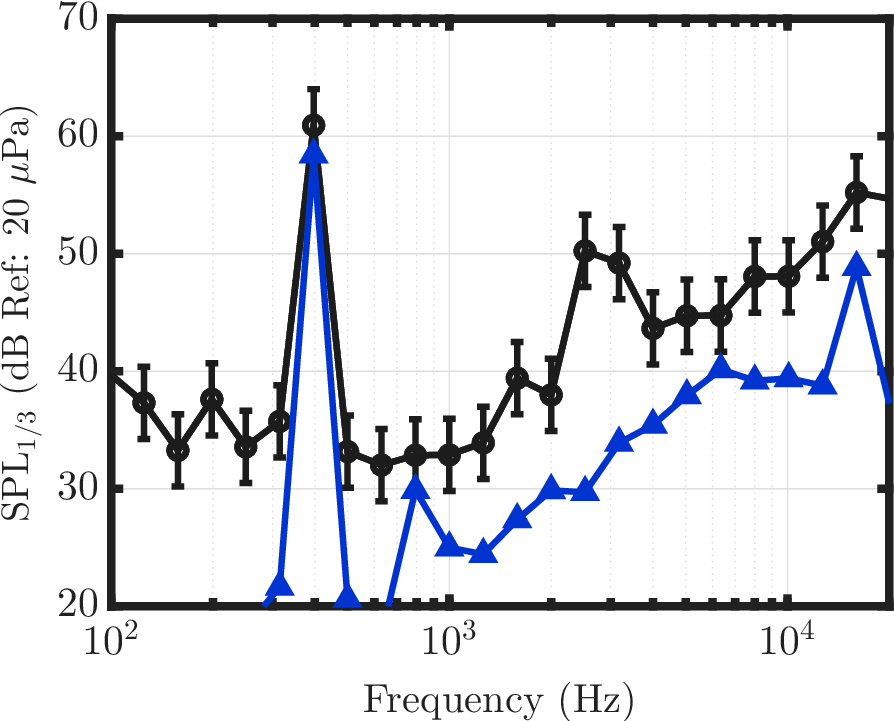}}
    
    \subfloat[\label{subfig: }]{\includegraphics[width = 0.32\textwidth]{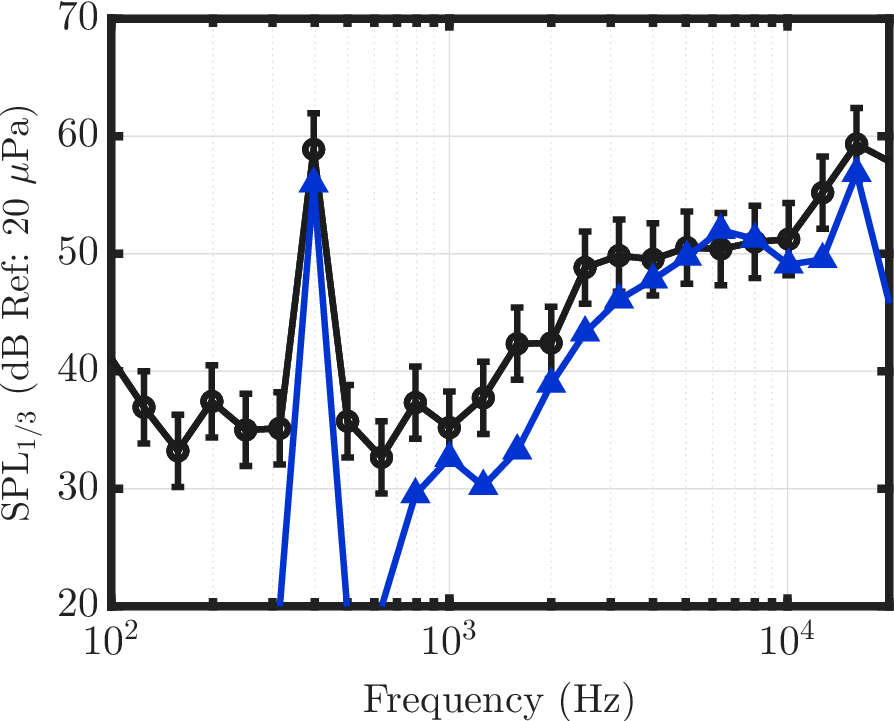}}
    \hspace{0.15 cm}
    \subfloat[\label{subfig: }]{\includegraphics[width = 0.32\textwidth]{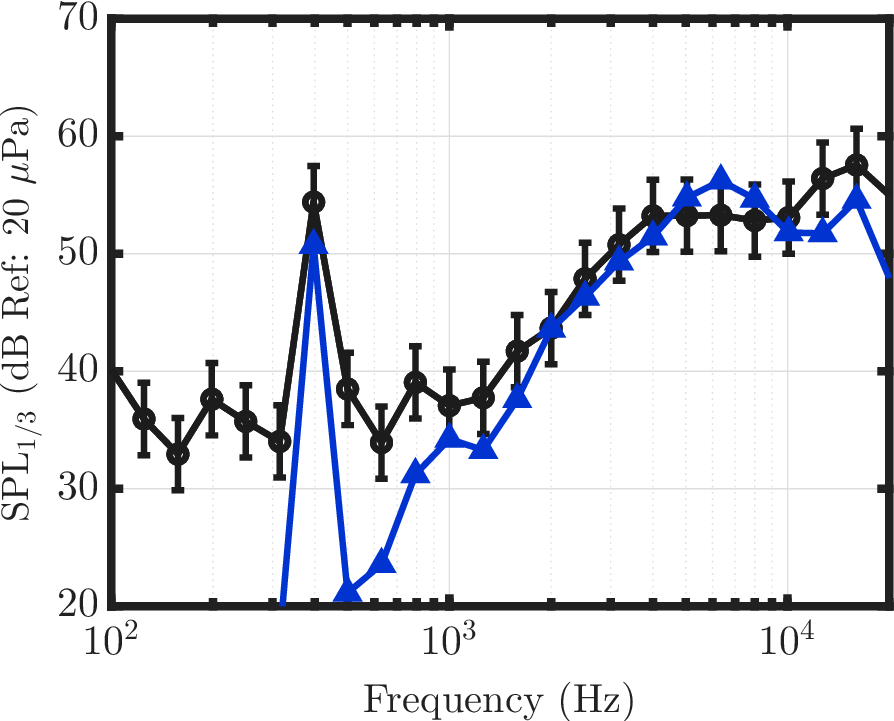}}
    \hspace{0.15 cm}
    \subfloat[\label{subfig: }]{\includegraphics[width = 0.32\textwidth]{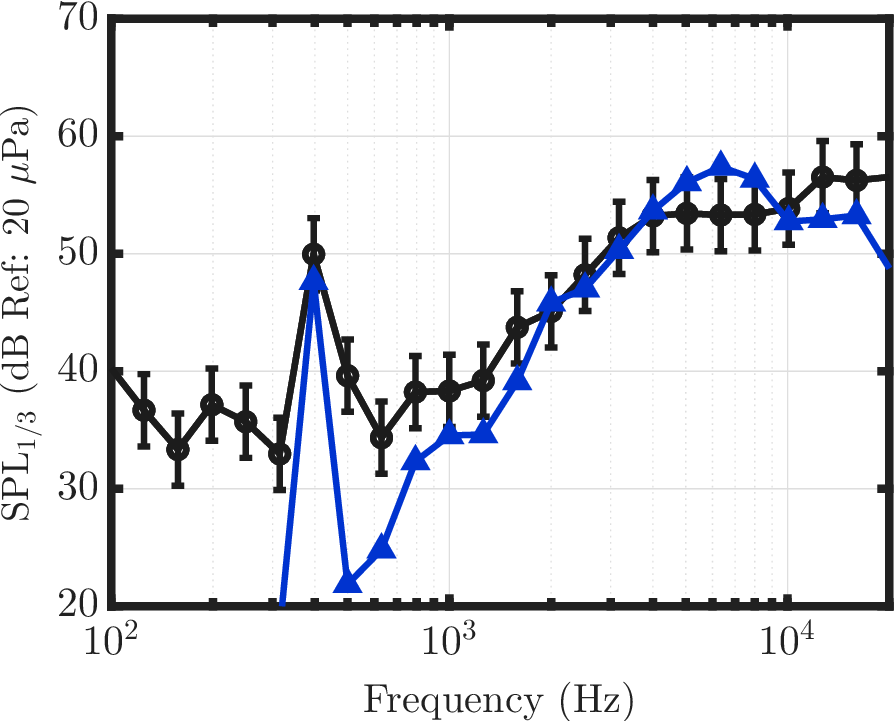}}
    \caption{One-third octave band far-field acoustic spectrum comparison of the combined HRLES/FW-H and UCD-QuietFly predictions against the experimental measurements of \citet{pettingill2021}: (a) mic 1, (b) mic 2, (c) mic. 3, (d) mic 4, (e) mic 5, (f) mic. 6.}
    \label{fig: ITR_SPL13_Validation}
\end{figure}

Figure~\ref{fig: ITR_SPL13_Validation} presents one-third-octave-band predictions compared against experimental measurements at the microphone locations listed in Table~\ref{tab: microphone_locations}, with $\pm 3~\mathrm{dB}$ uncertainty bounds. The one-third-octave-band SPL comparison provides a more robust and convenient validation, representing the total acoustic energy within each frequency band. The mid-frequency BSVI noise between $1$ and $10$~kHz agrees well with the experimental measurements at microphone locations $1$, $2$, $5$, and $6$, where the predictions fall within the error bars. However, the predicted SPL between $5$ and $10$~kHz lies at the boundary of the experimental uncertainty, consistent with the overpredicted higher-frequency tones observed in the narrowband spectrum. TBL-TE noise predictions are also included in Fig. \ref{fig: ITR_SPL13_Validation}, but do not noticeably affect the one-third-octave-band noise levels owing to their relatively low magnitude compared with the harmonic peaks.

While the overall trend and magnitude are well captured at most microphone locations, the predictions at microphone location 3 significantly underpredict the measured SPL, with the exception of the low-frequency peak associated primarily with first-BPF thickness noise. This observer is positioned directly in the plane of the rotor disk and is therefore not expected to experience significant loading noise, which dominates the higher-frequency signature. The measured SPL does exhibit a decrease at this location, but it remains significantly higher than the predicted values. The most likely explanation for this discrepancy is motor noise and other real-world artifacts present during the experiments: for example, structural vibrations arising from experimental rig interactions with the blades and motor, together with acoustic reflections off these extra components, could account for the significant high-frequency noise observed. Moreover, \citet{pettingill2021} reported isolated motor noise tones at these mid-frequencies. Appendix \ref{Appendix: Validation Cont.} provides additional information concerning the unexplained noise sources at microphone location 3.



\begin{figure}
    \centering
    \subfloat[\label{subfig: ITR_Narrowband_df45}]{\includegraphics[width = 0.45\textwidth]{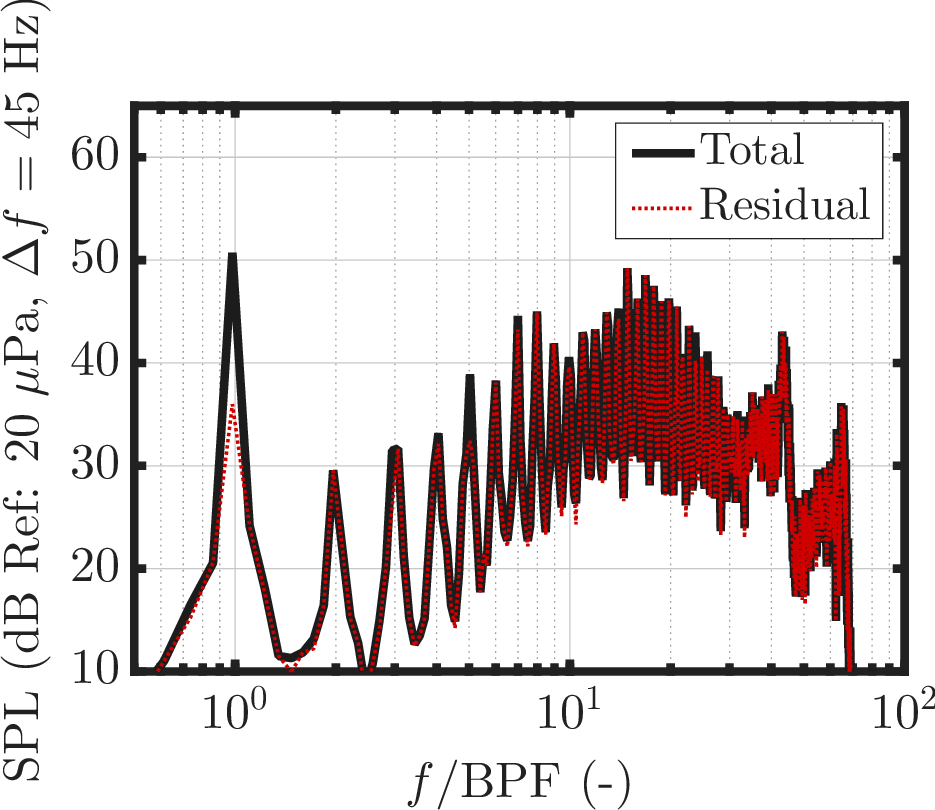}}
    \hspace{0.25 cm}
    \subfloat[\label{subfig: ITR_OneThird_Compare}]{\includegraphics[width = 0.45\textwidth]{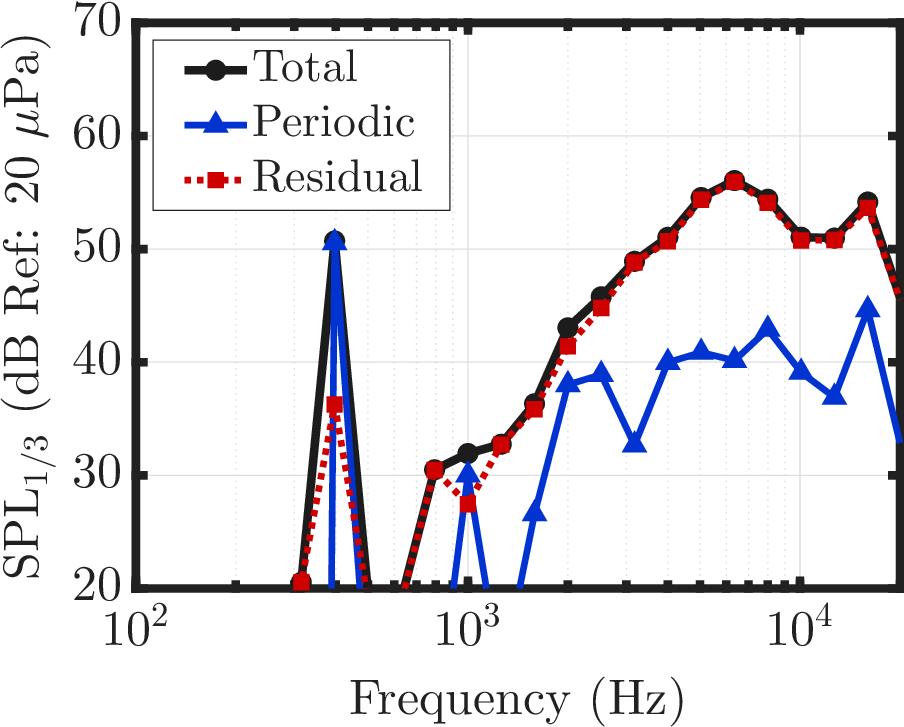}}
    \caption{Comparison of total, periodic, and residual far-field acoustic predictions for mic. 5 (HRLES/FW-H only, no TBL-TE contribution): (a) narrowband ($\Delta f = 45$~Hz) spectrum and (b) one-third-octave-band spectrum.}
    \label{fig: ITR_Comparison}
\end{figure}

Figures \ref{subfig: ITR_Narrowband_df45} and \ref{subfig: ITR_OneThird_Compare} present the HRLES/FW-H narrowband ($\Delta f=45$~Hz) and one-third-octave-band predictions, respectively, comparing the total, periodic, and residual pressure signals without experimental measurements. Both figures show that the residual signal's SPL accounts for most of the total predicted acoustic signature. These higher-harmonic tones correspond to the $1$--$10$~kHz range in Fig. \ref{subfig: ITR_OneThird_Compare}, where the residual and periodic noise sources differ in SPL by $10$--$15$~dB. This demonstrates that the BPF tones are primarily produced by stochastic loading and dominate the higher-frequency noise to which humans are more sensitive. In contrast, the residual noise is not significant at the first BPF, as expected, since the acoustics at the first BPF are dominated by deterministic phenomena such as periodic blade loading and thickness noise arising from the blade's kinematics.

\begin{figure}
    \centering
    {\includegraphics[width = 0.90\textwidth]{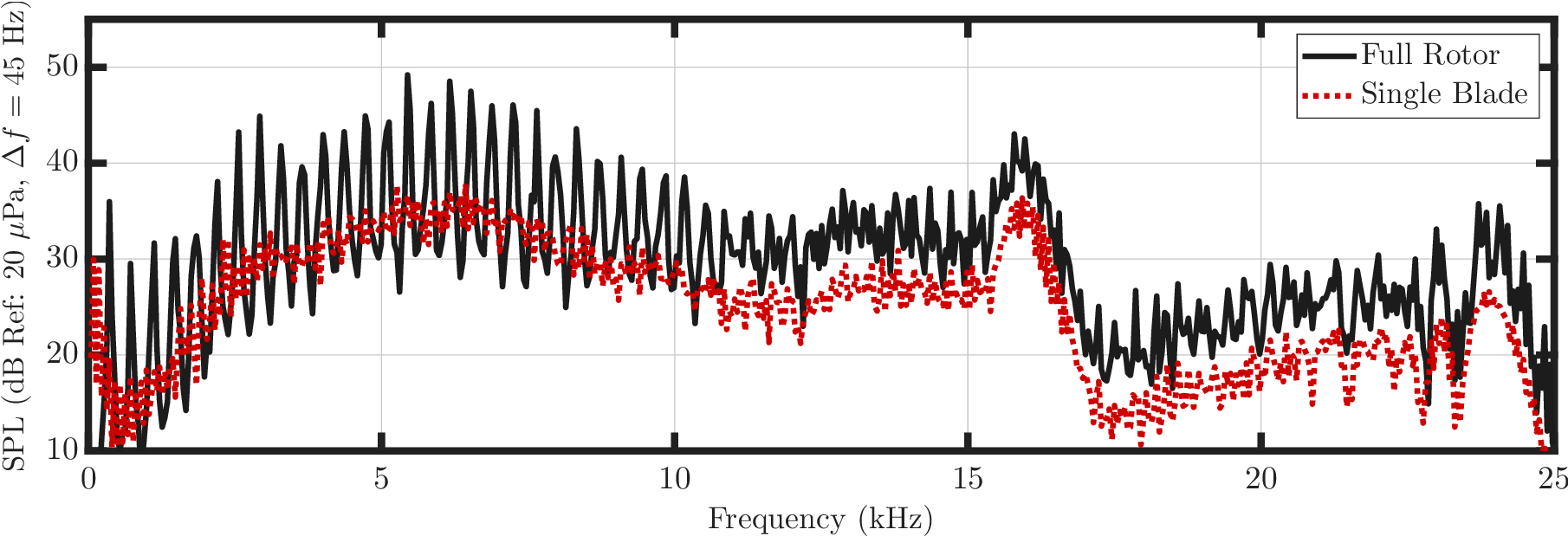}}    
    \caption{Residual far-field acoustic spectrum ($\Delta f = 45$~Hz) at mic. $5$, comparing HRLES/FW-H predictions for a single blade with those for the full four-bladed rotor.}
    \label{fig: BladeRotorComparison}
\end{figure}

A comparison between impermeable FW-H predictions for a single blade and for all four blades (full rotor) is provided in Fig. \ref{fig: BladeRotorComparison}. Both predictions are computed from the same HRLES simulation and time steps. UCD-QuietFly TBL-TE noise predictions are not included here, and the sound spectrum shown consists only of the residual acoustic component. As expected, the overall trends of both predictions agree, revealing dominant far-field noise between $1$--$10$~kHz. However, when the FW-H surface includes all four blades, this mid-frequency noise increases significantly at integer multiples of the BPF; relative to the single-blade prediction, the full-rotor prediction also exhibits valleys with lower SPLs. Thus, despite the aperiodicity of the acoustic pressure in the time domain---established through the phase-averaging procedure described in Section \ref{Acoustic Processing}---the sound spectrum exhibits quasi-tonal behavior, characterized by constructive interference at BPF harmonics and destructive interference in between. This effect is even more evident in Appendix \ref{Appendix: Validation Cont.}, which compares the FW-H-only predictions with the measurements at a bandwidth of $\Delta f = 20~\mathrm{Hz}$. Finally, above $10~\mathrm{kHz}$, the high-frequency noise is associated with amplified bluntness vortex shedding (BVS) noise \citep{WonLee2025, WonLee2026-Bevel} and does not exhibit this interference.

\begin{figure}
    \centering
    \includegraphics[width=0.95\linewidth]{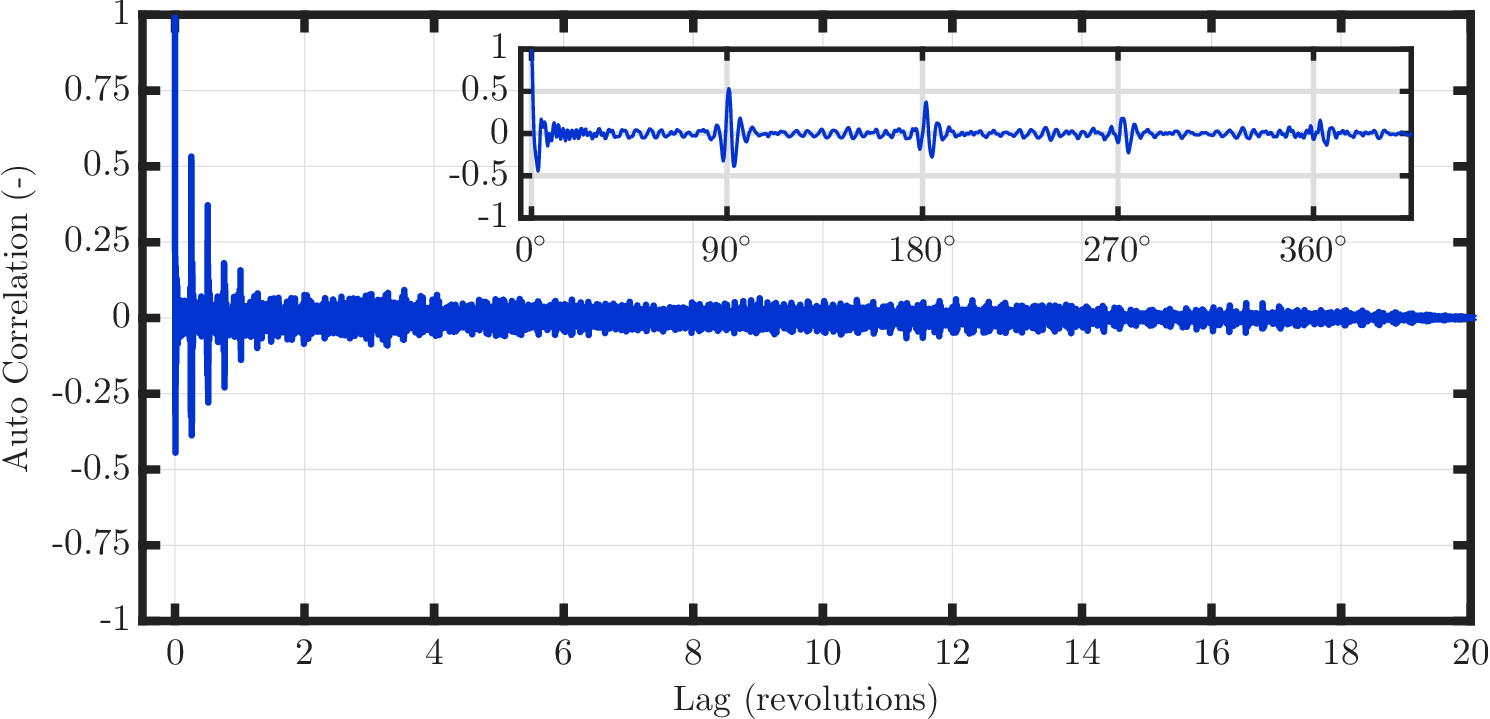}

    \caption{Autocorrelation of the HRLES/FW-H full rotor acoustic prediction plotted against time lag normalized by rotor period ($T_0 = 0.0109~\mathrm{s}$). The inset displays the autocorrelation spanning the first rotor period, with time lag conveyed in units of rotor azimuth angle.}
    \label{fig: AcousticAutoCorrelation}
\end{figure}

The constructive interference effect implies some degree of blade-to-blade correlation, which is confirmed by Fig. \ref{fig: AcousticAutoCorrelation}. The autocorrelation of the four-bladed FW-H prediction is plotted as a function of time lag normalized by the rotor period, with an expanded inset provided to show the first revolution in detail. The correlation clearly peaks at time lags corresponding to $90^\circ$ intervals, consistent with the $90^\circ$ blade spacing; importantly, however, this correlation becomes negligible beyond a $360^\circ$ time lag. 

This acoustic autocorrelation signature bears similarities to that of TI noise, which has been extensively studied by \citet{paterson&amiet1979}. Just as TI noise produces blade-to-blade-correlated haystack peaks at BPF harmonics through the repeated chopping of a single elongated, anisotropic turbulent eddy---while remaining fundamentally aperiodic from one rotor revolution to the next---the BSVI mechanism produces a comparable combination of short-time (sub-revolution) correlation and long-time (revolution-to-revolution) decorrelation. This analogy further supports the interpretation of BSVI as an intermittent, quasi-tonal noise source rather than a purely periodic one.

\begin{figure}
    \centering
    {\includegraphics[width = 0.90\textwidth]{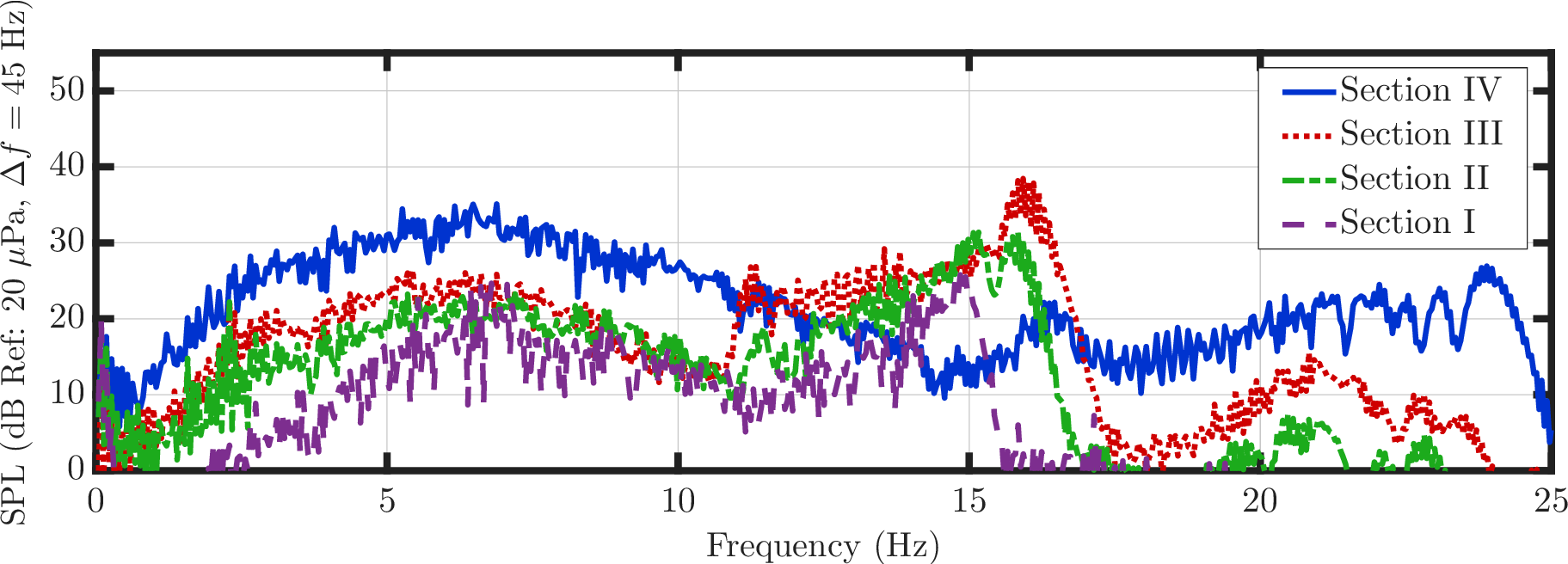}}
    \caption{Residual far-field acoustic spectrum ($\Delta f = 45$~Hz) at microphone $5$, comparing FW-H predictions for the four individual blade regions (I--IV) defined in Fig.~\ref{fig: ITR Geometry Blade}.}
    \label{fig: SingleBladeSections}
\end{figure}

Figure \ref{fig: SingleBladeSections} presents FW-H predictions for a single blade with impermeable surfaces corresponding to the different blade regions previously defined in Fig. \ref{fig: ITR Geometry Blade}. Region IV---spanning $0.9 < r/R \le 1.0$---exhibits significantly more noise between $1$--$10$~kHz, consistent with the unsteady loading shown in Fig. \ref{fig: dCndpsi_instant}. This provides further evidence that the dominant noise source underlying the higher-harmonic BPF peaks in this frequency range is the outboard loading caused by secondary vortex impingement: the relatively reduced SPL predicted for the other regions reinforces this conclusion, since Regions I and II encompass a larger area than III and IV yet produce less noise. Finally, although not the focus of this study, Regions II and III exhibit prevalent BVS noise above $10$~kHz \citep{WonLee2025,WonLee2026-Bevel}.

\begin{figure}
    \centering
    \includegraphics[width=0.90\linewidth]{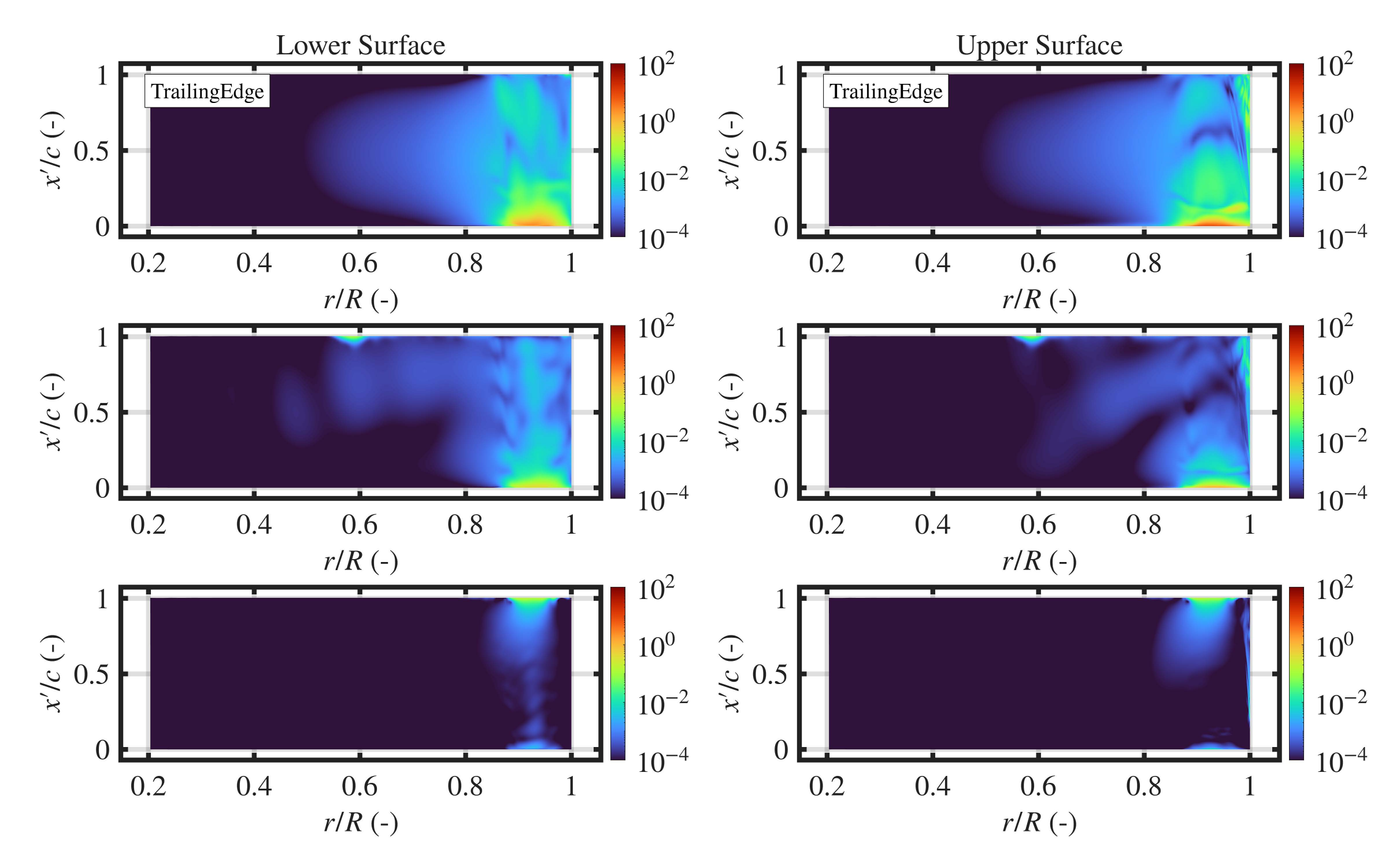}
   
    \caption{Untwisted blade contours colored by WPS magnitude ($\mathrm{Pa/Hz}$) taken at various frequencies: (top) $f = 5{,}000~\mathrm{Hz}$, (middle) $f = 10{,}000~\mathrm{Hz}$, and (bottom) $f = 17{,}000~\mathrm{Hz}$}
    \label{fig: WPS Contours}
\end{figure}

WPS spectrum contours for a single blade are provided in Fig. \ref{fig: WPS Contours} at several frequencies, showing both the lower (left) and upper (right) surfaces. The $x$ and $y$ coordinates are transformed to $x^\prime$ and $y^\prime$, defined as

\begin{equation} \label{eq: body coordinates}
    \begin{aligned}
        x^\prime &= x\cos\left(\frac{\theta_\mathrm{tip}}{r/R}\right) + y\sin\left(\frac{\theta_\mathrm{tip}}{r/R}\right) + 0.25c, \\
        y^\prime &= -x\sin\left(\frac{\theta_\mathrm{tip}}{r/R}\right) + y\cos\left(\frac{\theta_\mathrm{tip}}{r/R}\right),
    \end{aligned}
\end{equation}

\noindent where the blade is assumed to be aligned at zero-degree azimuth. This body-coordinate transformation is used to generate contours of an untwisted blade with its leading edge at $(x^\prime,y^\prime)=(0,0)$ for any blade section $r/R$.

Three frequencies are presented to highlight where pressure fluctuations occur, since these are directly related to the loading-noise terms in Eq. \eqref{eq: FF1A_loading}. At $f = 5.0~\mathrm{kHz}$ (top panel), the WPS contours show a maximum at the leading edge, between $0.86 \le r/R \le 0.98$, consistent with the earlier finding that this mid-frequency noise results from leading-edge interactions with secondary vortices. Figure \ref{fig: ITR Vorticity Iso Surface} showed that these structures convect past the blade, producing the reduced but still present WPS magnitudes downstream of the leading edge. A similar pattern is observed for the $f = 10.0~\mathrm{kHz}$ contour (middle panel), although the downstream loading is far more apparent on the lower surface than the upper surface, likely because smaller secondary vortices interact only with the lower surface of the blade owing to its positive pitch. Finally, the bottom panel shows the $f = 17.0~\mathrm{kHz}$ contour, well above the $1$--$10$~kHz range associated with BSVI; here, as expected, the noise is concentrated in a lobe near the trailing edge, consistent with the association of these higher-frequency sources with BVS and amplified BVS noise \citep{WonLee2025}.

\begin{figure}
    \centering
    \subfloat[\label{subfig: WPS plot fore}]{\includegraphics[width = 0.45\textwidth]{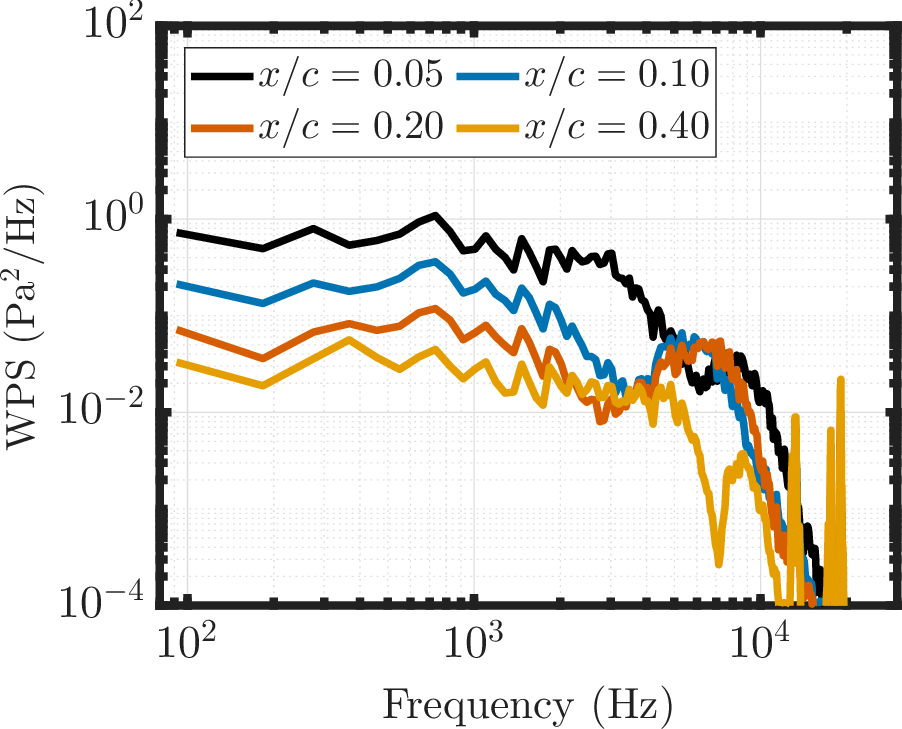}}
    \hspace{0.50 cm}
    \subfloat[\label{subfig: WPS plot aft}]{\includegraphics[width = 0.45\textwidth]{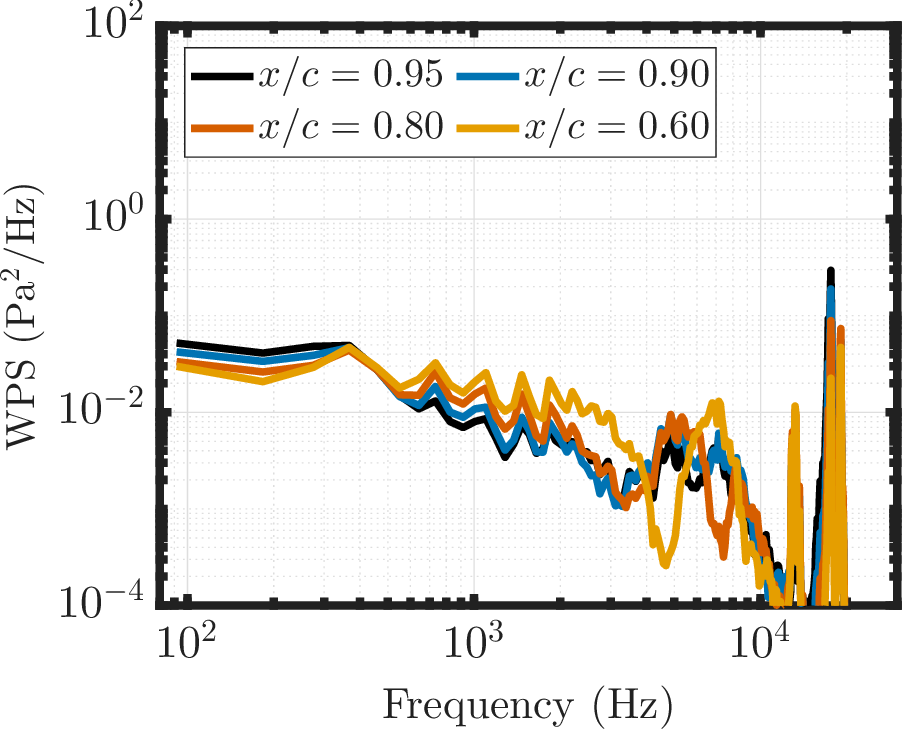}}

    \caption{Wall-pressure spectrum evaluated at the $r/R = 0.90$ radial section at various chordwise probes on the upper surface: (a) forward probes ($x/c = 0.05$--$0.40$) and (b) aft probes ($x/c = 0.60$--$0.95$).}
    \label{fig: WPS Plot}
\end{figure}

A more quantitative analysis is presented in Fig. \ref{fig: WPS Plot}, where the WPS is computed at various chordwise probes along the $r/R =0.90$ radial section on the upper surface. Figure \ref{subfig: WPS plot fore} shows that most pressure fluctuations over the forward half of this outboard section are associated with mid-frequency noise below $10$~kHz, with the WPS magnitude greatest at the $x/c = 0.05$ probe. However, between $4$ and $9$~kHz, the $x/c = 0.10$ and $0.20$ probes exhibit a notable hump in WPS magnitude that even exceeds that at $x/c = 0.05$. It is not fully understood why this higher-frequency hump appears slightly downstream of the leading edge, but it is shown later that these frequencies are more closely associated with the secondary vortex structures. The aft WPS probes are shown in Fig. \ref{subfig: WPS plot aft}, where a prominent peak is observed at approximately $12$~kHz that diminishes as the probe moves forward toward the leading edge; this peak is associated with amplified BVS, as discussed by \citet{WonLee2025}, consistent with the trailing-edge lobes presented in Fig. \ref{fig: WPS Contours}.

\subsection{BSVI Characterization and Statistics} \label{sec: statistics}

\begin{figure}
    \centering
    \includegraphics[width=0.95\linewidth]{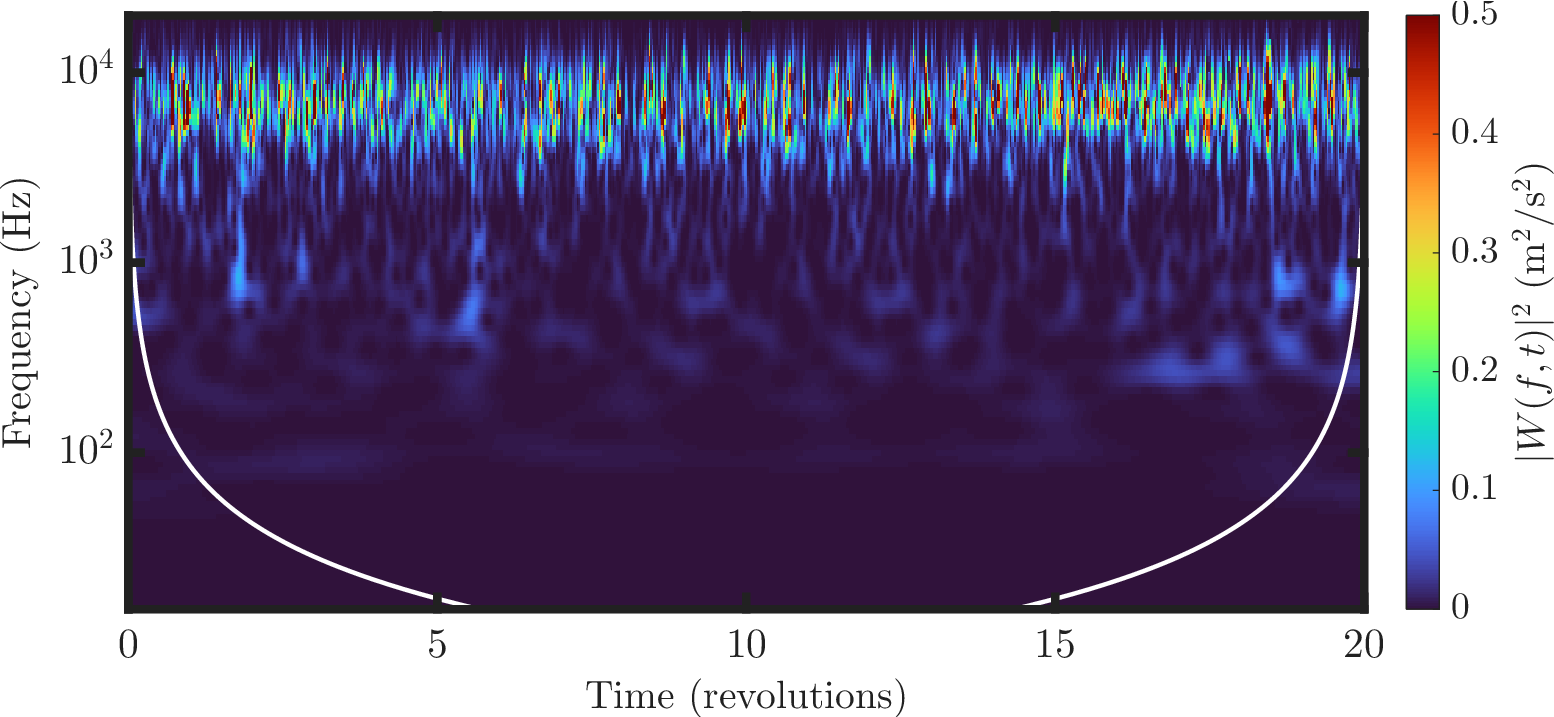}
    \caption{Upwash wavelet transform evaluated at the $r/R = 0.90$ radial section for a probe at $(x^\prime,y^\prime)= (-0.05c,0.00c)$.}
    \label{fig: wavelet}
\end{figure}

This section further characterizes the BSVI noise mechanism through analysis of the flow-field properties near a single blade as it rotates. Throughout this section, coordinates are expressed relative to the body coordinates defined in Eq. \eqref{eq: body coordinates}. Figure \ref{fig: wavelet} shows the scalogram computed from MATLAB's continuous wavelet transform of the zero-mean residual upwash fluctuation, sampled from a probe at $x^\prime/c = -0.05$ and $y^\prime/c = 0.0$ for the $r/R = 0.90$ radial section. This probe location was selected for its proximity to the BSVI region.

The scalogram shows that upwash fluctuations---which ultimately produce unsteady loading---occur intermittently in time and are most significant between $3.0$ and $10$~kHz. This intermittency explains why the mid-frequency peaks associated with BSVI noise appear in the residual signal rather than the periodic one. Notably, the higher-frequency unsteady upwash exhibits far greater magnitudes than the lower-frequency upwash between $0.5$ and $3.0$~kHz. The acoustic signature shown earlier in Fig. \ref{fig: BladeRotorComparison} also exhibits lower-frequency residual tones, namely the first $10$ BPFs, corresponding to frequencies below $3.6$~kHz. Although these tones are not as dominant as the higher-frequency ones, they are also analyzed to help identify their physical origin.

\begin{figure} 
    \centering
    \subfloat[\label{subfig: Upwash PSD Probe 1}]{\includegraphics[width = 0.45\textwidth]{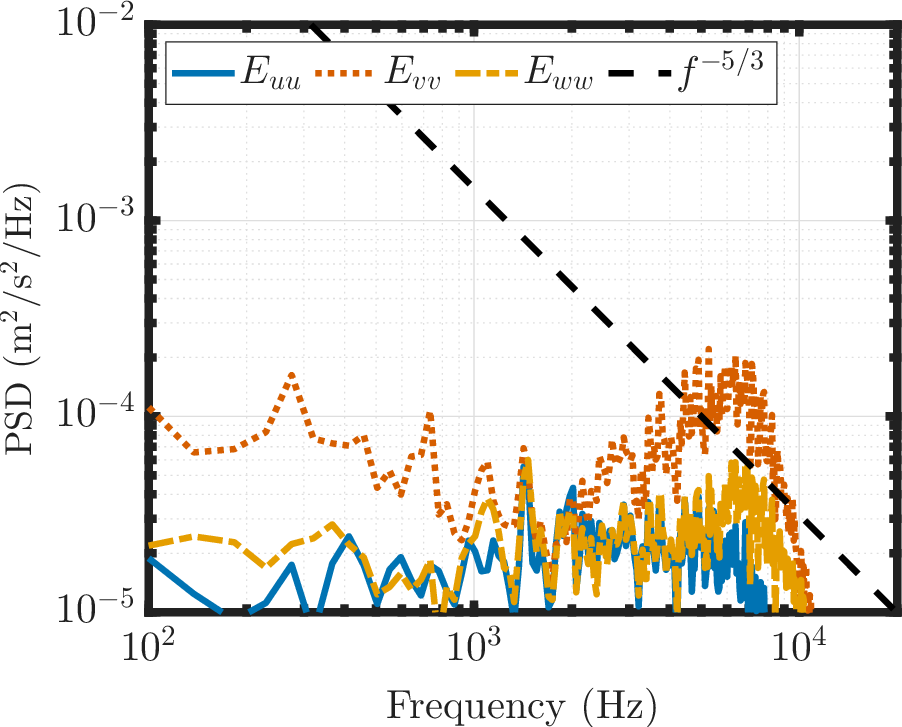}}
    \hspace{0.25 cm}
    \subfloat[\label{subfig: Upwash PSD Probe 2}]{\includegraphics[width = 0.45\textwidth]{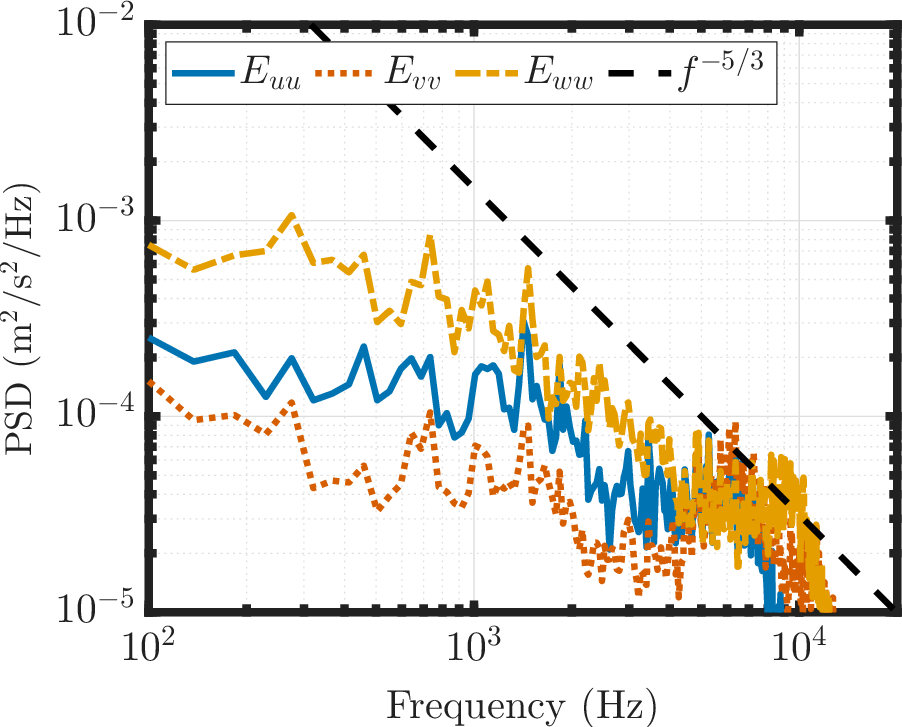}}
    \caption{Velocity fluctuation PSD evaluated at the $r/R = 0.90$ radial section for (a) a probe at $(x^\prime,y^\prime)= (-0.05c,0.00c)$ and (b) $(x^\prime,y^\prime)= (-0.05c,0.20c)$}
    \label{fig: Upwash PSD}
\end{figure}

Figure \ref{subfig: Upwash PSD Probe 1} shows the velocity PSD computed at the same location as the upwash scalogram in Fig. \ref{fig: wavelet}. The three velocity components are defined such that $u$ is aligned with the blade's tangential velocity vector, $v$ is normal to the rotor disk, and $w$ is aligned with the blade's radial direction. Notably, the upwash PSD ($E_{vv}$) is drastically different from $E_{uu}$ and $E_{ww}$, where it exhibits a notably higher magnitude and, between $3.0$--$10$~kHz, a large hump is observed, consistent with the upwash scalogram, which showed intermittent and significantly higher-magnitude upwash fluctuations at these frequencies. Furthermore, this comparison between the velocity PSDs reveals an interesting property of secondary vortices: they produce a highly anisotropic spectrum that does not decrease monotonically with frequency as one may expect from a traditional energy cascade. Indeed, the higher-frequency hump observed in $E_{vv}$, which is also slightly observed in $E_{ww}$, exhibits greater energy than the lower-frequency fluctuation due to the presence of convecting secondary vortex structures.  


Figure \ref{subfig: Upwash PSD Probe 2} shows the velocity PSD for a probe located below the blade within the PTV core. As with the previous probe, the differences among the $u$, $v$, and $w$ spectra reveal significant anisotropy. A higher-frequency hump in the $E_{vv}$ PSD is also observed, but it is far less pronounced than the probe immediately forward of the blade's leading edge. Additionally, each velocity PSD better aligns with the $-5/3$ scaling law expected from the energy cascade. Thus, the weaker upwash hump, together with this scaling, highlights the structural distinction between the velocity fluctuations within the PTV core and those in the surrounding secondary vortices.

\begin{figure}
    \centering
    \subfloat[\label{subfig: Velocity Probe 90 low}]{\includegraphics[width = 0.45\textwidth]{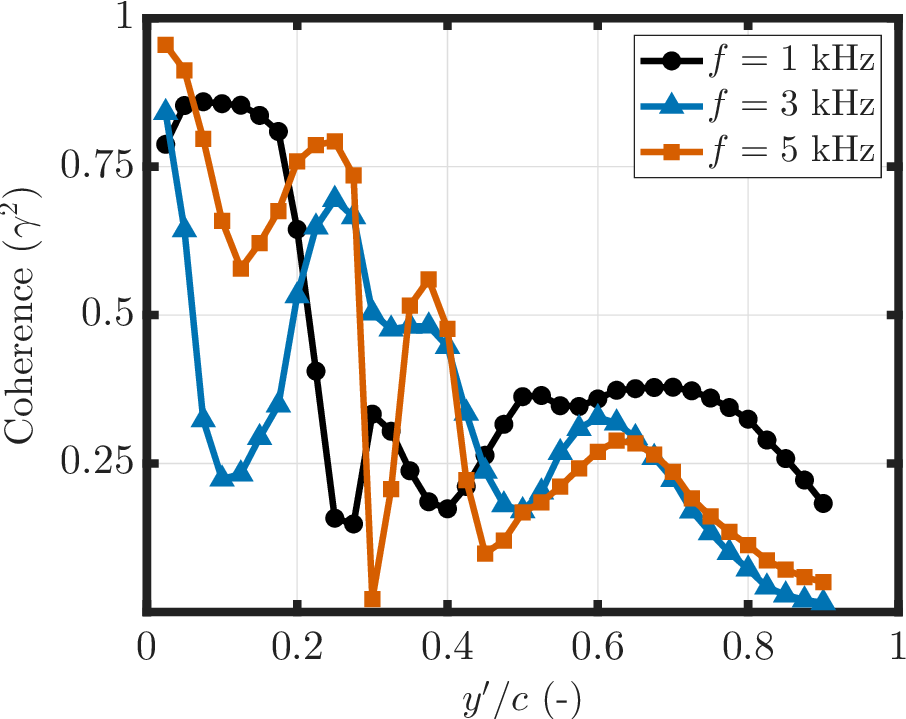}}
    \hspace{0.25 cm}
    \subfloat[\label{subfig: Velocity Probe 95 low}]{\includegraphics[width = 0.45\textwidth]{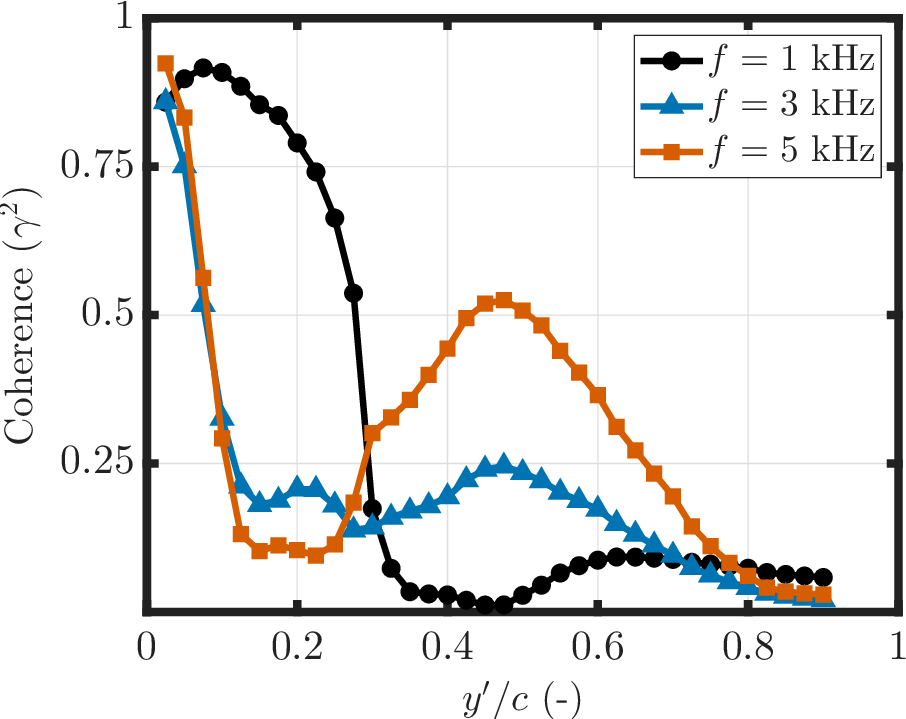}}

    \subfloat[\label{subfig: Velocity Probe 90 high}]{\includegraphics[width = 0.45\textwidth]{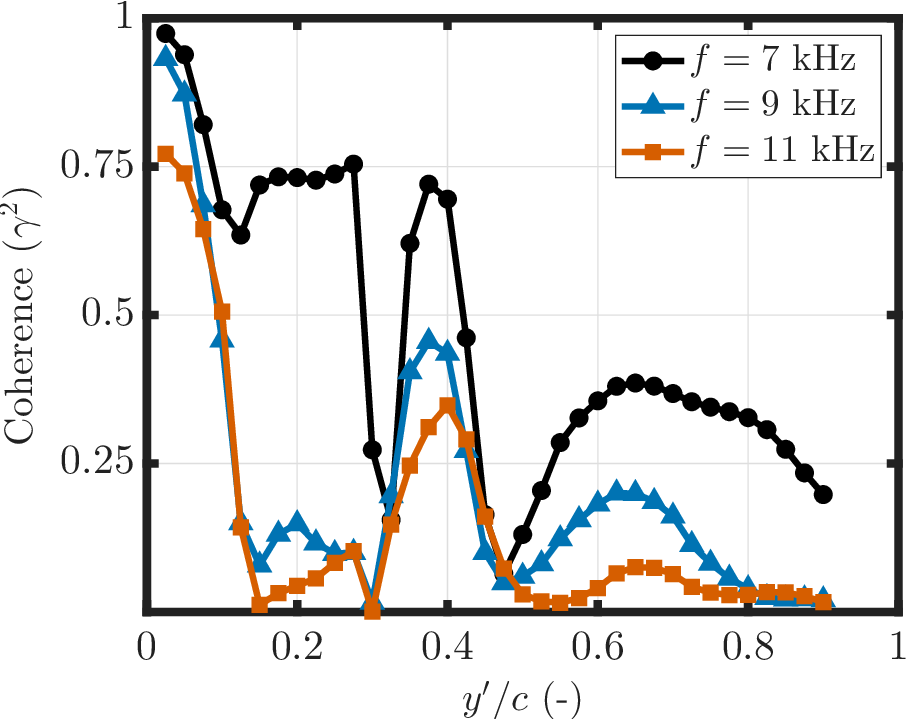}}
    \hspace{0.25 cm}
    \subfloat[\label{subfig: Velocity Probe 95 high}]{\includegraphics[width = 0.45\textwidth]{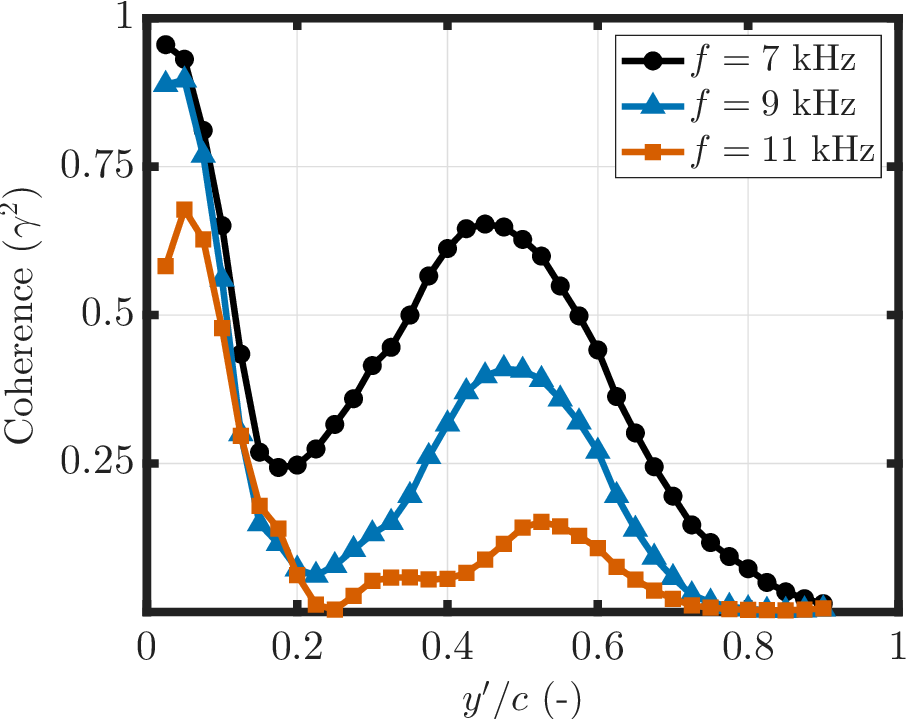}}

    \caption{Squared coherence between velocity fluctuations sampled at $x^\prime = -0.05c$ and the leading-edge pressure fluctuation, plotted as a function of $y^\prime/c$ below the rotor disk, for the $r/R = 0.90$ (left column) and $r/R = 0.95$ (right column) radial sections at lower ($f = 1$, $3$, $5$~kHz; top row) and higher ($f = 7$, $9$, $11$~kHz; bottom row) frequencies.}
    \label{fig: Velocity Probe Coherence}
\end{figure}

Figure \ref{fig: Velocity Probe Coherence} shows the squared coherence between velocity fluctuations at various probes located at $x^\prime = -0.5c$ and the pressure fluctuation on the blade's upper surface, for the $r/R = 0.90$ (left) and $0.95$ (right) radial sections. This coherence measures how closely a velocity fluctuation convecting through the domain relates to the unsteady loading at the blade's leading edge, which contributes to the dipole noise sources expressed in FF1A. The squared coherence is plotted as a function of probe location in the $y^\prime$ direction (i.e., below the rotor disk) for several frequencies. For both radial sections, the lower-frequency ($1.0$~kHz) coherence exhibits a single broad hump over $0.0 \le y^\prime/c \le 0.30$, whose maximum lies not at $y^\prime/c = 0.0$ (directly ahead of the leading edge) but slightly below it. Referencing the secondary vortex system portrayed in Fig. \ref{fig: Secondary Vortex System}, this hump is associated with the location of the first PTV, where its diameter spans this region. The higher-frequency coherence plots, by contrast, exhibit multiple narrower humps, consistent with the smaller-scale structures associated with secondary vortices.

Comparing these narrow humps in Figs.~\ref{subfig: Velocity Probe 90 low} and \ref{subfig: Velocity Probe 90 high} with the instantaneous secondary vortex system of Fig.~\ref{fig: Secondary Vortex System}, the squared coherence increases specifically at the regions occupied by the secondary vortices, and the multiple humps are indicative of their S-shape. Importantly, the high coherence between upper-surface blade pressure and the non-impinging upwash below the rotor further demonstrates that this leading-edge noise is generated by coherent worm-like vortices that extend below the rotor plane; this interpretation becomes more apparent in the subsequent SPOD analysis. The coherence distribution at $r/R = 0.95$ features similar behavior, but with only two broader peaks. These two peaks are a consequence of cutting an outboard and more diffuse region of the secondary vortices, which is evident in Fig.~\ref{fig: Secondary Vortex System}. These squared-coherence distributions support the conclusion that much of the mid-frequency noise between $3.0$--$10$~kHz is dominated by BSVI. Finally, there is a notable difference between the coherence distribution for the upwash within the PTV (at $1.0$ kHz) and the surrounding secondary vortex braids (between $3.0$--$10$ kHz), suggesting that the lower-BPF tones seen in Fig.~\ref{fig: BladeRotorComparison} are instead associated with stochastic velocity fluctuations caused by perpendicular BVI. This difference in spatial organization and spectral signature further distinguishes BSVI as a unique noise mechanism. Additional discussion contrasting perpendicular BVI and BSVI noise is included in Appendix~\ref{Appendix: Coarse Grid HRLES.}.

\begin{figure}
    \centering
    \subfloat[\label{subfig: }]{\includegraphics[width = 0.45\textwidth]{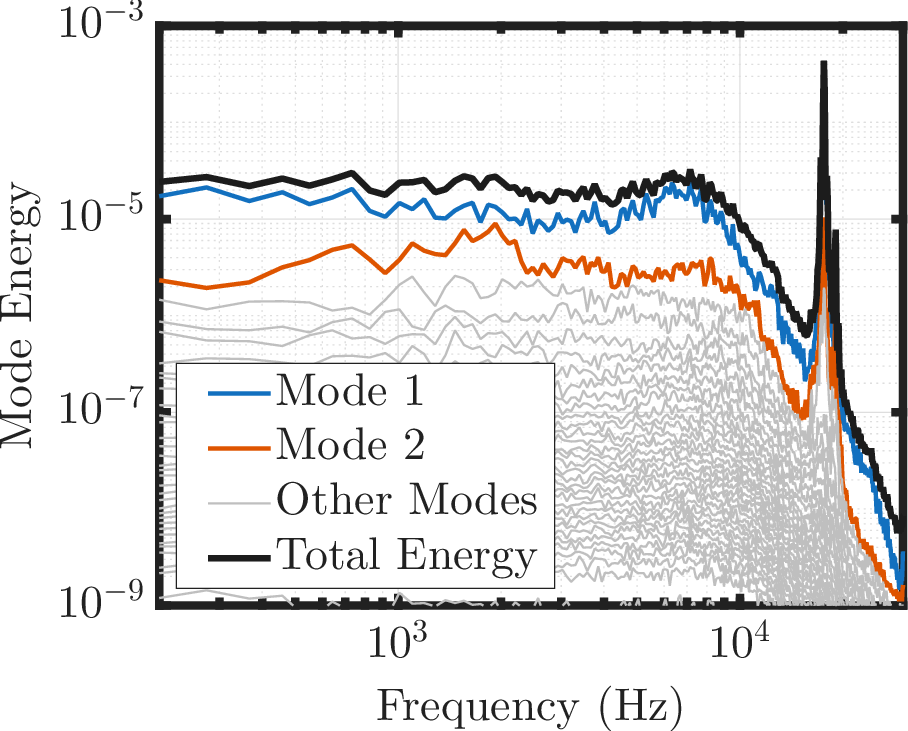}}
    \hspace{0.25 cm}
    \subfloat[\label{subfig: }]{\includegraphics[width = 0.45\textwidth]{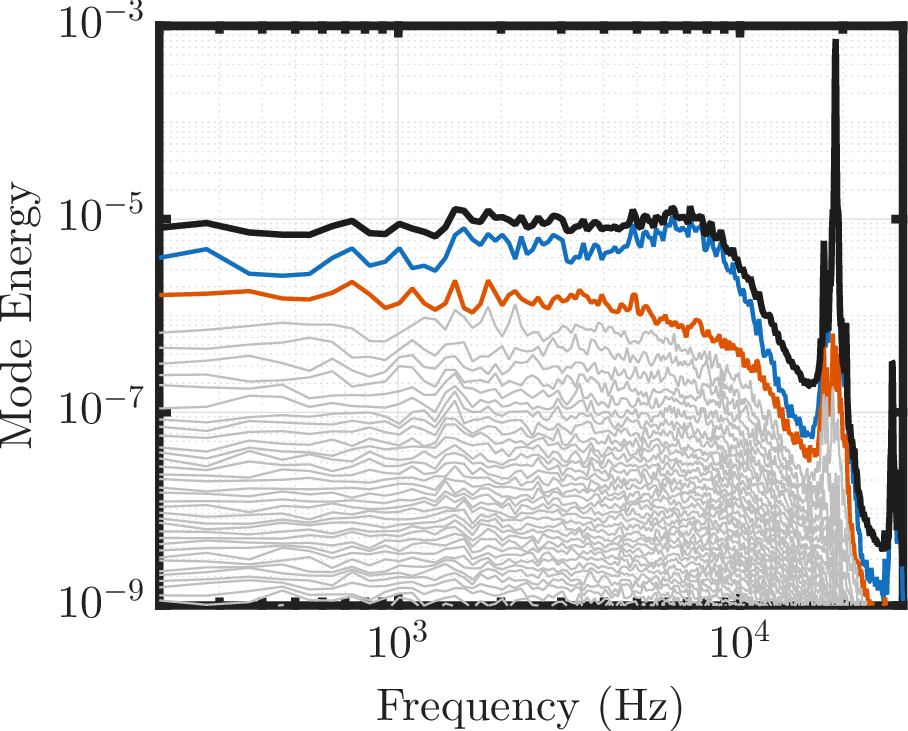}}
    \caption{SPOD mode energy spectra of the residual upwash fluctuation for the (a) $r/R = 0.90$ and (b) $r/R = 0.95$ radial sections, showing low-rank behaviour and a broad energetic hump over the $3.0$--$10$~kHz range associated with BSVI.}
    \label{fig: 2D SPOD Spectrum}
\end{figure}

SPOD analysis is conducted on the residual upwash fluctuations for the $r/R = 0.90$ and $0.95$ radial sections, using data from a single blade's NB grid. Data is sampled at $66$~kHz over $20$ time-accurate revolutions, using Welch's method with each block spanning a time period equivalent to one revolution and a $50\%$ overlap between blocks. Figure \ref{fig: 2D SPOD Spectrum} shows the mode energy spectrum for both sections, demonstrating low-rank behavior over the $3.0$--$10$~kHz range of interest and a slight hump consistent with the upwash PSD calculated for a single probe in Fig. \ref{subfig: Upwash PSD Probe 1}. This low-rank behavior indicates that the first mode is significantly more dominant than the others and can adequately represent the flow field of this 2D cross-section.

\begin{figure}
    \centering
    \subfloat[\label{subfig: 2D Mode roR090 f1000 m1}]{\includegraphics[width = 0.45\textwidth]{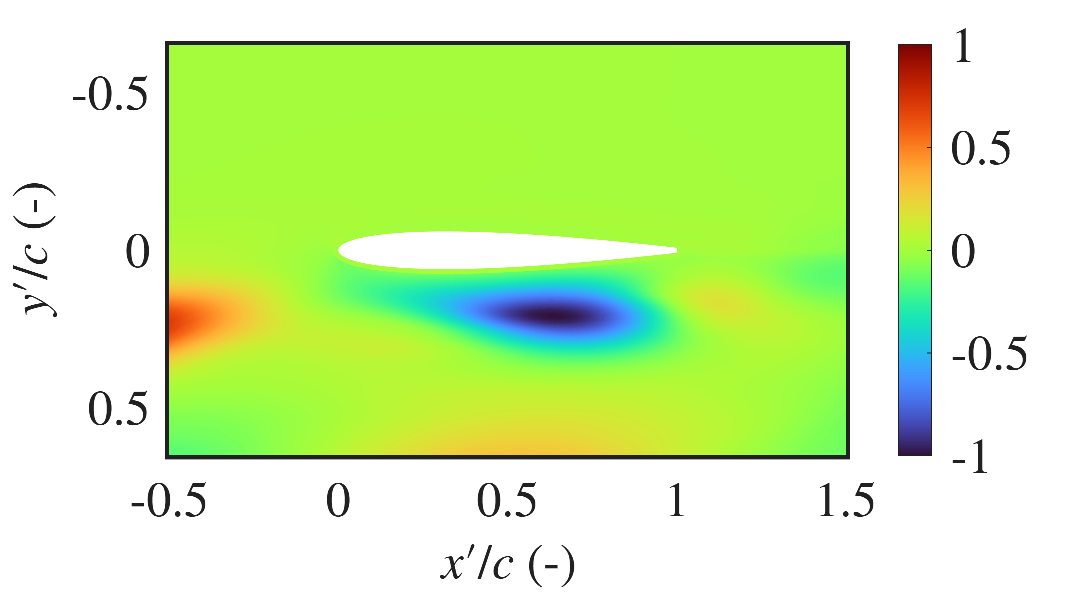}}
    \hspace{0.25 cm}
    \subfloat[\label{subfig: 2D Mode roR095 f1000 m1}]{\includegraphics[width = 0.45\textwidth]{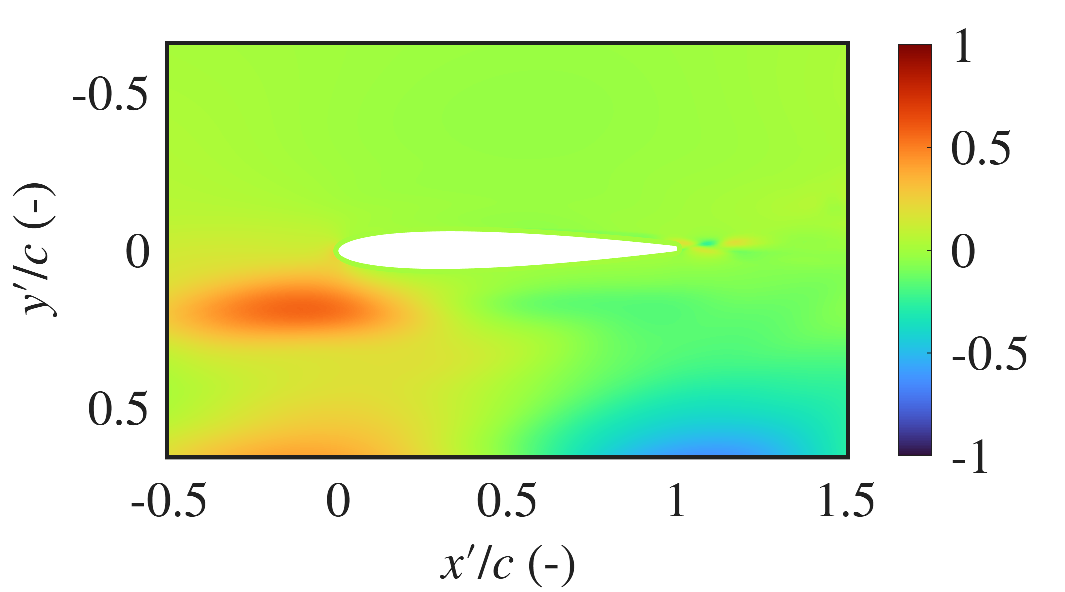}}
    
    \subfloat[\label{subfig: 2D Mode roR090 f5000 m1}]{\includegraphics[width = 0.45\textwidth]{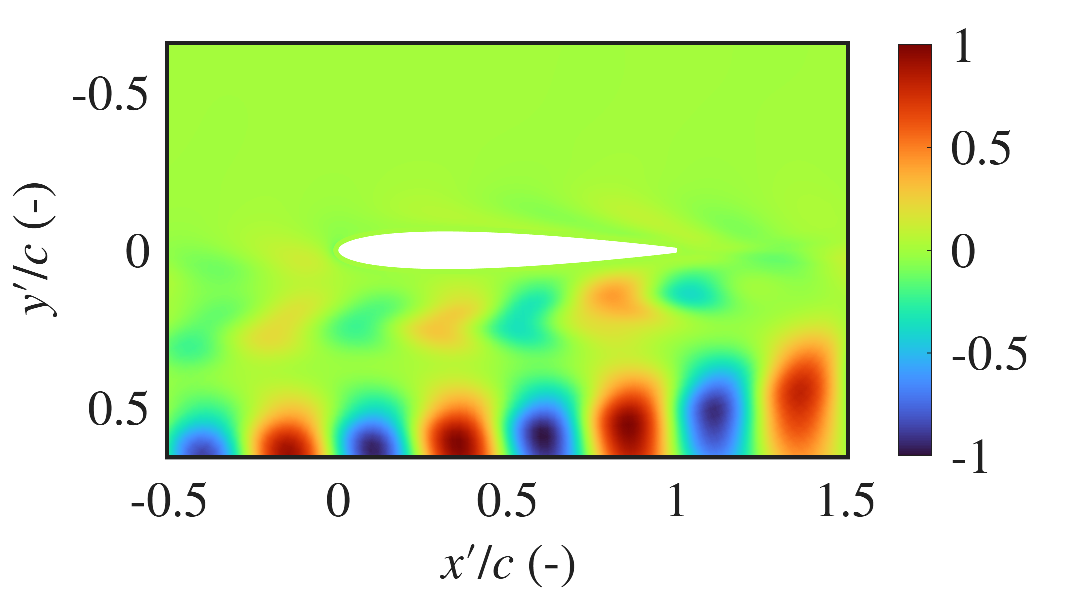}}
    \hspace{0.25 cm}
    \subfloat[\label{subfig: 2D Mode roR095 f5000 m1}]{\includegraphics[width = 0.45\textwidth]{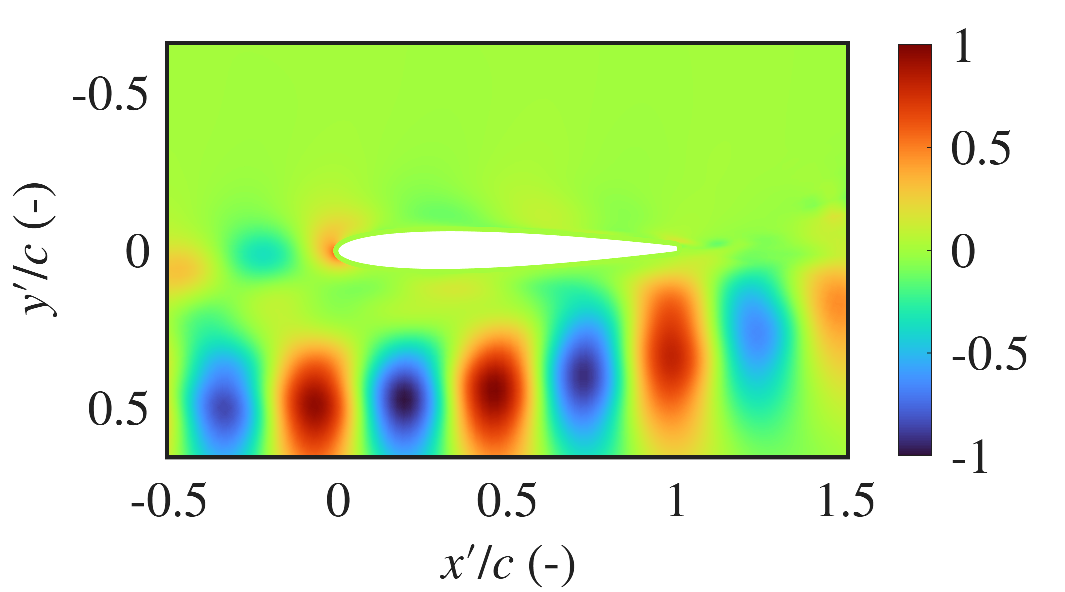}}
    
    \subfloat[\label{subfig: 2D Mode roR090 f7000 m1}]{\includegraphics[width = 0.45\textwidth]{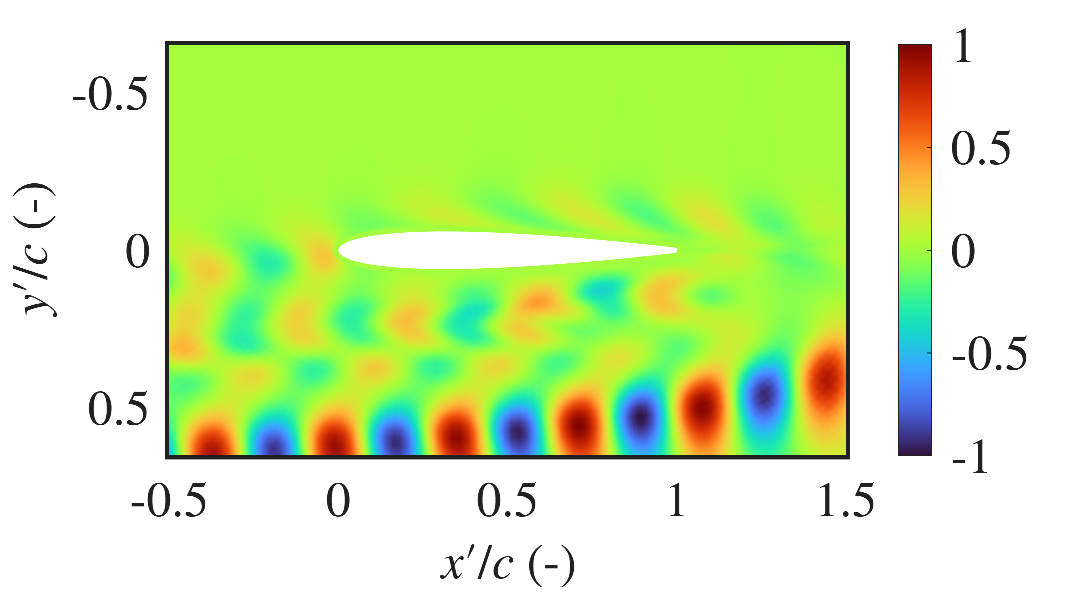}}
    \hspace{0.25 cm}
    \subfloat[\label{subfig: 2D Mode roR095 f7000 m1}]{\includegraphics[width = 0.45\textwidth]{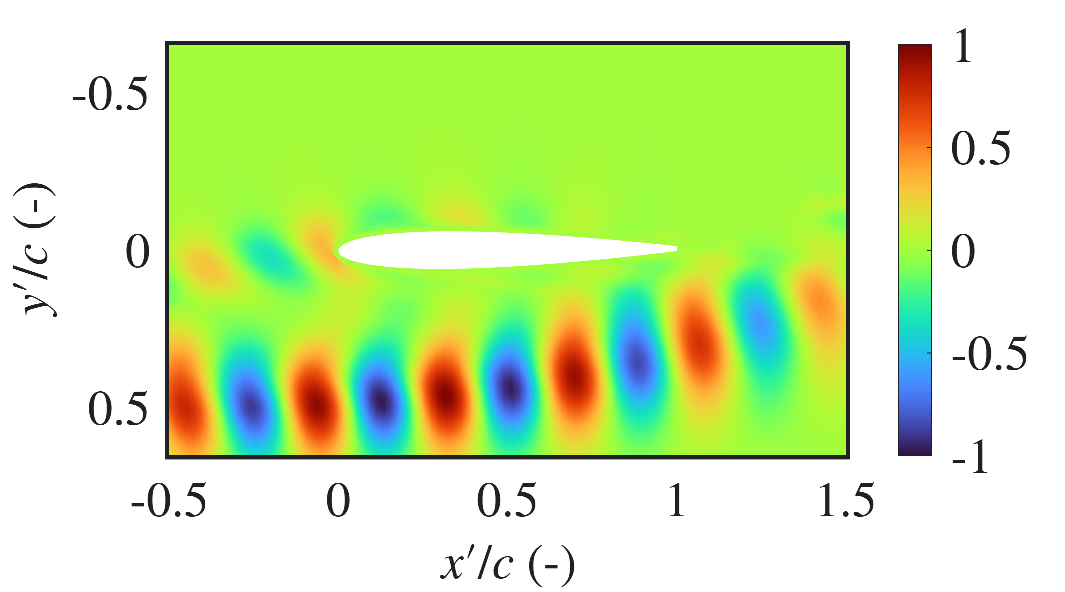}}
    \caption{2D SPOD mode shapes of the zero-mean residual upwash colored by the normalize modal value ($\phi/|\phi|$). Left and right figures display the $r/R = 0.90$ and $r/R = 0.95$ cross-sections, respectively. From top to bottom, the represented frequencies are $1.0$, $5.0$, and $7.0$ kHz.}
    \label{fig: 2D SPOD Mode Shapes}
\end{figure}

The first mode shapes for the $r/R = 0.90$ (left) and $0.95$ (right) radial sections are plotted in Fig. \ref{fig: 2D SPOD Mode Shapes}. The top panels correspond to the $1.0$~kHz mode, and the middle and bottom panels present the $5.0$ and $7.0$~kHz modes, respectively. Consistent with the coherence distribution, the dominant structure for the lower $1.0$~kHz mode appears as a relatively large-scale structure centered beneath the leading edge, in the same $y^\prime/c$ region as the first PTV core identified in the coherence hump of Fig.~\ref{subfig: Velocity Probe 90 low}. This indicates that the lower-frequency content and noise are primarily associated with the unsteady fluctuations within the PTV core rather than the secondary vortex braids. The higher-frequency modes, by contrast, exhibit smaller-scale structures distributed at various locations below the blade. In particular, Fig. \ref{subfig: 2D Mode roR090 f7000 m1} reveals four strips that align with the four humps in Fig. \ref{subfig: Velocity Probe 90 high}, while the dominant mode in Fig. \ref{subfig: 2D Mode roR095 f7000 m1} contains two distinct strips that align with the $r/R = 0.95$ velocity-pressure coherence in Fig. \ref{subfig: Velocity Probe 95 high}.

Notably, experiments by \citet{Wolf2019} showed that secondary vortices are highly intermittent and, at first glance, appeared to lack any consistent pattern. However, the dominant SPOD mode shapes in Fig. \ref{fig: 2D SPOD Mode Shapes} suggest that these structures nonetheless exhibit a degree of spatiotemporal coherence similar to the conceptual diagram presented in Fig. \ref{fig: BSVI Conceptual Diagram}, which is ultimately what produces the blade-to-blade correlation and acoustic tones of interest. Furthermore, the spatial regions occupied by these coherent structures also help explain why multiple humps appear in the squared-coherence distribution---a feature not immediately apparent from Fig. \ref{fig: Velocity Probe Coherence} alone, since it is not obvious why velocity sampled relatively far below the rotor should be coherent with the pressure fluctuation on the blade's upper surface. Given these novel findings, future experimental studies employing detailed time-resolved PIV and spatiotemporal analysis are needed to fully elucidate the nature of secondary vortices and BSVI. Such measurements would provide critical validation of the coherent flow structures identified here and further clarify their role in generating blade-to-blade correlation and the associated acoustic tones.

\section{Conclusions} \label{sec: Conclusions}

This study used HRLES coupled with an FW-H acoustic analogy to identify the source of the mid-frequency quasi-tonal noise reported by multiple small-scale hovering-rotor experiments, whose origin had remained uncertain. Applied to the four-bladed ITR of Pettingill et al. (2021), the present $2.5\%\,C_{\mathrm{tip}}$-resolution simulation reproduced the measured BPF-aligned harmonic tones and matched the one-third-octave-band SPL between 1.0 and 10.0 kHz to within the experimental uncertainty at four of the six far-field microphones, confirming that these tones arise from the rotor flow itself and not from enclosed-chamber wake recirculation.

Flow-field analysis identified \emph{blade secondary vortex interaction} (BSVI) as a significant noise-generating mechanism. Coherent, S-shaped secondary vortex structures form as the WSL is entrained into the PTVs of preceding blades, migrate upward, and are subsequently chopped by the following blades, producing stochastic loading over the outboard blade region ($r/R = 0.88$--$0.96$). Unlike conventional BWI involving more randomly distributed turbulence, these interactions occur with spatially organized, coherent vortical structures, producing aperiodic yet blade-to-blade correlated loading. The predicted far-field acoustic-pressure autocorrelation confirmed this behavior through peaks at time lags corresponding to the blade-passage interval. Moreover, these peaks decorrelated after a time lag of one full revolution due to the highly intermittent nature of secondary vortices. This aperiodicity and blade-to-blade correlation manifest as constructive interference at BPF harmonics and destructive interference in between, as highlighted in single-blade-versus-full-rotor acoustic comparisons.

Wall-pressure-spectrum, velocity-pressure coherence, and SPOD analyses of the residual flow field further identified two contributions within the mid-frequency band: a lower-frequency ($\sim 1$~kHz) component associated with stochastic fluctuations of the PTV core and a higher-frequency ($3$--$10$~kHz) component associated with the secondary vortices. The latter contribution is consistent with the BSVI mechanism identified here and differs from classical TI and BWI descriptions, which typically involve externally ingested or self-generated turbulent fluctuations rather than the organized secondary vortex structures analyzed in this rotor.


These findings support the characterization of BSVI as an aeroacoustic mechanism distinct from conventional BVI, BWI, and TI noise. Furthermore, the high-magnitude quasi-tones resulting from BSVI contributed substantially to the mid-to-high-frequency acoustic signature of the hovering rotor investigated here, coinciding with the frequency range where human hearing is most sensitive. Accurate prediction of BSVI requires scale-resolving CFD with sufficient spatial resolution to capture the secondary vortex structures and their interaction with the blades. Thus, an important direction for future work is the development of reduced-order or semi-analytical models capable of predicting BSVI noise at reduced computational cost. Additional high-fidelity computational and experimental studies across a wider range of rotor geometries, blade counts, and operating conditions---including collective pitch, rotation rate, and disk loading---would also help establish the parametric sensitivity of BSVI noise and clarify its relative importance compared with perpendicular BVI across the hovering-rotor design space relevant to eVTOL propulsors.

\begin{appendices}

\section{Acoustic Validation} \label{Appendix: Validation Cont.}

\begin{figure}
    \centering
    \subfloat[\label{subfig: }]{\includegraphics[width = 0.32\textwidth]{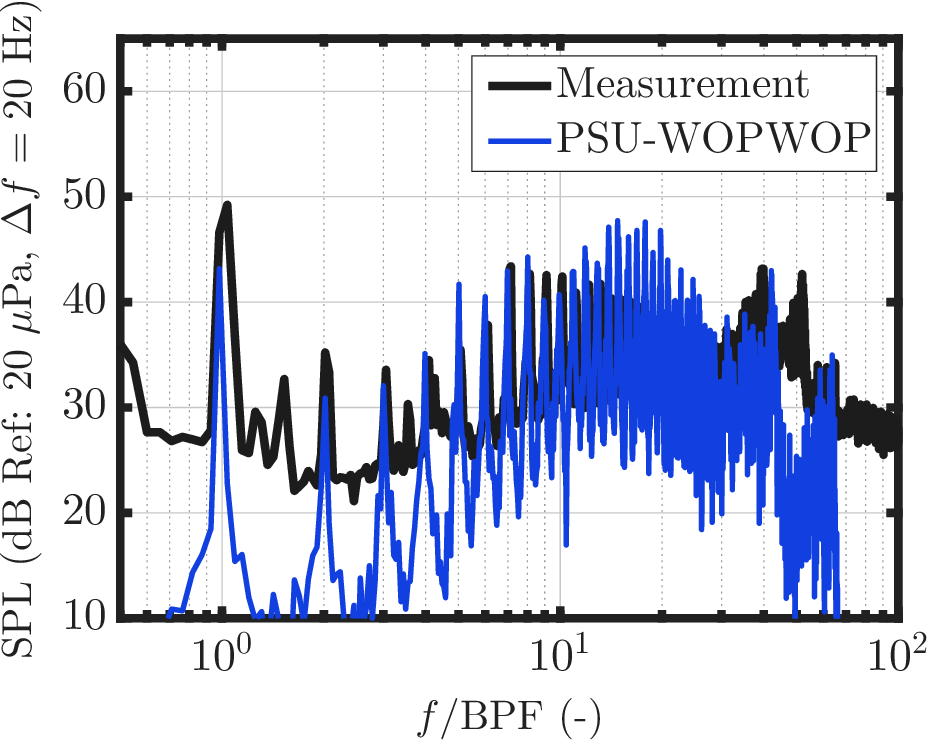}}
    \hspace{0.20cm}
    \subfloat[\label{subfig: }]{\includegraphics[width = 0.32\textwidth]{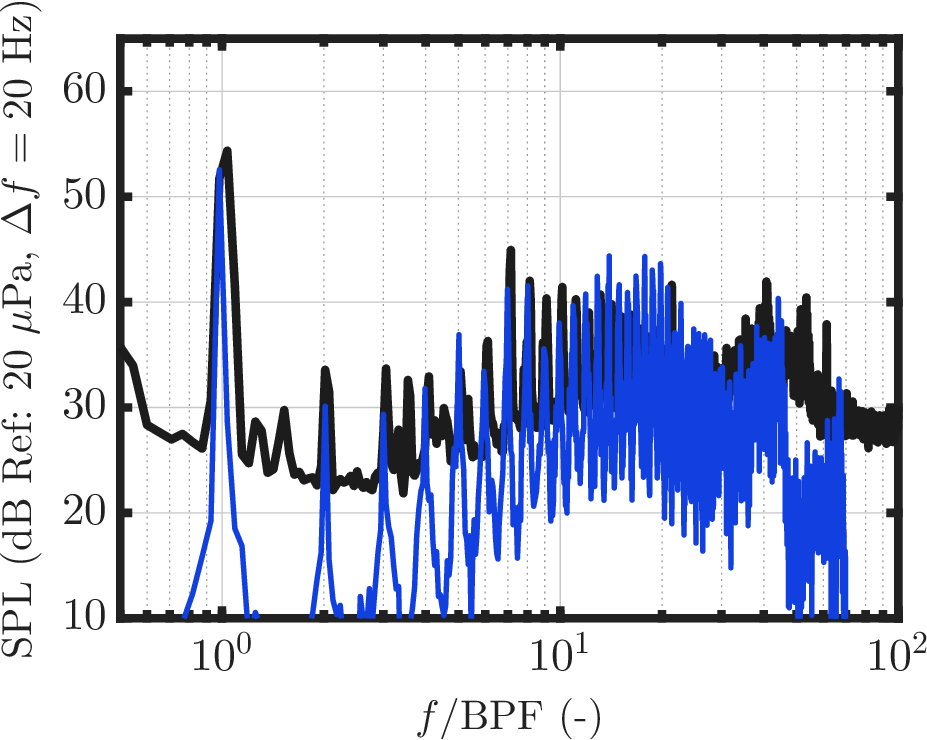}}
    \hspace{0.20cm}
    \subfloat[\label{subfig: narrowband cont. mic 3}]{\includegraphics[width = 0.32\textwidth]{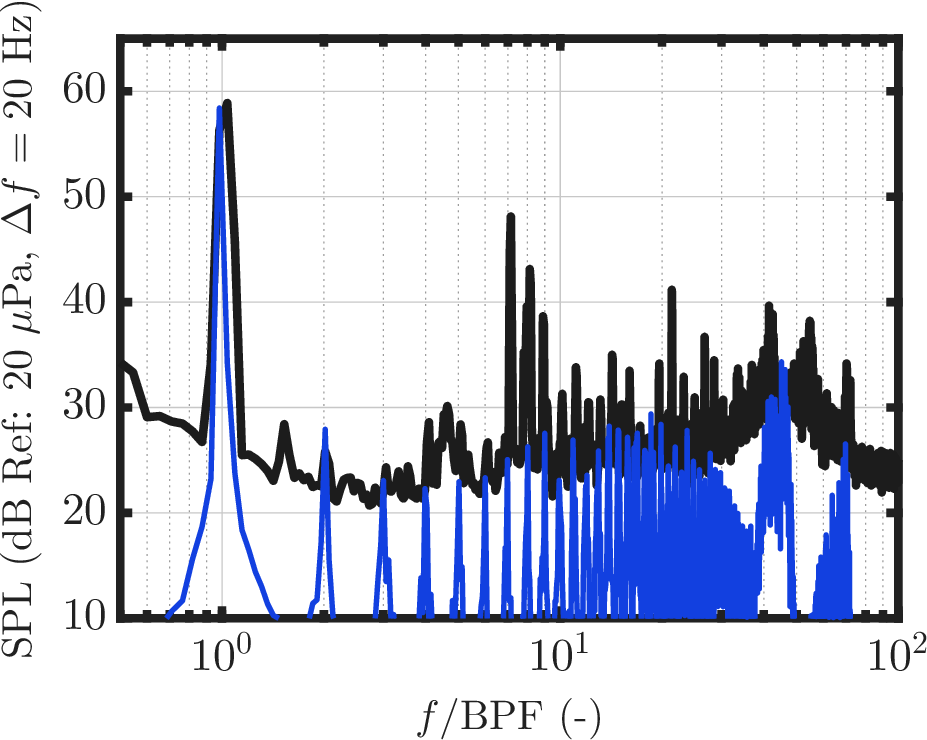}}

    \subfloat[\label{subfig: }]{\includegraphics[width = 0.32\textwidth]{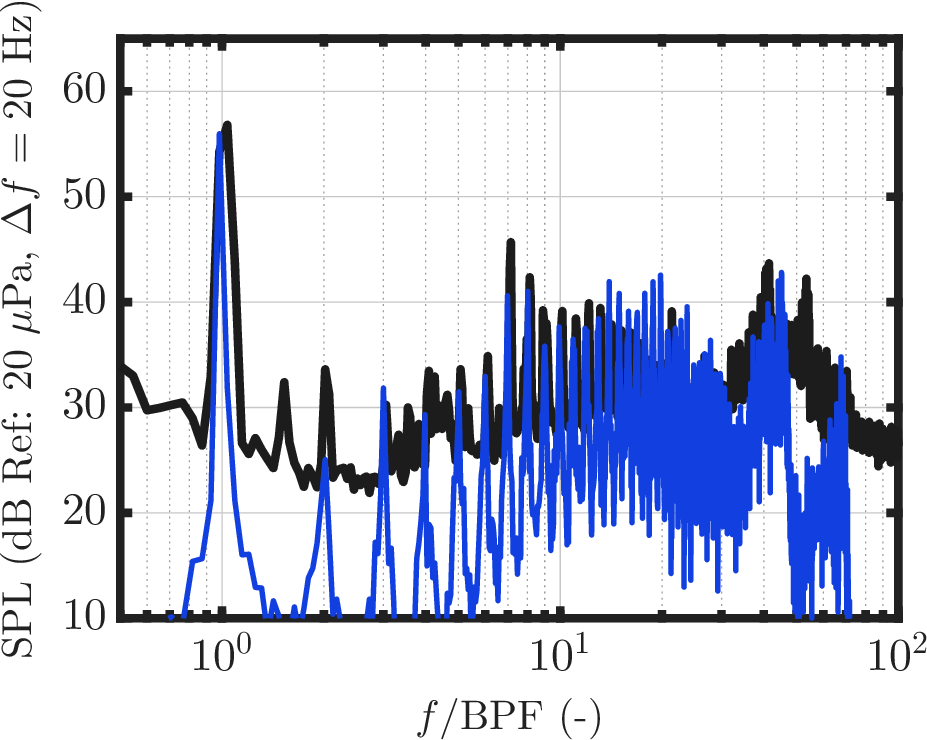}}
    \hspace{0.20cm}
    \subfloat[\label{subfig: }]{\includegraphics[width = 0.32\textwidth]{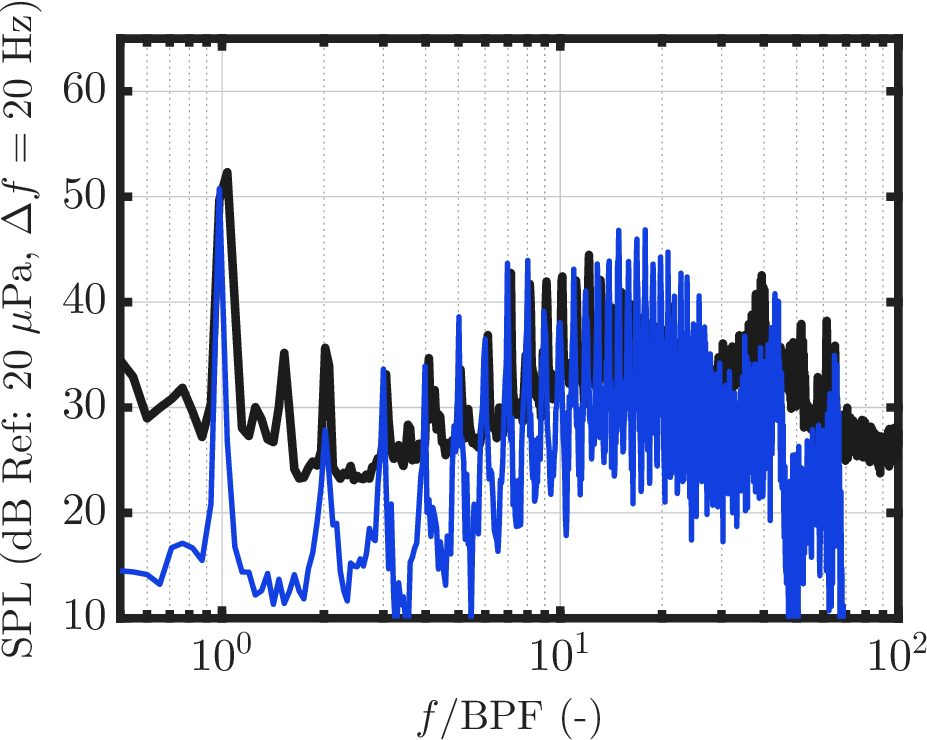}}
    \hspace{0.20cm}
    \subfloat[\label{subfig: }]{\includegraphics[width = 0.32\textwidth]{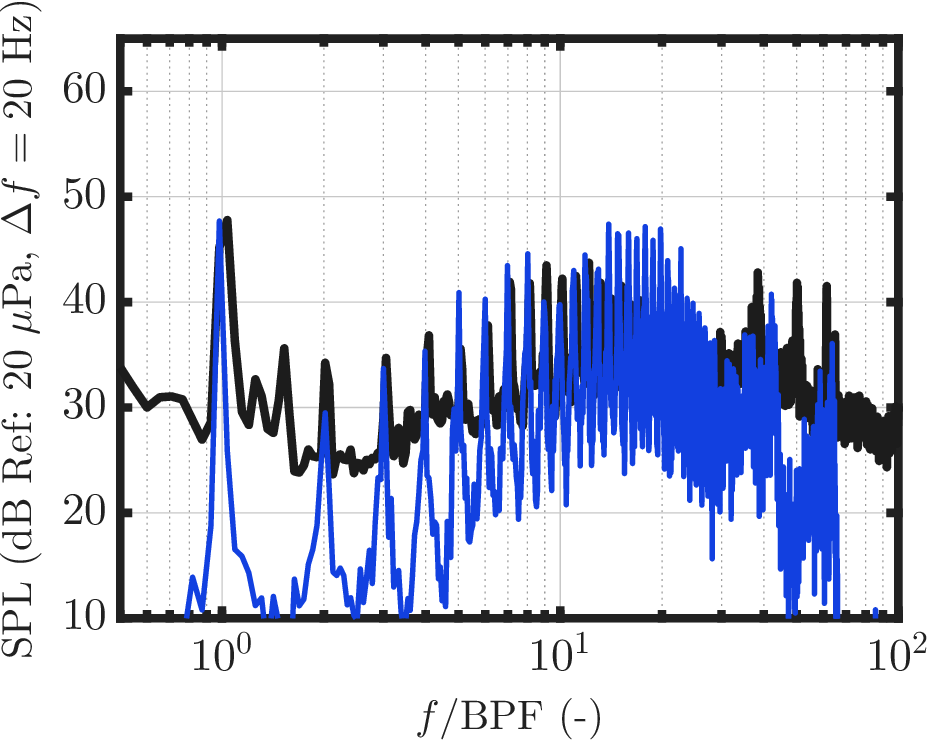}}

    \caption{Narrowband ($\Delta f = 20$~Hz) far-field acoustic spectra (PSU-WOPWOP only, no TBL-TE contribution) compared against experimental measurements at all six microphone locations: (a) mic. 1, (b) mic. 2, (c) mic. 3, (d) mic. 4, (e) mic. 5, (f) mic. 6.}
    \label{fig: narrowband validation cont.}
\end{figure}

Narrowband predictions from the HRLES/FW-H analysis are presented in Fig. \ref{fig: narrowband validation cont.} for all six microphone locations reported by \citet{pettingill2021}. Supplemental TBL-TE noise predictions from UCD-QuietFly are deliberately excluded here, highlighting the destructive interference---i.e., the valleys---associated with the blade-to-blade correlation discussed in Section~\ref{sec: Acoustics}. This quality of the HRLES/FW-H acoustic prediction is consistent with the analytical TI noise model of \citet{paterson&amiet1979}, who likewise demonstrated similar valleys that drastically underpredicted SPL while still predicting sharp quasi-tonal BPF peaks. The underprediction relative to experimental measurements is a result of the missing TBL-TE noise predictions. Future analytical work combining the intermittent upwash associated with secondary vortices with existing TI noise models may yield a less computationally intensive framework for modeling and designing rotors with reduced mid-to-high-frequency noise signatures.

At microphone location $3$, shown in Fig. \ref{subfig: narrowband cont. mic 3}, the predictions significantly underpredict the measured SPL, with the exception of the low-frequency peak associated primarily with first-BPF thickness noise. The exact source of this discrepancy is unclear, since this observer is positioned directly in the plane of the rotor disk and is therefore not expected to experience significant loading noise. The underprediction may indicate structural vibrations, blade bending,  or uncertainty in the microphone position during the experiments, any of which could explain why the nominally in-plane observer records elevated high-frequency noise levels. Motor noise from the experiments of \citet{pettingill2021} was also observed at these mid-frequencies; notably, the isolated motor noise measurements reported by \citet{pettingill2021} do not fully capture operational motor noise under load, since attaching the rotor blades and their rotation introduces system-wide vibrations and additional unaccounted-for noise. Consequently, despite the pronounced discrepancy at this specific observer location, the deviation is likely attributable to experimental artifacts related to this in-plane microphone placement. The strong agreement observed at all other microphone locations thus provides sufficient validation for the present HRLES approach.

\section{Coarse Grid Results} \label{Appendix: Coarse Grid HRLES.}

\begin{figure}
    \centering
    \includegraphics[width=0.80\linewidth]{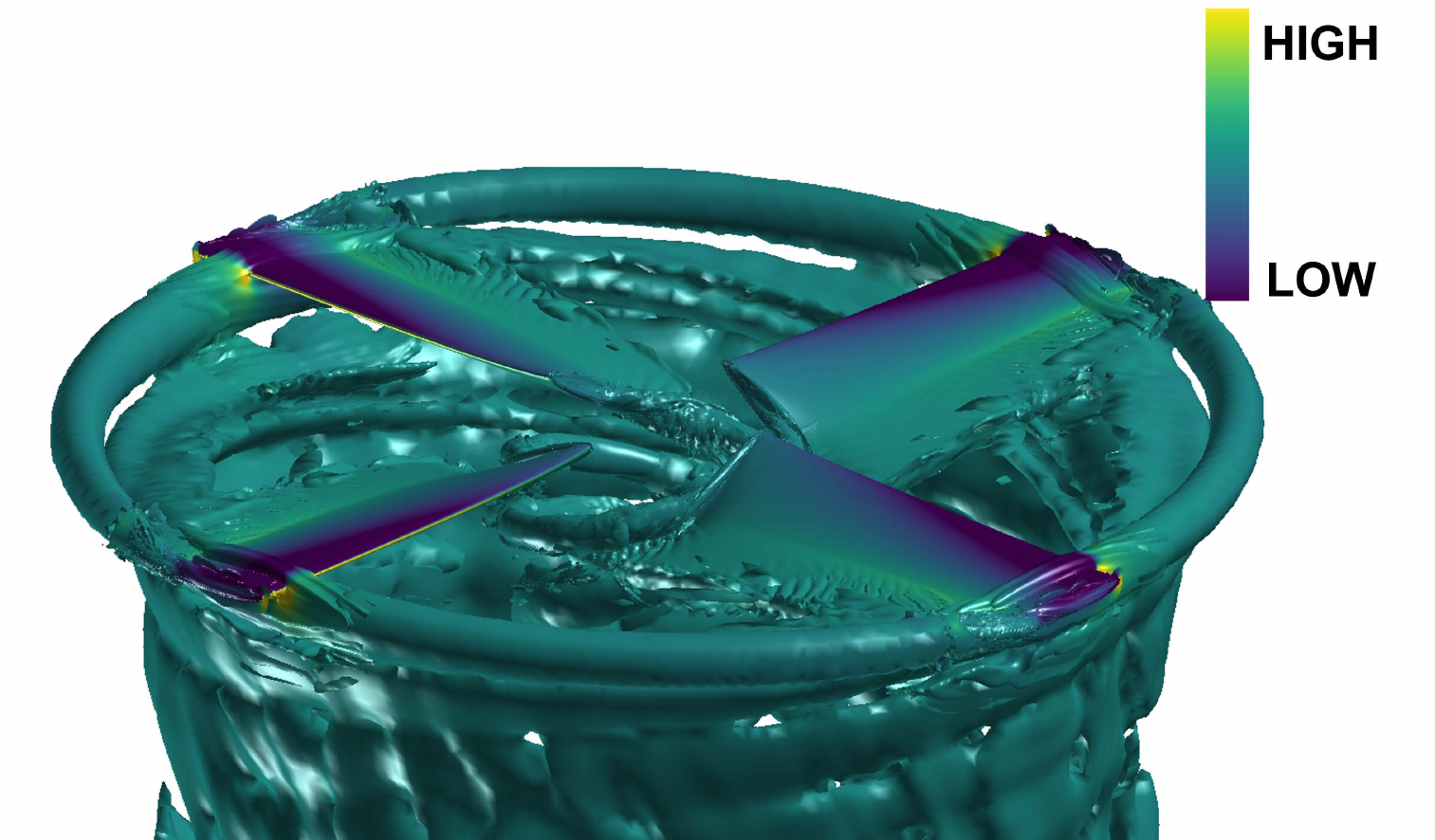}
    \caption{Iso-surface of vorticity magnitude ($|\omega| = 1000$ s$^{-1}$) colored by pressure coefficient for the coarse grid simulation, with color bar limits of -0.25 and 0.25.}
    \label{fig: ITR_Coarse_Grid_FlowVis}
\end{figure}

To assess the sensitivity of the BSVI noise mechanism to grid resolution, a supplementary HRLES computation is conducted on a coarsened OB grid using a target $10.0\%\,C_{\mathrm{tip}}$ spacing in the LES regions (in contrast to the $2.5\%\,C_{\mathrm{tip}}$ spacing used for the primary simulation of Section~\ref{sec: Methodology}). All other numerical settings are held fixed, including the fine NB grids, turbulence models, schemes, and the total number of revolutions. Additionally, \citet{Thurman2024} conducted a similar acoustic investigation using a target grid spacing of $5.0\%\,C_{\mathrm{tip}}$. Consistent with the findings of \citet{Chaderjian2023} and the grid-convergence trends reported by \citet{Bodling2024}, Fig.~\ref{fig: ITR_Coarse_Grid_FlowVis} reveals that the $10.0\%\,C_{\mathrm{tip}}$ grid spacing predicts larger, more diffuse PTVs and is unable to resolve the formation of the secondary vortex braids present in Fig.~\ref{fig: ITR Vorticity Iso Surface}. The absence of secondary vortices is attributed to premature WSL dissipation caused by the coarse grid spacing. Furthermore, the coarse grid is unable to resolve the small-scale fluctuations that drive vortex stretching, which ultimately leads to the formation of the distinct worm-like vortices referred to as secondary vortices. Importantly, \citet{Bodling2024} demonstrated agreement between HRLES predictions with a $3.0\%\,C_{\mathrm{tip}}$ grid spacing for the number of smaller-scale secondary vortices and the PIV measurements of \citet{Schwarz2022}. Therefore, the $2.5\%\,C_{\mathrm{tip}}$ grid spacing adopted in this study is considered sufficiently fine to accurately resolve these intermittent structures.

\begin{figure}
    \centering
    \includegraphics[width=0.95\linewidth]{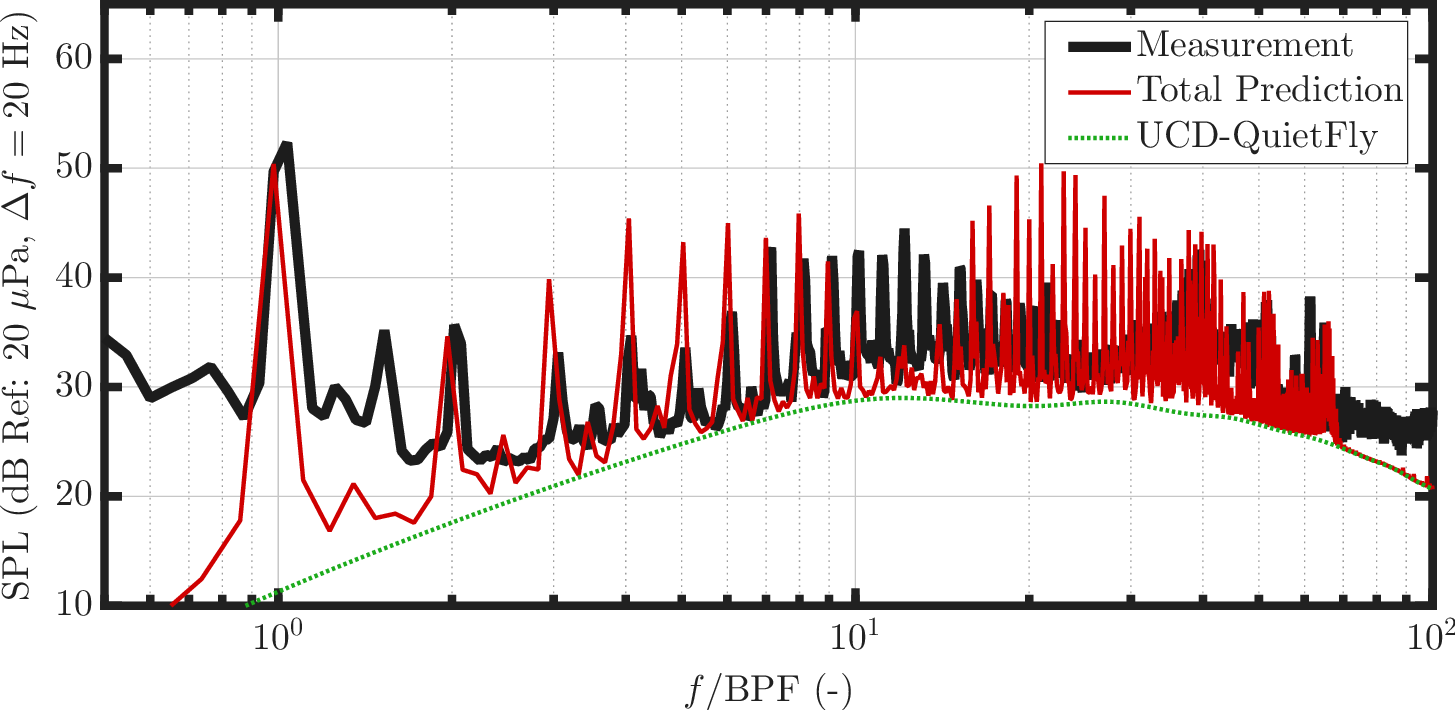}
    \caption{Narrowband ($\Delta f = 20$~Hz) far-field acoustic spectrum at microphone $5$: comparison of the combined coarse grid ($10.0\%\,C_{\mathrm{tip}}$ spacing) HRLES/FW-H and UCD-QuietFly total prediction against the experimental measurement of \citet{pettingill2021}.}
    \label{fig: ITR_Coarse_Grid_Acoustics}
\end{figure}

\begin{figure}
    \centering
    \includegraphics[width=0.95\linewidth]{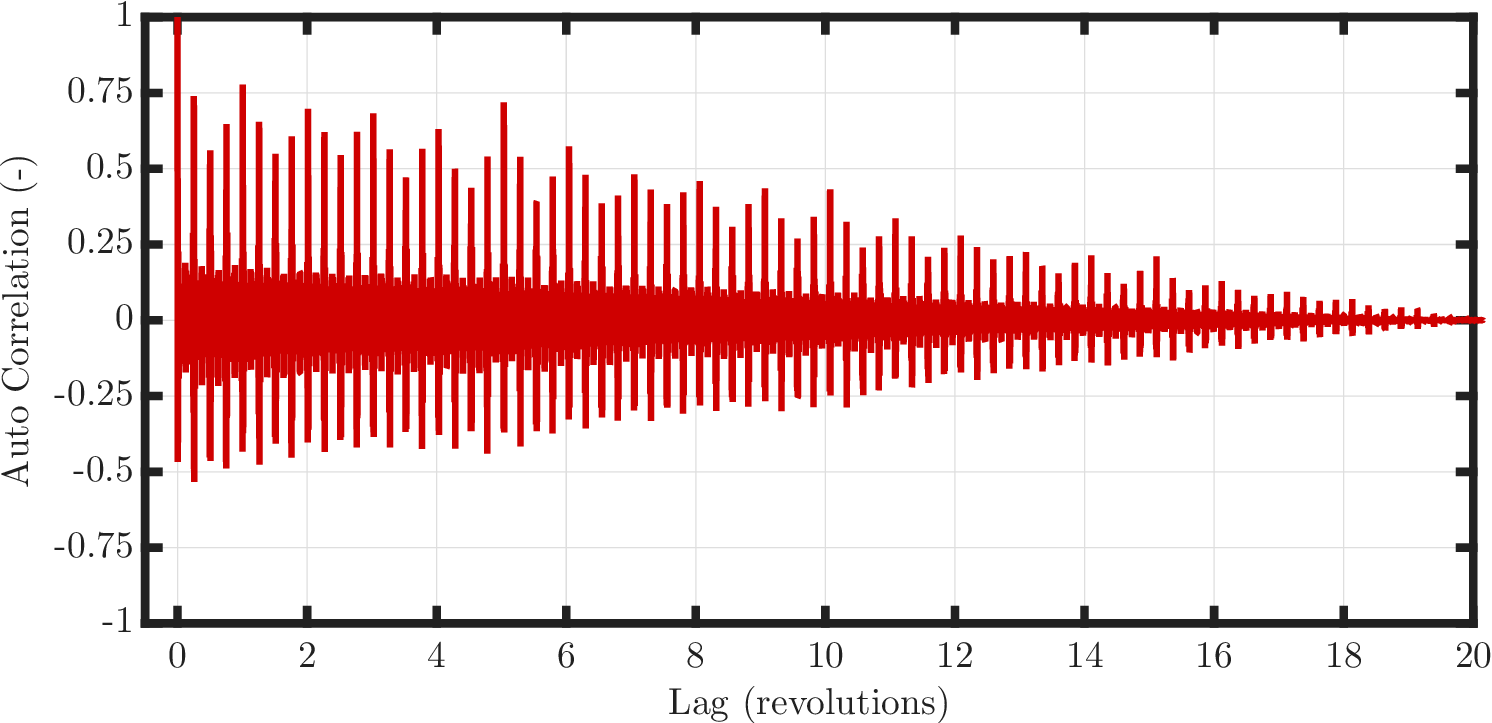}
    \caption{Autocorrelation of the coarse grid ($10.0\%\,C_{\mathrm{tip}}$ spacing) HRLES/FW-H far-field acoustic predictions for microphone location 5.}
    \label{fig: AcousticAutoCorrelation_Coarse}
\end{figure}

Figure~\ref{fig: ITR_Coarse_Grid_Acoustics} shows the microphone 5 FW-H predictions from the coarse HRLES calculations. Notably, mid-frequency peaks are still predicted, although BPFs 3--7 are overpredicted, BPFs 10--15 are drastically underpredicted, and non-physical tones extending up to the 40th harmonic overpredict the SPL relative to the measurements. The inability to predict the $10$--$15^\mathrm{th}$ BPFs is particularly interesting, as these tones correspond to the $3$--$5$~kHz frequencies associated with the secondary vortices discussed in Sec.~\ref{sec: Results}. However, what is most notable between the coarse- and fine-grid solutions is the blade-to-blade correlation demonstrated in Fig.~\ref{fig: AcousticAutoCorrelation_Coarse}. The coarse grid's autocorrelation exhibits peaks at every $90^\circ$ time lag across all revolutions, indicating a more periodic loading noise signature, explaining the mid-frequency BPF tones. However, this is significantly different from the blade-to-blade correlation observed in the fine HRLES solution (see Fig.~\ref{fig: AcousticAutoCorrelation}), which decorrelates after a single revolution. This difference highlights the unique intermittent nature of BSVI in Fig. \ref{fig: ITR Vorticity Iso Surface} when compared to the perpendicular BVI shown in Fig.~\ref{fig: ITR_Coarse_Grid_FlowVis}. Thus, despite the qualitative similarities observed in the coarse HRLES/FW-H acoustic spectrum---in particular, the mid-frequency quasi-tones---the underlying physical mechanism differs substantially from BSVI. 


\end{appendices}

\section{Acknowledgments}
This study was partially supported by the National Research Foundation of Korea (NRF) grant funded by the Korean Government (MSIT) (No.2021R1A5A1031868). The authors also gratefully acknowledge NASA Langley's Aeroacoustics Branch for providing experimental data on the ideally twisted rotor.

\section{Declaration of Interests}
The authors report no conflict of interest.

\bibliographystyle{jfm}
\bibliography{jfm}
\end{document}